\documentclass[12pt]{article}
\usepackage{ifpdf}
\usepackage{pgf}
\pgfdeclareimage{para}{fig-para}
\pgfdeclareimage{para-rechts}{fig-para-rechts}
\pgfdeclareimage{theta}{fig-theta}
\pgfdeclareimage{fixpot-s}{fig-fixpot-s}
\pgfdeclareimage{fixpot-s-1}{fig-fixpot-s-1}
\pgfdeclareimage{i-a}{fig-i-a}
\pgfdeclareimage{i-b}{fig-i-b}
\pgfdeclareimage{i-c}{fig-i-c}
\pgfdeclareimage{i-d}{fig-i-d}
\pgfdeclareimage{i-e}{fig-i-e}
\pgfdeclareimage{i-f}{fig-i-f}
\pgfdeclareimage{ii-a}{fig-ii-a}
\pgfdeclareimage{ii-b}{fig-ii-b}
\pgfdeclareimage{ii-c}{fig-ii-c}
\pgfdeclareimage{ii-d}{fig-ii-d}
\pgfdeclareimage{iii-a}{fig-iii-a}
\pgfdeclareimage{iii-b}{fig-iii-b}
\usepackage{caption}
\usepackage{amsmath}
\usepackage{amstext}
\usepackage{amsthm}
\usepackage{amssymb}
\usepackage{amsopn}
\usepackage{mathrsfs} 
\usepackage{dsfont} 
\usepackage[bottom]{footmisc}
\usepackage{indentfirst}
\usepackage{upref}
\usepackage{cite}
\usepackage{units}
\usepackage{fancybox}
\usepackage
{floatflt}
\numberwithin{equation}{section}

\renewcommand{\baselinestretch}{1.409}
\DeclareMathOperator{\tr}{Tr}
\DeclareMathOperator{\ee}{e}
\DeclareMathOperator{\re}{Re}
\DeclareMathOperator{\im}{Im}
\DeclareMathOperator{\sign}{sign}
\DeclareMathOperator{\dimension}{dim}
\newcommand{\chib}{\chi_{\text{B}}}
\newcommand{\etan}{\eta_{\text{N}}}
\newcommand{\ki}{k_{\text{intl}}}
\newcommand{\kt}{k_{\text{term}}}
\newcommand{\kT}{k_{\text{T}}}
\newcommand{\cc}{^{\text{\tiny CREH}}}
\newcommand{\gf}{^{\text{\tiny GFP}}}

\newcommand{\bcc}{^{\text{\,{\tiny CREH}}}}
\newcommand{\rr}{^{\text{reg}}}
\newcommand{\sr}{^{\text{sing}}}
\newcommand{\dd}{\mathrm{d}}
\newcommand{\p}{\partial}
\newcommand{\h}{\widehat}
\newcommand{\w}{\widetilde}
\newcommand{\ov}{\overline}
\newcommand{\order}[1]{\mathcal{O} (#1)}
\newcommand{\vare}{\varepsilon}
\newcommand{\U}{\Upsilon}
\newcommand{\rhot}{\w \rho}
\newcommand{\rhoe}{\rho_{\text{eff}}}
\begin{document}
\begin{titlepage} 
\renewcommand{\baselinestretch}{1.1}
\small\normalsize
\begin{flushright}
MZ-TH/08-12
\end{flushright}

\vspace{0.1cm}

\begin{center}   

{\Large \textbf{Conformal sector of Quantum Einstein Gravity \\in the local potential approximation: \\non-Gaussian fixed point and a phase of \\[3mm]unbroken diffeomorphism invariance
}}

\vspace{1.4cm}
{\large M.~Reuter and H.~Weyer}\\

\vspace{0.7cm}
\noindent
\textit{Institute of Physics, University of Mainz\\
Staudingerweg 7, D--55099 Mainz, Germany}\\

\end{center}

\vspace*{0.6cm}
\begin{abstract}
\vspace{12pt}
We explore the nonperturbative renormalization group flow of Quantum Einstein Gravity (QEG) on an infinite dimensional theory space. We consider ``conformally reduced'' gravity where only fluctuations of the conformal factor are quantized and employ the Local Potential Approximation for its effective average action. The requirement of ``background independence'' in quantum gravity entails a partial differential equation governing the scale dependence of the potential for the conformal factor which differs significantly from that of a scalar matter field. In the infinite dimensional space of potential functions we find a Gaussian as well as a non-Gaussian fixed point which provides further evidence for the viability of the asymptotic safety scenario. The analog of the invariant cubic in the curvature which spoils perturbative renormalizability is seen to be unproblematic for the asymptotic safety of the conformally reduced theory. The scaling fields and dimensions of both fixed points are obtained explicitly and possible implications for the predictivity of the theory are discussed. Spacetime manifolds with $R^d$ as well as $S^d$ topology are considered. Solving the flow equation for the potential numerically we obtain examples of renormalization group trajectories inside the ultraviolet critical surface of the non-Gaussian fixed point. The quantum theories based upon some of them show a phase transition from the familiar (low energy) phase of gravity with spontaneously broken diffeomorphism invariance to a new phase of unbroken diffeomorphism invariance; the latter phase is characterized by a vanishing expectation value of the metric.
\end{abstract}
\end{titlepage}
%
%
%
%
%
%
\section{Introduction}\label{s1}
The formulation of a mathematically consistent quantum theory of gravity whose predictive power extends down to arbitrarily small distances continues to be one of the major challenges of modern theoretical physics \cite{kiefer}. Even though the first attempts at quantizing the gravitational field date back to the 1930s \cite{bron} the various approaches available today are still plagued by considerable technical and conceptual problems. On the conceptual side, the ``background independence'' \cite{A,R,T} of classical General Relativity  and of the yet to be constructed theory of quantum gravity is at the heart of many difficulties. In the asymptotic safety approach to quantum gravity \cite{wein,mr,percadou,oliver1,frank1,oliver2,oliver3,oliver4,souma,frank2,prop,oliverbook,perper1,codello,litimgrav,frankmach,creh1,oliverfrac,jan1,jan2,je1,neuge,max,livrev} the gravitational degrees of freedom are described by a metric $g_{\mu \nu}$, as in General Relativity, and the standard rules of quantum mechanics are applied to it. Nevertheless, because of the requirement of ``background independence'', the resulting quantum field theory of gravity has many features which are rather different from the familiar matter field theories on a fixed, non-dynamical spacetime. These unusual features emerge because the theory is supposed to dynamically explain the origin of the spacetime it ``lives'' on as a kind of ground state property, and as a result, in its construction no metric should be distinguished from any other a priori. Most investigations of the asymptotic safety scenario are based upon the effective average action for gravity \cite{mr}. In this approach the requirement of ``background independence'' in the above sense of the word is met, somewhat paradoxically, by means of the background field technique \cite{back}. Even though, formally, the gravitational average action obtains by a similar Wilsonian mode cutoff and coarse graining operation as its counterpart for ordinary matter field theories \cite{avact,ym,avactrev,ymrev} there is a conceptually crucial difference: The metric determines all physical (proper) length and mass scales; in particular it fixes the physical coarse graining scale which is to be attributed to the cutoff parameter $k$ built into the average action $\Gamma_k$. Hence in quantum gravity the problem is that the field that is ``averaged'' or ``coarse grained'' by itself has to define the physical meaning of this coarse graining operation. If one was to use a rigid metric for this purpose the parameter $k$ would have no physical interpretation; hence $\Gamma_k$ hardly could be used for ``renormalization group improving'' classical solutions or field equations \cite{bh,erick1,cosmo1,cosmofrank,cosmo2,entropy,esposito,h1,h2,h3,girelli,litim,mof}, for instance.

In a recent paper \cite{creh1}, henceforth referred to as [\,I\,], we explained in detail how the use of the background field technique, besides making $\Gamma_k$ a diffeomorphism invariant functional, can make the average action and its renormalization group (RG) flow ``background independent''\footnote{Here and in the following the term ``background independent'', put in quotation marks, means the absence of a preferred metric; in this sense the term is frequently used in loop quantum gravity \cite{A,R,T} or in the dynamical triangulation approach \cite{ajl1,ajl2,ajl34}, for instance. Referring to the background field formalism, no quotation marks will be put.} in the sense that nowhere in the construction any preferred (rigid) metric occurs. In order to disentangle this problem from the gauge fixing issues which are the usual motivation for the background method we considered the ``conformal reduction'' of Quantum Einstein Gravity (QEG) in which only the conformal factor of the metric is quantized and all other degrees of freedom are discarded. Performing the conformal reduction on top of the usual Einstein--Hilbert truncation one is led to quantize an action of the familiar $\phi^4$-type. As we saw in [\,I\,] the RG flow in this conformally reduced Einstein--Hilbert (or ``CREH'') truncation is different from that in a standard scalar matter field theory, the reason being that in the gravitational case the mode suppression term $\exp (- \Delta_k S)$ in the functional integral defining $\Gamma_k$ must have a specific background field dependence in order to comply with the requirement of ``background independence''. This extra field dependence of $\Delta_k S$ is absent in the matter field theory. As a consequence, gravity in the CREH approximation has an RG flow which is very different from that of a standard $\phi^4$-theory. Including the extra field dependence turns the merely logarithmic running of the cosmological constant (earlier discussed by Polyakov \cite{polyakov} and by Jackiw et al.\ \cite{jackiw}) into the expected $k^4$-running, for instance. (Similar observations had been made by Floreanini and Percacci \cite{floper} in a different theory.)

Remarkably, the RG flow of the CREH approximation turned out to be qualitatively identical to the full Einstein--Hilbert flow where all degrees of freedom are quantized. In particular it has a non-Gaussian fixed point (NGFP) suitable for the asymptotic safety construction. The counterpart of the CREH approximation on the side of the standard matter theories is Symanzik's asymptotically free $\phi^4$-theory with a negative quartic coupling \cite{syman,hist}. This theory has a potential which is unbounded below, and it is notorious for the resulting infrared instability. In the context of QEG this instability is not an issue, however. In the full theory it is presumably cured by higher derivative terms. In fact, it is known \cite{oliver2} that a $\int \! \sqrt{g\,} \, R^2$-term, added to the Einstein--Hilbert action, stabilizes the kinetic term of the conformal factor at high scales; cf.\ also ref.\ \cite{kincond}.

The upshot of the analysis in [\,I\,] was that the quantum fluctuations of the conformal factor alone seem to be typical of the full set of degrees of freedom in $g_{\mu \nu}$ as far as their impact on the RG flow is concerned. This suggests that also in more general truncations the conformal factor should play a representative role\footnote{For earlier work in a similar spirit see ref.\ \cite{narpad}.}. Since the scalar--like reduced theory is technically much simpler than full-fledged QEG it suggests itself as a tool which should allow us to probe regions of theory space which are too difficult computationally in the full theory.

In the present paper we start this program with the local potential approximation, or ``LPA'', which includes arbitrary non-derivative terms in the truncation ansatz. In fact, the CREH truncation (in $4$ Euclidean dimensions) obtains by inserting metrics of the form $g_{\mu \nu} = \phi^2 (x) \,\, \h g_{\mu \nu}$ into the Einstein--Hilbert action. Here $\h g_{\mu \nu}$ is a fixed reference metric. The result is a kinetic term for $\phi$ which, apart from its sign, is of the standard form $\propto \h g^{\,\mu \nu} \, \p_\mu \phi \, \p_\nu \phi$, as well as a potential with two terms, proportional to $\phi^2$ and $\phi^4$, respectively:
\begin{align}\label{1.1}
U_k\cc (\phi)
& =
- \frac{3}{4 \pi \, G_k} \,
\left( \tfrac{1}{12} \, \h R \, \phi^2 
- \tfrac{1}{6} \, \Lambda_k \, \phi^4 \right).
\end{align}
(Here $\h R$ is the curvature scalar of $\h g_{\mu \nu}$.) The idea of the LPA \cite{avactrev} is to retain the standard kinetic term from the CREH approximation but to allow for arbitrary non-derivative terms:
\begin{align}\label{1.2}
\Gamma_k [\phi=const]
& =
+ \int \!\!\dd^4 x \, \sqrt{\h g\,}~U_k (\phi).
\end{align}
The running potential $U_k (\phi)$ may develop any $\phi$-dependence; it is no longer restricted to be a quartic polynomial. The LPA has played an important role in the RG analysis of standard scalar theories, in particular in the context of spontaneous symmetry breaking and the approach to convexity \cite{avactrev}. It is therefore the obvious next step in the analysis of conformally reduced QEG. The relevant theory space is infinite dimensional now; the corresponding flow equations will include a partial differential equation for $U_k (\,\cdot\,)$.

Clearly the LPA cannot be a numerically precise approximation to the full theory, even if the fluctuations of $\phi$ have always a ``typical'' impact on the RG flow. Nevertheless it should provide us with some qualitative understanding and conceptual insights into the general properties of QEG.

The (relative!) technical simplicity of the reduced theory comes at a certain price, however. In general it will not be possible to relate terms in $U_k (\phi)$ to invariants $I [g_{\mu \nu}]$ depending on the full metric in a unique way. Typically there will be many invariants contributing to a given term in the potential when we insert $g_{\mu \nu} = \phi^2 \,\, \h g_{\mu \nu}$. Nevertheless, the LPA is sensitive to a certain projection (or ``shadow'') of the invariants $I [g_{\mu \nu}]$ which make up the general $\Gamma_k [g_{\mu \nu}]$. If $\h g_{\mu \nu}$ is the metric on a curved space, $U_k (\phi)$ encodes information about higher derivative terms, for instance. If we consider a series expansion in the curvature scalar, for instance,
\begin{align}\label{1.3}
\Gamma_k [g_{\mu \nu}]
& =
\sum_n c_n (k) \, \int \!\! \dd^4 x \, \sqrt{g\,}~R(g)^n
\end{align}
and choose $\h g_{\mu \nu}$ to be the metric on a unit $4$-sphere, then the corresponding potential $\Gamma_k [\phi=const] \, / \!\int \dd^4 x \, \sqrt{\h g\,}$ reads
\begin{align}\label{1.4}
U_k (\phi)
& =
\sum_n (12)^n \, c_n (k) \, \phi^{4-2n}.
\end{align}
The constant term in \eqref{1.4}, $n=2$, descends from the invariant $\int \dd^4 x \sqrt{g\,} ~R^2$, but also $\int \dd^4 x \sqrt{g\,} ~(R_{\mu \nu})^2$ or $\int \dd^4 x \sqrt{g\,} ~(R_{\mu \nu \sigma \rho})^2$ would contribute to the constant piece in $U_k (\phi)$, as would many non-local terms. Similarly, the monomial $\phi^{-2}$ receives contributions from $R^3$ and the Goroff--Sagnotti term $\int \dd^4 x \, \sqrt{g\,}~R_{\mu \nu \alpha \beta} \, R^{\alpha \beta}_{\phantom{\alpha \beta}\rho \sigma} \, R^{\rho \sigma \mu \nu}$ \cite{sagnotti}, among others.

Despite these limitations the experience with conventional field theories suggests that the LPA is a powerful tool for a first exploration of regions in theory space which are computationally inaccessible otherwise. In the present paper we shall employ it in order to analyze and illustrate several conceptual issues related to the renormalization group flow of $\Gamma_k$ which are as to yet too computationally demanding to be studied in the full theory. In particular we focus on the following three topics.
\paragraph*{(a)}
In the LPA we have the possibility to obtain information about the non-Gaussian fixed point of the gravitational average action on an \textit{infinite dimensional theory space}. We shall find that the RG flow on the space of (dimensionless) potential functions and Newton's constant admits both a Gaussian and non-Gaussian fixed point. We compute the fixed point potentials and perform a linear stability analysis in each case. The scaling dimensions of the Gaussian fixed point (GFP) will be seen to characterize the difference between a standard scalar and the conformal factor in a particularly clearcut way. The scaling fields and dimensions (critical exponents $\theta$) of the NGFP are of interest for the asymptotic safety construction.

The relevant scaling fields, those with $\re \theta >0$, decay for $k \to \infty$, while the irrelevant ones, with $\re \theta <0$, blow up in this limit. An asymptotically safe quantum theory is specified by any RG trajectory that hits the NGFP for $k \to \infty$. We define the \textit{ultraviolet (UV) critical manifold} $\mathscr{S}_{\text{UV}}$ to be the set of all points in theory space which are pulled into the NGFP under the inverse RG flow (i.\,e., for increasing $k$). In the simplest case $\mathscr{S}_{\text{UV}}$ is a smooth submanifold of theory space. Every complete RG trajectory which lies inside $\mathscr{S}_{\text{UV}}$ defines a possible quantum theory. Hence, with $\Delta_{\text{UV}} \equiv \dimension \mathscr{S}_{\text{UV}}$ the dimensionality of the critical manifold, there exists a $\Delta_{\text{UV}}$-parameter family of asymptotically safe quantum field theories; which one is actually realized in Nature must be decided by an experimental determination of exactly those $\Delta_{\text{UV}}$-parameters
whose values are not predicted by theory. If $\mathscr{S}_{\text{UV}}$ is indeed a smooth manifold, and the NGFP has no marginal scaling fields ($\re \theta =0$), then $\Delta_{\text{UV}}$ equals the dimensionality of a subspace of the tangent space to $\mathscr{S}_{\text{UV}}$ at the NGFP, namely the one spanned by the relevant scaling fields, i.\,e.\ those whose critical exponents $\theta$ have a positive real part. Therefore the number of relevant scaling fields is related to the degree of predictivity that can be achieved. (See \cite{livrev} for a more general discussion.)

Within the LPA we will be able to study $\mathscr{S}_{\text{UV}}$ near the NGFP in great detail. We shall find that the value of $\Delta_{\text{UV}}$ depends sensitively on the space of functions in which allowed scaling fields are supposed to ``live''. It turns out that in principle there could be infinitely many relevant ones, but as we shall see this question cannot be decided on the basis of the LPA alone.
\paragraph*{(b)}
Using numerical methods we shall analyze the RG evolution of $U_k (\phi)$ also in the nonlinear regime. We are particularly interested in phase transitions, triggered by a variation of $k$, in which the global minimum of $U_k (\phi)$ either jumps, or continuously evolves from a value $\phi \neq 0$ to $\phi =0$. We interpret this as a transition from the familiar phase of gravity with broken diffeomorphism invariance to a \textit{phase with unbroken diffeomorphism invariance}.

We are considering metrics of the form $g_{\mu \nu} = \phi^2 \,\, \h g_{\mu \nu}$ where $\phi$ is the expectation value of the quantum conformal factor and $\h g_{\mu \nu}$ is a non-degenerate classical reference metric. If the expectation value $\phi$, interpreted here as the minimum of $U_k$, is nonzero the expectation value of the metric, $g_{\mu \nu} \equiv \langle \gamma_{\mu \nu} \rangle$, is non-degenerate as well. But if $\phi =0$ then the metric has a vanishing expectation value. (More precisely, in view of $g_{\mu \nu} = e^a_\mu \, e^b_\nu \, \delta_{ab}$ and since $g_{\mu \nu} = \phi^2 \,\, \h g_{\mu \nu}$ depends on $\phi$ quadratically, it is the vielbein $e^a_\mu$ that has a vanishing expectation value.)

The situation where $\phi =0$ can be thought of as the restoration of a spontaneously broken symmetry \cite{witten,floper}. The relevant local symmetries are the spacetime diffeomorphisms. If the quantum state is such that $\langle \gamma_{\mu \nu} \rangle \neq 0$ the group of diffeomorphisms is spontaneously broken down to the stability group of $\langle \gamma_{\mu \nu} \rangle$. If $\langle \gamma_{\mu \nu} \rangle = \eta_{\mu \nu}$, say, this stability group is the Poincar\'e group. The broken phase of gravity is the one we are familiar with. It is conceivable though that the symmetry breaking effectively disappears at very small distances. This is exactly what happens in spontaneously broken Yang--Mills theories, in the standard model, say. If we study phenomena involving momentum scales far above the W-mass, the full $SU(2) \times U(1)$ symmetry appears unbroken. Technically this symmetry restoration can be inferred from the running Higgs potential $U_k^{\text{Higgs}}$ at $k \gg M_{\text{W}}$.

We are going to analyze the running potential of the conformal factor in the same spirit. We shall find that typical trajectories in $\mathscr{S}_{\text{UV}}$, for $k$ sufficiently large, describe a phase of unbroken diffeomorphism invariance in the sense that the metric has a vanishing expectation value there.
\paragraph*{(c)}
While the beta functions implied by the functional RG equations do not refer to any special metric they do depend on the \textit{topology} of the manifold which carries the metrics $\gamma_{\mu \nu}$, $g_{\mu \nu}$, $\ov{g}_{\mu\nu}$, and $\h g_{\mu \nu}$. We illustrate this phenomenon by discussing in parallel the two cases where the topology of the (Euclidean) spacetime manifold is that of a sphere $S^4$ and of flat space $R^4$, respectively. The beta functions (or rather, functionals) of $U_k (\,\cdot\,)$ are different in the two cases since they depend on the spectrum of the Laplace--Beltrami operator built from the corresponding $\h g_{\mu \nu}$.

The remaining sections of this paper are organized as follows. In Section \ref{s2} we review and expand on conformally reduced QEG; in particular we show how the background field method achieves ``background independence'', and we introduce two types of background--quantum field split symmetries. The LPA is introduced in Section \ref{s3} then and the corresponding flow equations are derived for the $S^4$ and the $R^4$ topologies. In Section \ref{s3s} we analyze various general properties of the LPA flow, and Section \ref{s4} is devoted to its fixed points; for both the Gaussian and the non-Gaussian fixed point we determine the fixed point potential and perform a linear stability analysis in order to determine the local tangent space to the UV critical manifold. In Section \ref{s5} we solve the partial differential equation for the running potential numerically and describe various scenarios for the phase transition to a phase of (conformally reduced) QEG with unbroken diffeomorphism invariance. The Conclusions are given in Section \ref{s6}.
%
%
%
%
%
%
\section{``Background Independence'' via Background Fields}\label{s2}
In the construction of the full-fledged gravitational average action \cite{mr} there are three different metrics that play a role: the microscopic metric, i.\,e.\ the integration variable in the path integral, $\gamma_{\mu \nu}$, its expectation value $g_{\mu \nu} \equiv \langle \gamma_{ \mu \nu} \rangle$, and the background metric $\ov{g}_{\mu \nu}$. In conformally reduced gravity [\,I\,] all three of them are assumed conformal to the same, fixed reference metric $\h g_{\mu \nu}$. We write the conformal factor of $\gamma_{\mu \nu}$ and $\ov{g}_{\mu \nu}$ as $\chi^{2 \nu} (x)$ and $\chib^{2\nu} (x)$ where, in $d$ dimensions, $\nu \equiv 2 / (d-2)$. The motivation for the exponent $2 \nu$ is that in this way the Einstein--Hilbert action provides a standard kinetic term for $\chi$.

In the approach of [\,I\,] the original path integral over $\gamma_{\mu \nu}$ is reduced to an integral over $\chi (x)$ only, and the resulting scalar--like theory is then reformulated using the background field method. The ``microscopic'' conformal factor $\chi$ is decomposed as $\chi \equiv \chib + f$ and the $\chi$-integral is replaced by an integral over $f$. We denote the expectation values of $f$ and $\chi$ by $\ov{f} \equiv \langle f \rangle$ and 
\begin{align}\label{2.20}
\phi & \equiv \langle \chi \rangle = \chib + \ov{f},
\end{align}
respectively. Thus the following metrics are to be distinguished:
\begin{align}
\label{2.21}
\gamma_{\mu \nu}
& =
\chi^{2 \nu} \,\, \h g_{\mu \nu}
= (\chib + f)^{2 \nu} \,\,  \h g_{\mu \nu}
\\
\label{2.22}
\ov{g}_{\mu\nu}
& =
\chib^{2 \nu} \,\, \h g_{\mu \nu}
\\
\label{2.23}
g_{\mu \nu}
& =
\left\langle (\chib + f)^{2\nu} \right\rangle \,\, \h g_{\mu \nu}
\\
\label{2.24}
\breve g_{\mu \nu} 
& =
\phi^{2\nu} \,\, \h g_{\mu \nu}
= \bigl( \chib + \langle f \rangle \bigr)^{2\nu} \,\, \h g_{\mu \nu}.
\end{align}
Since $2\nu \neq 1$ in general, we are performing a \textit{nonlinear} background split here which causes the fields $g_{\mu \nu}$ and $\breve g_{\mu \nu}$ to be different. In this respect the correspondence between the full and the conformally reduced theory is not perfect. In the full theory one performs the \textit{linear} background split $\gamma_{\mu \nu} = \ov{g}_{\mu \nu} + h_{\mu \nu}$, integrates over $h_{\mu \nu}$, and defines $\ov{h}_{\mu \nu} \equiv \langle h_{\mu \nu} \rangle$. In the latter case there is no distinction between $g_{\mu \nu}$ and $\breve g_{\mu\nu}$:
\begin{align}\label{2.25}
g_{\mu \nu} \equiv \langle \gamma_{\mu \nu} \rangle
= \ov{g}_{\mu \nu} + \langle h_{\mu \nu} \rangle
\equiv \breve g_{\mu \nu}.
\end{align}

In [\,I\,] we explained in detail why, both in the full and the reduced theory, the background field is indispensable if one wants to implement ``background independence'' in the sense of \cite{A,R,T} and give a physical interpretation to the RG parameter $k$. One expands the path integral in eigenmodes of the covariant Laplacian $\ov{\Box}$ which is constructed from $\ov{g}_{\mu \nu}$ and then introduces $k$ as a cutoff in the spectrum of $\ov{\Box}$. In this manner the typical length scale displayed by the $\ov{\Box}$-eigenmode with eigenvalue $-k^2$, the ``last'' mode integrated out, is an approximate measure for the proper (w.\,r.\,t.\ $\ov{g}_{\mu \nu}$) coarse graining scale.
The infrared (IR) cutoff is built into the path integral by adding an appropriate cutoff action $\Delta_k S [f;\chib]$ to the bare action. It is a quadratic form in $f$ which involves a $\chib$-dependent integral kernel $\mathcal{R}_k [\chib]$. (See [\,I\,] for further details.)

In the full theory, for vanishing ghost fields, the gravitational average action is a functional of the expectation value and the background field, respectively: $\Gamma_k [\,\ov{h}_{\mu \nu}; \ov{g}_{\mu \nu}] \equiv \Gamma_k [g_{\mu \nu}, \ov{g}_{\mu \nu}]$. In the reduced setting, $\Gamma_k$ depends likewise on $\ov{f}$ and $\chib$ or alternatively on $\phi \equiv \chib + \ov{f}$ and $\chib$: $\Gamma_k [\,\ov{f}; \chib] \equiv \Gamma_k [\phi, \chib]$. Even though it is kept fixed usually, it is sometimes important to keep in mind that $\Gamma_k$ also depends on the reference metric $\h g_{\mu \nu}$. So occasionally it will be helpful to include it in the list of arguments, writing $\Gamma_k [\,\ov{f}; \chib, \h g_{\mu \nu}]$ or $\Gamma_k [\phi, \chib, \h g_{\mu \nu}]$, respectively.

In [\,I\,] we derived the following FRGE for the reduced average action:
\begin{align}\label{2.frge}
k \p_k \, \Gamma_k [\,\ov{f}; \chib, \h g_{\mu \nu}]
& =
\frac{1}{2} \tr \left[ 
\left( \Gamma_k^{(2)} [\,\ov{f}; \chib, \h g_{\mu \nu}]
+ \mathcal{R}_k [\chib, \h g_{\mu \nu}] \right)^{-1} \,
k \p_k\, \mathcal{R}_k [\chib, \h g_{\mu \nu}]
\right].
\end{align}
This equation has the same structure as in a scalar theory, except that the properties of the cutoff operator $\mathcal{R}_k$ are different here. In [\,I\,] we explained in detail how it has to be constructed in order to be ``background independent'' and to take account of the fact that the metric itself determines the proper coarse graining scale.

The reference metric $\h g_{\mu \nu}$ is unphysical and, in fact, has no counterpart in the full theory. A Weyl rescaling of $\h g_{\mu \nu}$ can always be absorbed by a suitable redefinition of the conformal factor. This is formalized by saying that the ``physical'' metrics $\gamma_{\mu \nu}$, $\ov{g}_{\mu \nu}$, $g_{\mu \nu}$, and $\breve g_{\mu \nu}$ are invariant under the following background--quantum field split symmetries:
\begin{subequations}\label{2.100}
\begin{gather}\label{2.100a}
\h g_{\mu \nu}^{\,\prime}
=
\ee^{-2 \sigma (x)} \,\, \h g_{\mu \nu}
\end{gather}
\begin{align}
\label{2.100b}
& & 
f' & = \ee^{\sigma (x) / \nu} \, f,
& \chib' & = \ee^{\sigma (x) / \nu} \, \chib
& & 
\\
\label{2.100c}
& & 
\ov{f}^{\,\prime} & = \ee^{\sigma (x) / \nu} \, \ov{f},
& \phi' & = \ee^{\sigma (x) / \nu} \, \phi.
& & 
\end{align}
\end{subequations}
Since the decomposition of $\gamma_{\mu \nu}$ as a conformal factor times $\h g_{\mu \nu}$ is completely arbitrary, the quantum theory of the conformal factor must respect the split symmetry \eqref{2.100}. In particular we require the cutoff action $\Delta_k S [f; \chib, \h g_{\mu \nu}]$ to be invariant under the ``$\sigma$-transformations'' \eqref{2.100}, as we shall call them. This invariance has an important consequence for the solutions $\Gamma_k [\,\ov{f}; \chib, \h g_{\mu \nu}]$ of the FRGE \eqref{2.frge}: If $\Gamma_k$ is invariant at one value of $k$, it is so also at any other $k$. The flow does not generate non-invariant terms and, as a consequence, it is enough to include $\sigma$-invariant terms in the truncation ansatz.

Later on we shall fix $\h g_{\mu \nu}$ to be the metric of a round $4$-sphere with radius $\h r$, $S^4 (\h r\,)$, so that the corresponding line element reads, using standard coordinates,
\begin{align}
\h g_{\mu \nu} \, \dd x^\mu \, \dd x^\nu
& =
\h r^{\,2} \, \left[ \dd \zeta^2 + \sin^2 \zeta \,
\Bigl( \dd \eta^2 + \sin^2 \eta \, 
\left( \dd \theta^2 + \sin^2 \theta \, \dd \phi^2 
\right) \Bigr) \right].
\end{align}
If we insist on this form of the reference metric, the invariance under local $\sigma$-trans\-for\-ma\-tions is broken to a symmetry under global ones only. Global split transformations \eqref{2.100} with, by definition, $\sigma (x) =  \sigma = const$, preserve the form of this metric, but rescale the radius of the sphere according to
\begin{align}\label{2.radius}
\h r^{\,\prime} & = \ee^{-\sigma} \, \h r.
\end{align}

Once the conformal reduction is performed, we are left with a theory of the field $\chi (x)$. To quantize it, we perform a second, this time linear background--quantum field split by decomposing $\chi = \chib + f$. This decomposition is supposed to mimic the $\gamma_{\mu \nu} = \ov{g}_{\mu \nu} + h_{\mu \nu}$ decomposition in the full theory. Obviously $\chi$ and $\phi \equiv \chib + \ov{f}$ are trivially invariant under a second set of split transformations:
\begin{align}\label{2.101}
& & 
\begin{split}
f' & = f + \tau (x),
\\
\chib' & = \chib - \tau (x),
\end{split}
& 
\begin{split}
\ov{f}^{\,\prime} & = \ov{f} + \tau (x)
\\
\h g_{\mu \nu}^{\,\prime} & = \h g_{\mu \nu}.
\end{split}
& & 
\end{align}
While the ``$\sigma$-transformations'' have no analog in the full theory, the above ``$\tau$-trans\-for\-ma\-tions'' do have a counterpart: $h_{\mu \nu}' = h_{\mu\nu} + \tau_{\mu \nu}$, $\ov{h}_{\mu \nu}^{\,\prime} = \ov{h}_{\mu \nu} + \tau_{\mu \nu}$, $\ov{g}_{\mu \nu}^{\,\prime} = \ov{g}_{\mu \nu} - \tau_{\mu \nu}$.

It is to be emphasized that both in the full and the reduced theory $\Delta_k S$ and as a consequence the FRGE and its exact solutions $\Gamma_k$ are \textit{not} invariant under the respective $\tau$-transformations if $k \neq 0$. This implies that the background field cannot be ``shifted away'' by redefining the quantum field (integration variable). In a certain sense, the background field must have acquired a ``quasi physical'' meaning therefore. From our discussion in [\,I\,] of the special status of the metric, it defines all proper length and mass scales, including that of $k$, it is clear what this meaning is\footnote{The gauge fixing issues we leave aside here. They are not important for the present discussion.}: The background encodes the information about the physical interpretation of the mode cutoff, the proper length scale of the ``coarse grained'' domains in spacetime, in particular. Being a cutoff in the spectrum of $\ov{\Box}$, the parameter $k^{-1}$ is a ``proper'' (as opposed to coordinate) length with respect to the background metric. The $\tau$-symmetry is broken by the cutoff action $\Delta_k S$ which implements the mode suppression. For $k \to 0$ the functional $\Gamma_k$ approaches the standard effective action $\Gamma$. In this limit $\Gamma_k$ becomes invariant under the ``$\tau$-transformations''.

In [\,I\,] we solved the FRGE \eqref{2.frge} on the $2$-dimensional truncated theory space spanned by the ansatz
\begin{align}\label{2.crehtrunc}
\begin{split}
\Gamma_k [\,\ov{f}; \chib, \h g_{\mu \nu}]
& =
- \frac{3}{4 \pi \, G_k} \, \int \!\!\dd^4 x \, \sqrt{\h g\,}~
\bigg\{ \tfrac{1}{2} ( \chib + \ov{f} \,) \,
\left( -\h \Box + \tfrac{1}{6} \, \h R \right) \, 
( \chib + \ov{f} \,)
\\
& \phantom{{==}- \frac{3}{4 \pi \, G_k} \, \int \!\!\dd^4 x \, \sqrt{\h g\,}~
\bigg\{ }
- \tfrac{1}{6} \, \Lambda_k \, ( \chib + \ov{f} \,)^4
\bigg\}.
\end{split}
\end{align}
This functional is invariant under $x$-dependent $\sigma$- as well as $\tau$-transformations. The pertinent cutoff action $\Delta_k S [f; \chib, \h g_{\mu \nu}]$ is only $\sigma$-invariant, however. A generalization of \eqref{2.crehtrunc} will be encountered in the next section.
%
%
%
\section{The Local Potential Approximation}\label{s3}
Next we introduce the local potential approximation (LPA) to the conformally reduced theory of gravity discussed in the previous section and we derive the corresponding beta functions. We restrict the discussion to the case of a maximally symmetric reference metric $\h g_{\mu \nu}$. To start with we shall assume that it refers to a sphere $S^4(\h r\,)$; later on we shall also consider the flat space $R^4$. For clarity's sake we focus on $d=4$ spacetime dimensions here; the $d$-dimensional generalizations of various key formulas are collected in the Appendix.
\subsection{The Truncation Ansatz}\label{s3.1}
The LPA truncation ansatz has the same kinetic term as the CREH truncation \eqref{2.crehtrunc} but instead of the single quartic term proportional to $\Lambda_k$ we now allow for an arbitrary running potential $H_k$:
\begin{align}\label{3.1}
\begin{split}
\Gamma_k \bigl[ \,\ov{f}; \chib, \h g_{\mu \nu} 
\bigr( S^4 (\h r\,) \bigr) \bigr]
& =
- \frac{3}{4 \pi \, G_k} \, \int \!\!\dd^4 x \, \sqrt{\h g\,}~
\bigg\{ \tfrac{1}{2} ( \chib + \ov{f}\,) \,
\left( -\h \Box + \tfrac{1}{6} \, \h R \right) \, 
( \chib + \ov{f} \,)
\\
& \phantom{{==}- \frac{3}{4 \pi \, G_k} \, \int \!\!\dd^4 x \, \sqrt{\h g\,}~
\bigg\{ }
+ \h r^{\,-4} \, H_k \bigl( \h r \, ( \chib + \ov{f} \, )
\bigr) \bigg\}.
\end{split}
\end{align}
After the restriction to spherical reference metrics we must make sure that $\Gamma_k$ is invariant under $x$-independent $\sigma$-transformations \eqref{2.100} which change the radius $\h r$ according to \eqref{2.radius}. This explains the explicit factors of $\h r$ in \eqref{3.1}. They guarantee that $\Gamma_k$ is indeed invariant under global $\sigma$-transformations:
\begin{align}\label{3.1s}
\Gamma_k \bigl[ \ee^{\sigma} \ov{f}; \ee^{\sigma} \chib, \h g_{\mu \nu} 
\bigr( S^4 (\ee^{-\sigma} \,\h r\,) \bigr) \bigr]
& =
\Gamma_k \bigl[ \,\ov{f}; \chib, \h g_{\mu \nu} 
\bigr( S^4 (\h r\,) \bigr) \bigr].
\end{align}

The ansatz \eqref{3.1} happens to be invariant under the $\tau$-transformations \eqref{2.101}. This accidental $\tau$-invariance is due to the still comparatively simple form of the truncation ansatz; recall that the FRGE and the exact solutions do not have this symmetry. In fact, in a more general truncation we would assume that the potential depends on the two fields $\w \chi_{\text{B}} \equiv \h r \, \chib$ and $\w{\ov{f}} \equiv \h r \, \ov{f}$ separately, and not only via their sum $\w \phi \equiv \h r \, \phi$. Then $\Gamma_k$ would not be $\tau$-invariant in general. The simplifying assumption of a single--field LPA potential parallels analogous truncation steps in the full theory \cite{mr}: Starting from a general potential $H_k (\w \phi, \w \chi_{\text{B}} )$, without loss of generality, we may decompose it as $H_k (\w \phi, \w \chi_{\text{B}} ) = \ov{H}_k (\w \phi) + \h H_k (\w \phi, \w \chi_{\text{B}} )$ where $\ov{H}_k (\w \phi) \equiv H_k (\w \phi, \w \phi )$, and $\h H_k$ vanishes for $\w \phi = \w \chi_{\text{B}}$. Adopting the ansatz \eqref{3.1} amounts to neglecting $\h H_k$, i.\,e.\ to the approximation $H_k \approx \ov{H}_k$. With the potentials $\ov{H}_k$ and $\h H_k$ corresponding to the functionals $\ov{\Gamma}_k [g_{\mu \nu}]$ and $\h \Gamma_k [g_{\mu\nu}, \ov{g}_{\mu \nu}]$, respectively, this is the same kind of ``single--field approximation'' one often makes in full QEG, see Section 3 of ref.\ \cite{mr}. However, even though we take $H_k$ to depend on one field only, special solutions of the type \eqref{3.1} are in fact general enough to demonstrate the qualitative consequences of the $\tau$-invariance breaking $\chib$-dependences which enter the flow equation via $\mathcal{R}_k$ and are dictated by ``background independence'', see [\,I\,] for a first example.

It will be convenient to combine the $\h R$-term which stems from $\int\!\sqrt{g\,} \, R$ with the potential $H_k$. On the $4$-sphere $\h R = 12 / \h r^{\,2}$ is a constant. Therefore, if we introduce the total potential as
\begin{align}\label{3.2}
F_k (\w \phi)
& \equiv
\w \phi \,^2 + H_k (\w \phi)
\end{align}
the average action assumes the form
\begin{align}\label{3.3}
\begin{split}
\Gamma_k \bigl[ \,\ov{f}; \chib, \h g_{\mu \nu} 
\bigr( S^4 (\h r\,) \bigr) \bigr]
& =
- \frac{3}{4 \pi \, G_k} \, \int \!\!\dd^4 x \, \sqrt{\h g\,}~
\bigg\{ - \tfrac{1}{2} ( \chib + \ov{f}\,) \,
\h \Box \, ( \chib + \ov{f} \,)
\\
& \phantom{{==}- \frac{3}{4 \pi \, G_k} \, \int \!\!\dd^4 x \, \sqrt{\h g\,}~
\bigg\{ }
+ \h r^{\,-4} \, F_k \bigl( \h r \, ( \chib + \ov{f} \, )
\bigr) \bigg\}.
\end{split}
\end{align}
We are going to insert this ansatz into the FRGE \eqref{2.frge}. Under the trace on its RHS we need the Hessian $\Gamma_k^{(2)}$. It reads
\begin{align}\label{3.4}
\begin{split}
& \Gamma_k^{(2)} \bigl[ \,\ov{f}; \chib, \h g_{\mu \nu} 
\bigr( S^4 (\h r\,) \bigr) \bigr]_{xy}
\\
& \phantom{{==}} =
- \frac{3}{4 \pi \, G_k} \, 
\Bigl[ - \h \Box_x 
+ \h r^{\,-2} \, F_k'' \bigl( \h r \, \bigl( \chib (x)+ \ov{f} \,(x) \bigr)
\bigr) \Bigr] \, \frac{\delta (x-y)}{\sqrt{\h g (x) \,}\,}.
\end{split}
\end{align}
Here and in the following a prime denotes a derivative with respect to the argument. Comparing \eqref{3.3} to \eqref{1.2} we see that $U_k (\phi) \equiv - \tfrac{3}{4 \pi \, G_k} \, \h r\,^{-4} \, F_k (\h r \, \phi)$.

The ansatz \eqref{3.3} contains two scale dependent parameters, a single constant, namely $G_k$, and a function of the field, $F_k (\,\cdot\,)$. We determine their respective beta functions by evaluating both sides of the FRGE with the ansatz inserted for special field configurations which project out the relevant monomials in the action. In order to find $k \p_k \, F_k (\,\cdot\,)$ we set $\ov{f} =0$ and $\chib = const$, compute the trace on the RHS of the FRGE as a function of the $x$-independent variable $\chib$, and equate the result to  $k \p_k \, F_k (\chib)$. The beta function for Newton's constant obtains from the kinetic term. We set $\chib=const$ and leave $\ov{f}$ nonzero and $x$-dependent. We expand the trace to second order in $\ov{f}$ and second order in the derivatives; the $(\p \ov{f}\,)^2$-piece is proportional to $k \p_k\, (1/G_k)$ then.

As for both of these calculations a constant background field is sufficient we assume $\chib (x) = const \equiv \chib$ from now on.

Before continuing a remark about dimensions might be in order here. In the above equations we use units such that the coordinates $x^\mu$ are dimensionless. Hence we have the canonical mass dimensions
\begin{gather}
[\gamma_{\mu \nu}] = [g_{\mu \nu}] = [\ov{g}_{\mu \nu}]
= [\h g_{\mu \nu}] = -2
\nonumber \\
\label{3.6}
[f] = [\,\ov{f}\,] = [\chib] = [\phi] = 0
\\
\nonumber
[H_k] = [F_k] = -2, \qquad [\h r\,]=-1.
\end{gather}
\subsection{The Cutoff Operator}\label{s3.2}
In [\,I\,] we explained in detail why it is crucial to define $k$ as a cutoff in the spectrum of $\ov{\Box}$ rather than $\h \Box$, say. (Here $\ov{\Box}$ and $\h \Box$ are the Laplace--Beltrami operators related to $\ov{g}_{\mu \nu} \equiv \chib^2 \,\, \h g_{\mu \nu}$ and $\h g_{\mu \nu}$, respectively.)
At least for truncations of the type considered here this can be achieved by designing the cutoff operator $\mathcal{R}_k$ in such a way that upon adding it to $\Gamma_k^{(2)}$ is leads to the replacement
\begin{align}\label{3.7}
(- \ov{\Box} \,) 
& \to
(- \ov{\Box} \,) + k^2 \, R^{(0)} \bigl( - \tfrac{\ov{\Box}}{k^2\,} \bigr).
\end{align}
Since, for $\chib=const$, $\ov{\Box} = \chib^{-2} \, \h \Box$, this substitution rule amounts to
\begin{align}\label{3.8}
(-\h \Box)
& \to 
(-\h \Box) + \chib^2 \, k^2 \,R^{(0)} \bigl( - \tfrac{\h \Box}{\chib^2 \, k^2\,} \bigr).
\end{align}
In our case $\Gamma_k^{(2)}$ is given by \eqref{3.4}. Therefore we define $\mathcal{R}_k$ such that, in operator notation,
\begin{align}\label{3.9}
\Gamma_k^{(2)} + \mathcal{R}_k
& =
- \frac{3}{4 \pi \, G_k} \, 
\Bigl[ - \h \Box + \chib^2 \, k^2 \,R^{(0)} \bigl( - \tfrac{\h \Box}{\chib^2 \, k^2\,} \bigr)
+ \h r^{\,-2} \, F_k'' \bigl( \h r \, \bigl( \chib + \ov{f} \, (x) \bigr)
\bigr) \Bigr].
\end{align}
As a result, we must set
\begin{align}\label{3.10}
\mathcal{R}_k [\chib; \h g_{\mu \nu}]
& =
- \frac{3}{4 \pi \, G_k} \, \chib^2 \, k^2 \,
R^{(0)} \bigl( - \tfrac{\h \Box}{\chib^2 \, k^2\,} \bigr).
\end{align}
Note the factors of $\chib^2$ on the RHS of eq.\ \eqref{3.10}. They are typical of conformally reduced gravity and would not appear in the standard quantization of a scalar field. Besides the different transformation behavior under diffeomorphisms, they are the most important feature which distinguishes the conformal factor from scalar matter fields. They lead to very significant modifications of the RG flow [\,I\,]. The cutoff action corresponding to \eqref{3.10} reads:
\begin{align}\label{3.11}
\Delta_k S [f; \chib, \h g_{\mu \nu}]
& =
- \frac{3\, \chib^2 \, k^2}{8 \pi \, G_k}\,
\int \!\! d^4 x \, \sqrt{\h g\,}~
f (x) \, R^{(0)} \bigl( - \tfrac{\h \Box}{\chib^2 \, k^2\,} \bigr)
\, f(x).
\end{align}
This functional is indeed invariant under global $\sigma$-transformations, as it should be:
\begin{align}\label{3.12}
\Delta_k S [\ee^\sigma f; \ee^\sigma \chib, \ee^{-2\sigma} \h g_{\mu \nu}]
& =
\Delta_k S [f; \chib, \h g_{\mu \nu}].
\end{align}
Note, however, that $\Delta_k S$ is \textit{not} invariant under $\tau$-transformations:
\begin{align}\label{3.13}
\Delta_k S [f+\tau; \chib-\tau, \h g_{\mu \nu}]
& \neq
\Delta_k S [f; \chib, \h g_{\mu \nu}].
\end{align}

At this point the flow equation assumes the form
\begin{align}\label{3.14}
\begin{split}
& k \p_k \,\Gamma_k \bigl[ \,\ov{f}; \chib=const, \h g_{\mu \nu} 
\bigr( S^4 (\h r\,) \bigr) \bigr]
\\
& \phantom{{==}} =
\frac{1}{2} \tr \left[
\left( - \h \Box + \chib^2\, k^2 \, 
R^{(0)} \bigl( - \tfrac{\h \Box}{\chib^2 \, k^2\,} \bigr)
+ \h r^{\,-2} \, F_k'' \bigl( \h r \, ( \chib + \ov{f} \, )
\bigr) \right)^{-1}
\right.
\\
& \phantom{{====}\frac{1}{2} \tr \Biggl[}
\times G_k \, k \p_k \left\{
G_k^{-1} \, \chib^2 \, k^2 \, 
R^{(0)} \bigl( - \tfrac{\h \Box}{\chib^2 \, k^2\,} \bigr) \right\}
\biggr]
\end{split}
\end{align}
It is possible to completely eliminate the $\h r$-dependence from this equation. (This is exactly as it should be since in the exact formulation, with all degrees of freedom retained, $\h g_{\mu \nu}$ and $\h r$ have no counterparts.) To see this, we introduce the metric on the unit $4$-sphere, $\h g_{\mu \nu}^{\,(1)} \equiv \h g_{\mu \nu} \bigl( S^4 (1) \bigr)$, so that we may write
\begin{align}\label{3.15}
\h g_{\mu \nu} \bigl( S^4 (\h r\,) \bigr)
& =
\h r^{\,2} \,\, \h g_{\mu \nu}^{\,(1)}.
\end{align}
Note that $\h g_{\mu \nu}^{\,(1)}$ is dimensionless, while $\h g_{\mu \nu}$ has the mass dimension $-2$. Writing $\h \Box^{(1)}$ and $\h g^{\,(1)}$ for the Laplacian and the determinant corresponding to $\h g_{\mu \nu}^{\,(1)}$ we have, respectively, $\h \Box = \h r^{\,-2} \, \h \Box^{(1)}$, and $\sqrt{\h g\,} = \h r^{\,4} \, \sqrt{\h g^{\,(1)}\,}.$
When we refer all quantities to the unit metric the average action reads
\begin{align}\label{3.16}
\begin{split}
\Gamma_k \bigl[ \,\ov{f}; \chib, \h g_{\mu \nu} 
\bigr( S^4 (\h r\,) \bigr) \bigr]
& =
- \frac{3}{4 \pi \, G_k} \, \int \!\!\dd^4 x \, \sqrt{\h g^{\,(1)}\,}~
\bigg\{ - \tfrac{1}{2} ( \h r \, \ov{f}\,) \,
\h \Box^{(1)} \, ( \h r \, \ov{f} \,)
\\
& \phantom{{==}- \frac{3}{4 \pi \, G_k} \, \int \!\!\dd^4 x \, \sqrt{\h g{\,(1)}\,}~
\bigg\{ }
+ F_k ( \h r \, \chib + \h r \, \ov{f} \, ) \bigg\}.
\end{split}
\end{align}
We may use this form of the functional on the LHS of eq.\ \eqref{3.14}. In the same notation, its RHS assumes the form
\begin{align}\label{3.17}
\begin{split}
& \frac{1}{2} \tr \left[
\left( - \h \Box^{(1)} + (\h r \, \chib)^2\, k^2 \, 
R^{(0)} \bigl( - \tfrac{\h \Box^{(1)}}{(\h r \, \chib)^2 \, k^2\,} \bigr)
+ F_k'' ( \h r \, \chib + \h r \, \ov{f} \, ) \right)^{-1}
\right.
\\
& \phantom{{=}\frac{1}{2} \tr \Biggl[}
\times G_k \, k \p_k \left\{
G_k^{-1} \, (\h r \, \chib)^2 \, k^2 \, 
R^{(0)} \bigl( - \tfrac{\h \Box^{(1)}}{(\h r \, \chib)^2 \, k^2\,} \bigr) \right\}
\biggr].
\end{split}
\end{align}
Notice that $\chib$, $\ov{f}$, and $\h r$, both in \eqref{3.16} and \eqref{3.17}, appear only in the $\sigma$-invariant combinations $\h r \, \chib$ and $\h r \, \ov{f}$. It is therefore natural to introduce new fields, in terms of which $\h r$ disappears completely:
\begin{align}\label{3.18}
\w \chi_{\text{B}} \equiv \h r \, \chib, \qquad
\w{\ov{f}} \equiv \h r \, \ov{f}, \qquad
\w \phi \equiv \h r \, \phi.
\end{align}
In analogy with $\phi = \chib + \ov{f}$ we defined $\w \phi = \w \chi_{\text{B}} + \w{\ov{f}}$. The new variables have the dimension of a length, while the old ones were dimensionless. The interpretation of $\w \chi_{\text{B}}$, say, is clear: From $\ov{g}_{\mu \nu} = \chib^2 \,\, \h g_{\mu \nu}$ it follows that
\begin{align}\label{3.19}
\ov{g}_{\mu \nu} = 
(\h r \,\chib)^2 \,\, \h g_{\mu \nu}^{\,(1)}
= \w \chi_{\text{B}} \,\, \h g_{\mu \nu}^{\,(1)}.
\end{align}
Hence $\w \chi_{\text{B}}$ is nothing but the radius of the sphere with the metric $\ov{g}_{\mu \nu}$. Likewise $\w \phi$ is the radius of a $4$-sphere with metric $\breve g_{\mu \nu}$.

To summarize: We have seen that we can eliminate the reference metric from the flow equation determining the RG flow of $G_k$ and $F_k$. Therefore the corresponding beta functions cannot depend on it. Since the new variables are singlets under $\sigma$-transformations, the flow equation is invariant under the split transformations which express the arbitrariness of the reference metric.

Having convinced ourselves that the flow equations have the correct invariance properties we shall now simplify our notation and omit the tilde from the new fields or, what is the same, we shall stick to the old fields but set $\h r =1$ in all formulas. Then $\h g_{\mu \nu} \equiv \h g_{\mu \nu}^{\,(1)}$ and $\h \Box \equiv \h \Box^{(1)}$ so that we can omit the superscript ``$(1)$'' henceforth. The table of canonical dimensions changes accordingly; we have, along with $[x^\mu]=0$,
\begin{gather}
[\gamma_{\mu \nu}] = [g_{\mu\nu}] = [\,\ov{g}_{\mu \nu}] = -2,
\qquad [\h g_{\mu \nu}] =0
\nonumber \\
\label{3.20}
[f] = [\,\ov{f}\,] = [\chib] = [\phi] = -1
\\
\nonumber
[H_k] = [F_k] = -2.
\end{gather}

With these simplifications the flow equation which we have to evaluate further reads
\begin{align}\label{3.21}
\begin{split}
& - \frac{3}{4 \pi} \, \int \!\!\dd^4 x \, \sqrt{\h g\,}~
\biggl( - \tfrac{1}{2} \, k \p_k \, G_k^{-1} \, \ov{f} \,
\h \Box \ov{f}
+ k \p_k \, \left \{ G_k^{-1} F_k ( \chib + \ov{f} \, ) \right\} \biggr)
\\
& \phantom{{==}} =
\chib^2\, k^2 \, \tr \Biggl[
\left( - \h \Box + \chib^2\, k^2 \, 
R^{(0)} \bigl( - \tfrac{\h \Box}{\chib^2 \, k^2\,} \bigr)
+ F_k'' \bigl( \chib + \ov{f}\, (x) \bigr) \right)^{-1}
\\
& \phantom{{====} \chib^2\, k^2 \, \tr \Biggl[}
\times \bigg\{\left( 1 - \tfrac{1}{2} \etan \right) \, 
R^{(0)} \bigl( - \tfrac{\h \Box}{\chib^2 \, k^2\,} \bigr) 
- \Bigl( - \frac{\h \Box}{\chib^2 \, k^2\,} \Bigr) \,
{R^{(0)}}' \bigl( - \tfrac{\h \Box}{\chib^2 \, k^2\,} \bigr) 
\bigg\}
\Biggr].
\end{split}
\end{align}
Here we used the anomalous dimension related to Newton's constant, defined as in \cite{mr},
\begin{align}\label{3.22}
\etan & \equiv k \p_k \, \ln G_k.
\end{align}

We shall see that the beta functions depend on the \textit{topology} of the background manifold. We shall analyze the cases $S^4$ and $R^4$ in turn. For $S^4$, we derived eq.\ \eqref{3.21} where $\h g_{\mu \nu}$ refers to the unit sphere now. For $R^4$ it is easy to see that the corresponding flow equations also can be obtained from \eqref{3.21} but with $\h g_{\mu \nu}$ interpreted as the flat metric $\h g_{\mu \nu} = \delta_{\mu \nu}$.
\subsection{The RG Equation for $\boldsymbol{F_k}$}\label{s3.3}
\subsubsection{The $\boldsymbol{S^4}$ Topology}\label{s.3.3.1}
We begin by deriving the flow equation for the potential $F_k (\phi)$. As explained above, we evaluate \eqref{3.21} for $\ov{f}=0$ to this end, retain the full dependence on the (constant) field $\chib = \phi$, however. (Recall that $\phi \equiv \chib + \ov{f}$ so that $\phi = \chib$ now.) We perform the trace on the RHS of \eqref{3.21} in the eigenbasis of $\h \Box$, the Laplace--Beltrami operator on the unit $S^4$. Its eigenvalues $-\mathcal{E}_n$ and their degeneracies $D_n$, $n=0,1,2,\cdots$, are given by \cite{rubin}
\begin{align}\label{3.23}
& & \mathcal{E}_n & = n (n+3),
& D_n & = \tfrac{1}{6} \, (n+1) (n+2) (2n+3).
& &
\end{align}
With $\sigma_4 \equiv \int \dd^4 x \sqrt{\h g\,} = 8 \pi^2/3$ the volume of $S^4(1)$, this leads us to
\begin{align}\label{3.24}
- \frac{3 \, \sigma_4}{4 \pi \, G_k} \, \Bigl( k \p_k F_k (\phi)
- \etan \, F_k (\phi) \Bigr)
& =
T_1 (\phi) + T_2 (\phi)
\end{align}
with the two spectral sums
\begin{align}\label{3.25}
\begin{split}
T_1 & \equiv
\varphi^2 \, \sum_{n=0}^\infty D_n \,
\frac{R^{(0)} (\mathcal{E}_n / \varphi^2) 
- \bigl( \mathcal{E}_n / \varphi^2 \bigr) \, 
{R^{(0)}}' (\mathcal{E}_n / \varphi^2)}
{\mathcal{E}_n + \varphi^2 \, R^{(0)} (\mathcal{E}_n / \varphi^2)
+ F_k'' (\phi)}
\\
T_2 & \equiv
- \frac{1}{2} \, \etan \, 
\varphi^2 \, \sum_{n=0}^\infty D_n \,
\frac{R^{(0)} (\mathcal{E}_n / \varphi^2)}
{\mathcal{E}_n + \varphi^2 \, R^{(0)} (\mathcal{E}_n / \varphi^2)
+ F_k'' (\phi)}.
\end{split}
\end{align}
Here we employed the abbreviation
\begin{align}\label{3.26}
\varphi & \equiv k \phi.
\end{align}
Note that $\varphi$ is dimensionless as $\phi$ has the dimension of a length.

In principle \eqref{3.24} gives the beta function of $F_k$ for an arbitrary shape function $R^{(0)}$. In order to simplify the eigenvalue sums it is convenient to use the optimized shape function \cite{opt}, however, the same that was also used in [\,I\,]:
\begin{align}\label{3.27}
R^{(0)} (z) & = \left( 1-z \right) \, \theta (1-z).
\end{align}
With this choice, the final result for the flow equation of the potential reads
\begin{align}\label{3.28}
k \p_k \, F_k (\phi) 
& =
\etan \, F_k (\phi)
- \frac{G_k}{2 \pi} \,
\frac{\left( 1 - \etan/2 \right) \, k^2 \, \phi^2 \, \rho (k \, \phi)
+ \tfrac{1}{2} \, \etan \, \rhot (k \, \phi)}
{k^2 \, \phi^2 + F_k'' (\phi)}.
\end{align}
Here $\rho$ and $\rhot$ are two new, simpler spectral functions, for an arbitrary $\h g_{\mu \nu}$ defined by
\begin{align}\label{3.29}
& & 
\rho (\varphi)
& \equiv
\tr \Bigl[ \theta (\varphi^2 + \h \Box) \Bigr],
& \rhot (\varphi)
& \equiv
\tr \Bigl[ (-\h \Box)
 \, \theta (\varphi^2 + \h \Box) \Bigr].
& & 
\end{align}
In the case at hand, for the sphere,
\begin{align}\label{3.30}
& &
\rho (\varphi)
& =
\sum_{n=0}^\infty D_n \, \theta (\varphi^2 - \mathcal{E}_n),
& \rhot (\varphi)
& =
\sum_{n=0}^\infty \mathcal{E}_n \, D_n \, \theta (\varphi^2 - \mathcal{E}_n).
& & 
\end{align}
These functions are of the form
\begin{align}\label{3.31}
& & 
\rho (\varphi) & = J_4 \bigl( n_{\text{max}} (\varphi) \bigr),
& \rhot (\varphi) & = 
\w J_4 \bigl( n_{\text{max}} (\varphi) \bigr).
& & 
\end{align}
By definition, $n_{\text{max}} (\varphi)$ is the largest integer $n=0,1,2,\cdots$ such that $\mathcal{E}_n = n (n+3) < \varphi^2$. Here we introduced the finite sums
\begin{align}\label{3.jsums}
& & 
J_4 (N) & \equiv \sum_{n=0}^N D_n,
& \w J_4 (N) & \equiv \sum_{n=0}^N \mathcal{E}_n \, D_n.
& & 
\end{align}
They can be worked out explicitly:
\begin{subequations}\label{3.32}
\begin{align}\label{3.32a}
J_4 (N)
& = \tfrac{1}{12} \, N^4 + \tfrac{2}{3} \, N^3 + \tfrac{23}{12} \, N^2
+ \tfrac{7}{3} \, N + 1
\\
\label{3.32b}
\w J_4 (N)
& =
\tfrac{1}{18} \, N^6 + \tfrac{2}{3} \, N^5 \, + \tfrac{55}{18} \, N^4
+ \tfrac{20}{3} \, N^3 + \tfrac{62}{9} \, N^2 + \tfrac{8}{3} \, N.
\end{align}
\end{subequations}
Since the largest contributing eigenvalue has the quantum number
\begin{align}
n_{\text{max}} (\varphi) & \approx
\begin{cases} 
\varphi & \text{for } \varphi \gg 1
\\
0 & \text{for } \varphi \ll 1
\end{cases}
\end{align}
the results \eqref{3.32} imply the limits
\begin{subequations}\label{3.34}
\begin{align}\label{3.34ab}
\rho (\varphi) & \approx
\begin{cases} 
\tfrac{1}{12} \, \varphi^4 & \text{for } \varphi \gg 1
\\
1 & \text{for } \varphi \ll 1
\end{cases}
\\
\label{3.34cd}
\rhot (\varphi) & \approx
\begin{cases} 
\tfrac{1}{18} \, \varphi^6 & \text{for } \varphi \gg 1
\\
0 & \text{for } \varphi \ll 1
\end{cases}
\end{align}
\end{subequations}
In particular we see that
\begin{align}\label{3.35}
& & 
\rho(0) & = 1,
& \rhot (0) & = 0.
& &
\end{align}

On $S^4$ the spectrum of $\h \Box$ is completely discrete. As a consequence, $\rho (\varphi)$ and $\rhot (\varphi)$ have discontinuities at the $\varphi$-values at which $n_{\text{max}} (\varphi)$ jumps. When one analyzes the differential equation \eqref{3.28} smooth functions $\rho (\varphi)$ and $\rhot (\varphi)$ clearly would be easier to deal with. For this reason we henceforth adopt a certain \textit{smoothing procedure} which replaces the original functions $\rho$ and $\rhot$ by smooth interpolating functions. We shall again denote them by $\rho$ and $\rhot$. The way how this interpolation or smoothing is done is in no way unique. Choosing a specific smoothing procedure has the same conceptual status as choosing a particular cutoff function $R^{(0)}$: It amounts to specifying how precisely the transition from the high to the low momentum regime takes place, i.\,e.\ how the modes start getting suppressed when their eigenvalue passes the threshold value given by $k$. Observable quantities derived from the RG flow must be independent of both $R^{(0)}$ and the smoothing procedure.

It is convenient to assume the smooth $\rho$ and $\rhot$ of polynomial form. We get a function with the correct behavior for $\varphi \ll 1$ and $\varphi \gg 1$ if 
\begin{align}\label{3.36}
\rho (\varphi)
& =
\sum_{k=0}^4 a_k \, \varphi^k,
\qquad \text{with} \quad a_0=1, \quad a_4 = 1/12
\end{align}
and similarly for $\rhot$. A good fit to the $\rho (\varphi)$ of \eqref{3.31} is obtained with $a_1 \approx -0.0118142$, $a_2 \approx -0.0832909$, $a_3 \approx -0.000333389$. For many purposes the following approximation is perfectly sufficient:
\begin{align}\label{3.366}
& & 
\rho (\varphi) & = 1 + \tfrac{1}{12} \, \varphi^4,
& 
\rhot (\varphi) & = \tfrac{1}{18} \, \varphi^6.
& & 
\end{align}

In particular when one searches for fixed points the flow equation is needed in dimensionless form. For this reason we represent $F_k$ as
\begin{align}\label{3.37}
F_k (\phi) & \equiv k^{-2} \, \, Y_k (k\,\phi).
\end{align}
Here $Y_k$ is a dimensionless function of a dimensionless argument $k \, \phi \equiv \varphi$, i.\,e.\
\begin{align}\label{3.38}
Y_k (\varphi) & = k^2 \, F_k (\varphi / k).
\end{align}
The RG equation for $Y_k$ follows from \eqref{3.28}:
\begin{align}\label{3.39}
k \p_k \, Y_k (\varphi) 
& =
(2+\etan) \, Y_k (\varphi) - \varphi \, Y_k' (\varphi)
- \frac{g_k}{2 \pi} \,
\frac{\left( 1 - \etan/2 \right) \, \varphi^2 \, \rho (\varphi)
+ \tfrac{1}{2} \, \etan \, \rhot (\varphi)}
{\varphi^2 + Y_k'' (\varphi)}.
\end{align}
Here $g_k \equiv k^2 \, G_k$ is the dimensionless Newton constant. In terms of $Y_k$, the original potential is given by $U_k (\phi) = - \tfrac{3}{4 \pi \, g_k} \, Y_k (k\,\phi)$.
\subsubsection{The $\boldsymbol{R^4}$ Topology}\label{s.3.3.2}
If the manifold on which the metrics $\h g_{\mu \nu}$ and $\ov{g}_{\mu \nu}$ are defined has the topology of a plane we may use eq.\ \eqref{3.21} as well, but with $\h g_{\mu \nu} = \delta_{\mu \nu}$. The operator $\h \Box$ is easily diagonalized in a plane wave basis then, and the spectral functions \eqref{3.29} are found to be
\begin{subequations}\label{3.40}
\begin{align}\label{3.40a}
\rho (\varphi) & = \left( \frac{\varphi^4}{12} \right)
\, \Bigl( \text{$\int \dd^4 x$} \Bigr)/\sigma_4
\\
\label{3.40b}
\rhot (\varphi) & = \left( \frac{\varphi^6}{18} \right)
\, \Bigl( \text{$\int \dd^4 x$} \Bigr)/\sigma_4.
\end{align}
\end{subequations}
The resulting flow equation for the dimensionful potential reads
\begin{align}\label{3.41}
k \p_k \, F_k (\phi)
& =
\etan \, F_k (\phi)
- \frac{G_k}{24 \pi} \, \left( 1 - \tfrac{1}{6} \, \etan \right) \,
\frac{\phi^6 \, k^6}{\phi^2 \, k^2 + F_k'' (\phi)}.
\end{align}
For the dimensionless $Y_k$ one finds the completely explicit partial differential equation
\begin{align}\label{3.42}
k \p_k \, Y_k (\varphi)
& =
(2+\etan) \, Y_k (\varphi) - \varphi \, Y_k' (\varphi)
- \frac{g_k}{24 \pi} \, \left( 1 - \tfrac{1}{6} \, \etan \right) \,
\frac{\varphi^6}{\varphi^2 + Y_k'' (\varphi)}.
\end{align}

The $R^4$ equations \eqref{3.41} and \eqref{3.42} coincide exactly with the corresponding ones for $S^4$, eqs.\ \eqref{3.28} and \eqref{3.39}, if in the latter the $\varphi \gg 1$-approximations from \eqref{3.34ab} and \eqref{3.34cd} are used\footnote{Note that the first factor on the RHS of \eqref{3.40a} and \eqref{3.40b}, respectively, coincides exactly with the $S^4$ results \eqref{3.34ab} and \eqref{3.34cd} valid for $\varphi \gg 1$. The second factor $\int \dd^4 x / \sigma_4$ takes the different volume normalizations in the two cases into account.} for \textit{all} values of $\varphi$. This was to be expected, of course, because for $\varphi \gg 1$ the spectral sums of $S^4$ are dominated by very many densely spaced eigenvalues which form a quasi-continuum. The difference between $R^4$ and $S^4$ is most pronounced for small values of $\varphi$. Here the finite volume of $S^4$ and the resulting discreteness of the spectrum strongly modifies the spectral density $\rho$. In the continuum it is proportional to $\varphi^4$, while on the sphere it approaches $\rho (0) =1$ for $\varphi \ll 1$. Note that $\varphi \equiv k \, \phi$ can be interpreted as the radius of the sphere with metric $\ov{g}_{\mu\nu}$, measured in cutoff units. For $\varphi \ll 1$ the trace on the RHS of the FRGE is dominated by a single eigenvalue, namely the zero mode of $\h \Box$ which has $n=0$. For the $R^4$ topology, on the other hand, the $\rho \sim \varphi^4$-behavior extends down to $\varphi=0$.

In [\,I\,] we explained in detail how the special status of the metric affects the RG equations and why those for conformally reduced gravity are different from the corresponding flow equations in scalar matter field theories. It is therefore instructive to compare the partial differential equations \eqref{3.41} and \eqref{3.42} to their scalar counterparts. In a standard scalar theory, the analog of eq.\ \eqref{3.41} for $F_k$, say, would have the last factor on its RHS replaced by $k^6 / \bigl( k^2 + F_k'' (\phi) \bigr)$. It does not contain the crucial extra factors of $\phi$ which originate from the $\chib$-dependence of $\mathcal{R}_k$. It is clear that these extra factors of $\phi$ modify the RG evolution of the function $F_k (\phi)$ quite significantly.

Within the present framework, the CREH approximation of [\,I\,] amounts to the following more restrictive ansatz for $Y_k$:
\begin{align}\label{3.43}
Y_k\cc (\varphi)
& =
c_0 \, \varphi^2 - \tfrac{1}{6} \, \lambda_k \, \varphi^4.
\end{align}
Here
\begin{align}\label{3.44}
c_0 & = 
\begin{cases}
0 & \text{for } R^4
\\
1 & \text{for }S^4
\end{cases}
\end{align}
and $\lambda_k \equiv \Lambda_k / k^2$ is the dimensionless cosmological constant. We can derive its RG equation
\begin{align}\label{3.444}
k \p_k \, \lambda_k = \beta_\lambda^{\text{\,CREH}} (g_k, \lambda_k)
\end{align}
by inserting \eqref{3.43} into \eqref{3.39} or \eqref{3.42} and comparing the coefficients of $\varphi^4$. From either equation one obtains
\begin{align}\label{3.45}
\beta_\lambda^{\text{\,CREH}} (g_k, \lambda_k)
& =
- (2-\etan) \, \lambda_k
+ \frac{g_k}{4 \pi} \, \left( 1 - \tfrac{1}{6} \, \etan \right) \,
\frac{1}{1 - 2 \, \lambda_k}.
\end{align}
This is exactly the beta function found in [\,I\,].
\subsection{The RG Equation for $\boldsymbol{G_k}$}\label{s3.4}
In order to close the system of equations we need an equation for $\p_k \, g_k$ or, equivalently, the anomalous dimension:
\begin{align}\label{3.46}
k \p_k \, g_k
& =
\Bigl[ 2 + \etan \bigl( g_k, [Y_k] \bigr) \Bigr] \, g_k.
\end{align}
In the case at hand, $\etan$ is a \textit{function} of $g_k$ and a \textit{functional} of $Y_k$. The desired expression for $\etan$ is obtained from eq.\ \eqref{3.21} by fixing a constant value of $\chib$ and performing a derivative expansion of the functional trace up to order $(\p \ov{f}\,)^2$.
\subsubsection{The $\boldsymbol{R^4}$ Topology}\label{s.3.4.1}
For $\h g_{\mu \nu} = \delta_{\mu \nu}$ the result for $\etan$ can be inferred from the calculation in Appendix A of [\,I\,]. Expressed in dimensionful quantities, it reads
\begin{align}\label{3.47}
\etan
& =
- \frac{G_k}{6 \pi} \,
\left[ \frac{F_k''' (\phi_1)}{k \, \phi_1} \right]^2 \,
\h \Sigma_4 \! \biggl( \frac{F_k'' (\phi_1)}{k^2 \, \phi_1^2} \biggr)
\, \left[ 1 + \frac{G_k}{12 \pi} \,
\left[ \frac{F_k''' (\phi_1)}{k \, \phi_1} \right]^2 \, 
\w \Sigma_4 \!\biggl( \frac{F_k'' (\phi_1)}{k^2 \, \phi_1^2} \biggr)
\right]^{-1}.
\end{align}
Here $\h \Sigma_4$ and $\w \Sigma_4$ are threshold functions defined in [\,I\,]. They depend on $R^{(0)}$. For the optimized shape function \eqref{3.27} one has, for instance,
\begin{align}\label{3.48}
& & 
\h \Sigma_4 (w) & = \frac{1}{4} \, \frac{1}{(1+w)^4\,},
& \w \Sigma_4 (w) & = 0.
& & 
\end{align}
In \eqref{3.47} the derivatives of $F_k$ are evaluated at a \textit{fixed} field value, $\phi_1$. It is the value of $\chib$ about which we perform the expansion of \eqref{3.21}. There is a certain arbitrariness as to the choice of $\chib \equiv \phi_1$ in $\etan$. It is due to the fact that the truncation used allows for a field independent wave function renormalization $Z_{Nk} \propto 1 / G_k$ only, see Appendix A of [\,I\,]. In standard scalar calculations $\phi_1$ is usually identified with the minimum of the potential \cite{avactrev}. Notice that since $\phi_1$ is fixed, $\etan$ is indeed a functional of $F_k$ and not a function of the field.

With the optimized shape function $\etan$ assumes the simple form
\begin{align}\label{3.49}
\etan
& =
- \frac{G_k}{24 \pi} \,
\left[ \frac{F_k''' (\phi_1)}{k \, \phi_1} \right]^2 \,
\left[ 1 + \frac{F_k'' (\phi_1)}{k^2 \, \phi_1^2} \right]^{-4}.
\end{align}
We see that $\etan$ is negative if $G_k >0$, for any choice of $\phi_1$ and for any potential function $F_k (\,\cdot\,)$. This means that the gravitational interaction is \textit{anti}screening: As we increase $k$, Newton's constant decreases according to $k \p_k \, G_k = \etan \, G_k$.

In terms of the dimensionless quantities $g_k$ and $Y_k$ the anomalous dimension \eqref{3.49} reads, with $\varphi_1 \equiv k \, \phi_1$,
\begin{align}\label{3.50}
\etan \bigl( g_k, [Y_k] \bigr)
& =
- \frac{g_k}{24 \pi} \,
\frac{\Bigl[ \varphi_1^3 \, Y_k''' (\varphi_1) \Bigr]^2}
{\Bigl[ \varphi_1^2 + Y_k'' (\varphi_1) \Bigr]^4\,}.
\end{align}
If we insert the CREH form of the running potential, $Y_k\cc (\varphi) = - \tfrac{1}{6} \, \lambda_k \, \varphi^4$, we obtain from \eqref{3.50}:
\begin{align}\label{3.51}
\etan\cc (g_k, \lambda_k)
& = 
- \frac{2}{3 \pi} \, \frac{g_k \, \lambda_k^2}{\left( 1-2 \, \lambda_k
\right)^4\,}.
\end{align}
This is precisely the anomalous dimension of the CREH approximation which was derived in [\,I\,]. (In [\,I\,] the above $\etan\cc$ was denoted $\etan^{\text{(kin)}}$.) Note that \eqref{3.51} is independent of the expansion point $\varphi_1$: exactly when $Y_k \propto \varphi^4$ it drops out from the formula for $\etan$.

When we solve the coupled flow equations later on we set $\varphi_1 \to \infty$ for the expansion point. The motivation is that the relevant solutions $Y_k (\varphi)$ turn out to behave as $Y_k \propto \varphi^4$ for $\varphi \to \infty$. Hence with this choice $\etan$ is given by the CREH formula \eqref{3.51} which is independent of the precise value of $\varphi_1$ as long as it is sufficiently large. Other choices of $\varphi_1$ are conceivable, for instance minima of the potential. It turns out, however, that at least in the particular interesting UV fixed point regime $Y_k$ has no stationary points at all.
\subsubsection{The $\boldsymbol{S^4}$ Topology}\label{s.3.4.2}
Also in the $S^4$ case we shall send $\varphi_1$ to infinity. The corresponding sphere has a very large radius then and we can expect that $\etan$ is well approximated by the $R^4$ result. For this reason we are going to employ \eqref{3.51} also in the spherical case.
We defer a more precise treatment, including an evaluation of $\etan$ on a sphere of finite radius, to a future publication.
%
%
%
%
%
%
\section{General Properties of the LPA Flow}\label{s3s}
The RG flow on the infinite dimensional theory space with coordinates $\bigl( g, Y(\,\cdot\,) \bigr)$ is described by the coupled equations
\begin{subequations}\label{3s.1}
\begin{align}\label{3s.1a}
k \p_k \, Y_k (\varphi)
& =
\beta_Y \bigl( g_k, Y_k (\varphi) \bigr)
\\
\label{3s.1b}
k \p_k \, g_k
& =
\beta_g \bigl( g_k, [Y_k] \bigr)
= \Bigl[ 2 + \etan \bigl( g_k, [Y_k] \bigr) \Bigr] \, g_k.
\end{align}
\end{subequations}
The beta function for the potential, $\beta_Y$, is given by the RHS of \eqref{3.39} for $S^4$ and \eqref{3.42} for $R^4$. The anomalous dimension will be used in the form \eqref{3.50}. Its functional dependence on the potential is in terms of derivatives of $Y_k$ evaluated at the fixed expansion point $\varphi_1$ which we shall leave unspecified for the time being.

Alternatively we may use the flow equations in dimensionful form. The partial differential equation for $F_k (\phi) \equiv k^{-2} \, Y_k (k \, \phi)$ is given in \eqref{3.28} for $S^4$, and in \eqref{3.41} for $R^4$; in either case $G_k$ evolves according to $k \p_k \, G_k = \etan G_k$ with the $\etan$ of \eqref{3.49}.

In this section we are going to discuss various general properties of the coupled RG equations.
\subsection{The $\boldsymbol{\varphi \to \infty}$ Asymptotics}\label{s3s.1}
Let us separate off the CREH form of the potential, $Y_k\cc (\varphi) = c_0 \, \varphi^2 - \tfrac{1}{6} \, \lambda_k \, \varphi^4$, from the function $Y_k$:
\begin{align}\label{3s.2}
Y_k (\varphi) & \equiv \Delta Y_k (\varphi) + Y_k\cc (\varphi).
\end{align}
Exploiting \eqref{3.444} we can derive a flow equation for the ``correction term'' $\Delta Y_k$. For $S^4$ we obtain
\begin{subequations}\label{3s.3}
\begin{gather}\label{3s.3a}
k\p_k \, \Delta Y_k (\varphi) 
- (2+\etan) \, \Delta Y_k (\varphi) 
+ \varphi \, \Delta Y_k' (\varphi)
=
\etan \, \varphi^2
- \frac{g_k}{2 \pi} \, \mathbf{B}_k (\varphi).
\end{gather}
Here we abbreviated
\begin{align}\label{3s.3b}
\begin{split}
\mathbf{B}_k (\varphi)
& \equiv
\left[\rho (\varphi) - \frac{\etan}{2} \, \left( \rho (\varphi)
- \frac{\rhot (\varphi)}{\varphi^2\,} \right)
\right] \,
\left( 1 - 2 \, \lambda_k 
+ \frac{2 + \Delta Y_k'' (\varphi)}{\varphi^2\,} \right)^{-1}
\\
& \phantom{{==}}
- \left[ \frac{\varphi^4}{12} 
- \frac{\etan}{2} \, \frac{\varphi^4}{36} \right] \,
\left( 1 - 2 \, \lambda_k \right)^{-1}.
\end{split}
\end{align}
\end{subequations}
What prevents us from setting $\Delta Y_k =0$ is the inhomogeneity on the RHS of eq.\ \eqref{3s.3a}. It controls the type of terms which are generated during the RG running. It is important to notice that, in the limit $\varphi \to \infty$, this inhomogeneity simplifies under certain conditions. In fact, from \eqref{3.34} we know that $\rho \to \varphi^4/12$ and $(\rho - \rhot / \varphi^2) \to \varphi^4/36$ when $\varphi \to \infty$. Therefore we learn from \eqref{3s.3b} that \textit{$\mathit{\mathbf{B}_k (\varphi)}$ vanishes for $\mathit{\varphi \to \infty}$ if $\mathit{\Delta Y_k'' (\varphi) / \varphi^2 \to 0}$ for $\mathit{\varphi \to \infty}$}.
The latter condition means that the asymptotic growth of $\Delta Y_k (\varphi)$ is weaker than $\propto \varphi^4$. Let us assume that at some initial value of $k$ the potential $\Delta Y_k (\varphi)$ does indeed grow more slowly than $\varphi^4$. Then the RHS of \eqref{3s.3a}, for large values of $\varphi$, is proportional to $\varphi^2$ and therefore does not generate any components (monomials) in $\Delta Y_k$ that would grow $\propto \varphi^4$ or faster. This implies that \textit{if initially $\mathit{\Delta Y_k (\varphi)}$ contains no terms growing like $\mathit{\varphi^4}$ or faster then no such terms will be generated by the flow}. 

While eqs.\ \eqref{3s.3} were written down for $S^4$ it is obvious that this statement holds for both $S^4$ and $R^4$.

If $\Delta Y_k$ grows more slowly than $\varphi^4$ then, at any $k$, the large-$\varphi$ asymptotics of the full potential is always given by the CREH potential or, more specifically, by its cosmological constant term since the $\varphi^2$-piece is subdominant:
\begin{align} \label{3s.4}
Y_k (\varphi) & \to - \tfrac{1}{6} \, \lambda_k \, \varphi^4
\qquad \text{for } \varphi \to \infty.
\end{align}
\subsection{Structure of the Initial Value Problem}\label{s3s.2}
In the case at hand the specification of the truncated theory space involves choosing an appropriate space of functions on which $Y_k$ evolves, in particular boundary conditions must be defined for $Y_k$. The above discussion shows that it is consistent with the differential equation to partially define the theory space by the requirement that $Y_k (\varphi)$ does not grow faster than $\varphi^4$ for $\varphi \to \infty$. In the following we shall adopt this choice. It is further motivated by the RG fixed points we shall find in the next section. In particular a NGFP suitable for the asymptotic safety construction and all corresponding ``physical'' trajectories (those hitting it for $k \to \infty$) are inside the space of functions growing not stronger than $\varphi^4$.

As we mentioned already, the asymptotics \eqref{3s.4} motivates the choice $\varphi_1 \to \infty$ for the expansion point in the calculation of $\etan$. We shall adopt this choice in the following. It has the technical advantage that now $\etan$ is no longer the complicated functional of $Y_k$ given in \eqref{3.50}, but rather assumes the simple CREH form \eqref{3.51}. It depends on $g_k$ and only one particular characteristic of $Y_k$, namely the coefficient $\lambda_k$ appearing in its large-$\varphi$ asymptotics \eqref{3s.4}.

With the above two choices made, solving the initial value problem for the system (\ref{3s.1}a,b) proceeds as follows:
\begin{enumerate}
\item Fix, at the initial scale $\ki$, a Newton constant $g_{\ki}$ and a potential $Y_{\ki} (\varphi)$.
\item Extract the number $\lambda_{\ki} \equiv 
- 6 \, \lim_{\varphi \to \infty} Y_{\ki} (\varphi) / \varphi^4$ from the asymptotic behavior of the initial potential.
\item Solve the two coupled ordinary differential equations of the CREH truncation
\begin{align}\label{3s.5}
\begin{split}
k \p_k \, g_k\cc
& =\Bigl[ 2 + \etan\cc \bigl( g_k\cc, \lambda_k\cc \bigr) \Bigr] \, g_k\cc
\\
k \p_k \, \lambda_k\cc
& =\beta_\lambda\bcc \bigl( g_k\cc, \lambda_k\cc \bigr)
\end{split}
\end{align}
subject to the initial conditions $g_{\ki}\cc = g_{\ki}$, $\lambda_{\ki}\cc = \lambda_{\ki}$.
In \eqref{3s.5}, $\etan\cc$ and $\beta_\lambda\bcc$ are given by eqs.\ \eqref{3.51} and \eqref{3.45}, respectively.
\item Use the solution of \eqref{3s.5} to express the anomalous dimension as the following explicit function of $k$:
\begin{align} \label{3s.6}
\etan (k) & \equiv \etan\cc \bigl( g_k\cc, \lambda_k\cc \bigr).
\end{align}
\item Insert the solution $g_k = g_k\cc$ of \eqref{3s.5} and the anomalous dimension $\etan \equiv \etan (k)$ of \eqref{3s.6} into the partial differential equation for $Y_k$, eq.\ \eqref{3.39}, and solve it with the initial condition specified.
\end{enumerate}
Note that by this procedure the equations for $g_k$ and $Y_k$ get decoupled; at step (v) we are dealing with an equation for $Y_k$ alone, albeit one which contains complicated $k$-dependent coefficient functions.
\subsection{A Robustness Property of the CREH Approximation}\label{s3s.3}
The structure of the initial value problem for the LPA of conformally reduced gravity sheds an interesting light on the simpler CREH truncation. Both of them retain only the simplest derivative term, $(\p \varphi)^2$. The former allows for an arbitrary potential $Y_k (\varphi)$, the latter only for the $\varphi^4$-monomial. In this sense the LPA represents a refinement of the CREH approximation.

Usually when one refines a truncation ansatz by adding further field monomials to it the projection of some RG trajectory ``living'' in the new, higher dimensional space onto the smaller space of the old truncation equals only approximately (or in the case of badly chosen spaces, not at all) an old trajectory in the smaller space, as computed with the simpler beta functions of the lower dimensional flow. Stated the other way around, if we compute the flow of a small set of couplings in a simple truncation, then refine the truncation, and finally project the higher dimensional trajectories thus obtained onto the old, smaller space, we find that the projections of the higher dimensional trajectories will in general not coincide with the lower dimensional ones. (But they will approximately if the truncations are reliable.)

The above discussion of the initial value problem shows that the LPA of the conformal factor, seen as a refinement of the CREH approximation, is a notable exception to this rule: The projection of the RG trajectories in the infinite dimensional $\bigl( g, Y (\,\cdot\,) \bigr)$-space onto the $2$-dimensional $(g, \lambda)$-space coincides exactly with the trajectories computed from the two beta functions $\beta_g$ and $\beta_\lambda$ appropriate for the smaller space. Generalizing the space of potential functions from the $1$-dimensional line $\{ \lambda \, \varphi^4 \}$ to the infinite dimensional $\{Y (\,\cdot\,) \}$ has no impact on the RG evolution of $g_k$ and $\lambda_k$. This indicates once more that the Einstein--Hilbert truncation is particularly robust under the inclusion of further invariants.

This robustness property has, among others, the following implication. We know \cite{frank1} that some trajectories of the Einstein--Hilbert truncation, those of Type IIIa, terminate at a finite scale, and the same was found to be true for the corresponding CREH trajectories [\,I\,]. Because the LPA trajectories $\bigl( g_k, Y_k (\,\cdot\,) \bigr)$ are generalizations of CREH trajectories $(g_k, \lambda_k)$ in the sense explained above, the breakdown of a CREH trajectory at some $\kt$ implies that the corresponding LPA trajectory terminates at the same scale $k = \kt$. Generalizing the $\varphi^4$-term to an arbitrary function $Y_k (\varphi)$ seems not to help in extending the validity of the truncation towards the IR. This result is somewhat surprising since in the case of scalar matter field theories trajectories which break down at a finite scale within a polynomial truncation (those in the broken phase) typically can reach $k=0$ when the LPA is employed.
\subsection{The $\boldsymbol{\varphi \to 0}$ Behavior}\label{s3s.4}
Next we analyze the partial differential equation for $Y_k (\varphi)$ in the limit $\varphi \to 0$. We first focus on the inhomogeneous term of the flow equation, the last term on the RHS of \eqref{3.39}. We may assume that for $\varphi \to 0$ the function $\rhot (\varphi)$ vanishes at least proportional to $\varphi^2$. Hence, in the numerator of the inhomogeneous term, $(1 - \etan/2) \, \varphi^2 \, \rho (\varphi) + \tfrac{1}{2} \, \etan \, \rhot (\varphi) = \order{\varphi^2}$. In order to estimate the denominator $\varphi^2 + Y_k'' (\varphi)$ we decompose $Y_k$ according to $Y_k (\varphi) = Y_k\rr (\varphi) + Y_k\sr (\varphi)$ where the ``regular'' part $Y_k\rr$ is defined by the condition $( Y_k\rr)'' (\varphi) < \infty$ for $\varphi \to 0$, while for the ``singular'' part $Y_k\sr$ the second derivative blows up for vanishing field: $(Y_k\sr)'' (\varphi) \to \infty$ for $\varphi \to 0$. Let us assume a behavior of the form $(Y_k\sr)'' (\varphi) =  a / \varphi^\mu$ with constants $a \neq 0$ and $\mu >0$. Since, by definition, $(Y_k\rr)''$ is $\order{1}$ or vanishes even this implies the following structure of the inhomogeneous term:
\begin{align*}
\frac{\order{\varphi^2}}{\varphi^2 + \order{1} + a \, \varphi^{-\mu}\,}
& =
\frac{\order{\varphi^{2+\mu}}}
{\varphi^\mu \left[ \varphi^2 + \order{1} \right] + a}
= \order{\varphi^{2+\mu}}.
\end{align*}
We conclude that the inhomogeneity in the flow equation vanishes $\propto \varphi^{2+\mu}$, i.\,e.\ with a power larger than $2$, when $\varphi \to 0$. Therefore, in a small-$\varphi$ expansion it contains only terms whose second $\varphi$-derivative vanishes for $\varphi \to 0$ at least like $\varphi^\mu$. Hence, upon inserting $Y_k = Y_k\rr + Y_k\sr$ into \eqref{3.39}, we see that there are no terms in the inhomogeneity that would match those in $Y_k\sr$. As a result, the ``sing'' part of the potential has no source term; its RG equation is homogeneous, implying that \textit{no ``sing'' terms are generated if they are absent originally}:
\begin{align}\label{3s.10}
& & & & 
Y_k\sr & = 0
& & \Longrightarrow
& k \p_k \, Y_k\sr & = 0.
& & & & 
\end{align}

Note that the attributes ``regular'' and ``singular'' refer to the second $\varphi$-derivative of $Y_k$, not the potential itself. As an example consider $Y (\varphi) = \varphi^m$ with an exponent $m \in \mathds{R}$. Then $Y'' (\varphi) = m \, (m-1) \, \varphi^{m-2}$ so that $Y$ is ``regular'' if $m=0$, or, $m=1$, or $m \geq 2$; otherwise it is ``singular''.

An argument similar to the above one has also been used in \cite{frankmach}, in a different physical regime though ($R \to 0$, corresponding to $\varphi \to \infty$ here).
%
%
%
\section{The Fixed Points}\label{s4}
In this section we continue the analysis of the RG equations \eqref{3s.1} by searching for fixed points $\bigl( g_\ast, Y_\ast (\,\cdot\,) \bigr)$, i.\,e.\ solutions of $\beta_Y \bigl( g_\ast, Y_\ast (\varphi) \bigr) =0= \beta_g \bigl( g_\ast, [Y_\ast] \bigr)$. Depending on whether $\beta_g=0$ is achieved by $g_\ast=0$ or $\etan \bigl( g_\ast, [Y_\ast] \bigr) =-2$ we call them, in a slight abuse of language, a Gaussian fixed point or a non-Gaussian fixed point, respectively. (For a more precise discussion of the distinction see \cite{livrev}.)
\subsection{The Gaussian Fixed Point ($\boldsymbol{R^4}$ and $\boldsymbol{S^4}$)}\label{s4.1}
We look for fixed points with $g_\ast =0$. As a consequence, the second condition $\beta_Y=0$ boils down to 
\begin{align}\label{4.2}
\left( 2 + \eta_\ast \right) \, Y_\ast (\varphi)
- \varphi \, Y_\ast' (\varphi) & = 0.
\end{align}
This equation obtains from \eqref{3.39} as well as from \eqref{3.42}, so it holds for both $S^4$ and $R^4$. Here $\eta_\ast \equiv \etan \bigl( g_\ast, [Y_\ast] \bigr)$ is the anomalous dimension at the fixed point. By eq.\ \eqref{3.50} it vanishes: $\eta_\ast =0$. The differential equation \eqref{4.2} is trivial to solve then, with the result
\begin{align}\label{4.3}
& & 
Y_\ast\gf (\varphi) & = c\, \varphi^2,
& g_\ast\gf & = 0.
& & 
\end{align}
Here $c$ is an arbitrary constant of integration.

As $c$ in not fixed by the equations it might seem that we found a $1$-parameter family of fixed points. However, let us recall the CREH form of the potential: $Y_k\cc (\varphi) = c_0 \, \varphi^2 - (\lambda_k /6) \, \varphi^4$. Its quadratic term $c_0 \, \varphi^2$ is linked to the kinetic term by a local Weyl rescaling, a ``$\sigma$-transformation''. (The $c_0 \, \varphi^2$-term follows from the $\tfrac{1}{6} \, \h R$-term in \eqref{3.1} upon inserting $\h g_{\mu \nu}$; both the $\tfrac{1}{6} \, \h R$- and the $\h \Box$-term emerge from $\int \!\sqrt{g\,} \, R$ when the conformal factor is separated off.) If we impose the condition of \textit{local} $\sigma$-invariance we must pick the value $c = c_0$ therefore; then $Y_\ast$ can be identified with (a part of) the conformal reduction of $\int \!\sqrt{g\,} \, R$. Thus $c=0$ for $R^4$ and $c=1$ for $S^4$. It should be noted here that local $\sigma$-invariance really amounts to an additional assumption. After insisting on a $\h g_{\mu \nu}$-metric of the $S^4$-type the split symmetry was broken down to \textit{global} $\sigma$-transformations, and they are not strong enough to forbid the generation of $\varphi^2$-terms unrelated to the kinetic term and $\int \!\sqrt{g\,} \, R$. For conceptual clarity we shall leave $c$ arbitrary in the following, but clearly $c = c_0$ is the most sensible choice.
\subsection{Linear Stability Analysis}\label{s4.2}
Given an arbitrary fixed point $\bigl( g_\ast, Y_\ast (\,\cdot\,) \bigr)$, not necessarily a Gaussian one, we can explore its stability properties by linearizing the flow in its vicinity according to
\begin{subequations}\label{4.4}
\begin{gather}\label{4.4a}
g_k = g_\ast + \delta g_k
\\
\label{4.4b}
Y_k (\varphi) = Y_\ast (\varphi) + \delta Y_k (\varphi).
\end{gather}
\end{subequations}
For the ``tangent vector'' $\bigl( \delta g_k, \delta Y_k (\,\cdot\,) \bigr)$ we make the ansatz
\begin{subequations}\label{4.5}
\begin{gather}\label{4.5a}
\delta g_k =
\vare \, y_g \, \left( \frac{k_0}{k} \right)^\theta
\equiv \vare \, y_g \, \ee^{-\theta t}
\\
\label{4.5b}
\delta Y_k (\varphi) =
\vare \, \U (\varphi) \, \left( \frac{k_0}{k} \right)^\theta
\equiv \vare \, \U (\varphi) \, \ee^{-\theta t}
\end{gather}
\end{subequations}
Here $k_0$ is an arbitrary fixed scale, $t \equiv \ln (k / k_0)$ denotes the ``RG time'',  and $\vare$ is an infinitesimal parameter. We shall determine $\bigl( \delta g_k, \delta Y_k (\,\cdot\,) \bigr)$ by expanding the flow equation to first order in $\vare$. Then $\bigl( y_g, \U (\,\cdot\,) \bigr)$ is an eigenvector (``scaling field'') of the corresponding infinite dimensional stability matrix with the negative eigenvalue (``critical exponent'') $\theta$.
Scaling fields with $\re \theta >0$ ($\re \theta <0$) are said to be relevant (irrelevant). They grow (are damped) along the RG flow, i.\,e.\ when $k$ is lowered.

By definition, the fixed point's UV critical manifold $\mathscr{S}_{\text{UV}}$, or synonymously, its unstable manifold, consists of all points $\bigl( g, Y (\,\cdot\,) \bigr)$ which are pulled into the fixed point by the inverse flow. The tangent space to $\mathscr{S}_{\text{UV}}$ at the fixed point is spanned by the relevant scaling fields. Hence $\Delta_{\text{UV}} \equiv \dimension \mathscr{S}_{\text{UV}}$ equals the number of relevant scaling fields, i.\,e.\ the number of positive $\re \theta$'s. (See \cite{livrev} for a more precise discussion.)

The scaling fields and dimensions are to be determined from the following two equations which are obtained by varying \eqref{3s.1a} and \eqref{3s.1b}, respectively. The first one is
\begin{subequations}\label{4.6}
\begin{gather}\label{4.6a}
\left[ 2 + \theta + \eta_\ast \right] \, \delta g_k
+ g_\ast \, \delta \etan =0
\end{gather}
and the second equation reads
\begin{align}\label{4.6b}
\begin{split}
&
\left[ 2 + \theta + \eta_\ast \right] \, \delta Y_k (\varphi) 
- \varphi \, \delta Y_k' (\varphi)
+ \delta \etan \, Y_\ast (\varphi)
\\
& \phantom{{=}} =
\frac{\delta g_k}{2 \pi} \,
\frac{\left( 1 - \eta_\ast/2 \right) \, \varphi^2 \, \rho (\varphi)
+ \tfrac{1}{2} \, \eta_\ast \, \rhot (\varphi)}
{\varphi^2 + Y_\ast'' (\varphi)}
\\
& \phantom{{===}}
- \frac{g_\ast}{4 \pi} \,\, \delta \etan\,\,
\frac{\varphi^2 \, \rho (\varphi) - \rhot (\varphi)}
{\varphi^2 + Y_\ast'' (\varphi)}
\\
& \phantom{{====}}
+ \frac{g_\ast}{2 \pi} \,
\frac{\left( 1 - \eta_\ast/2 \right) \, \varphi^2 \, \rho (\varphi)
+ \tfrac{1}{2} \, \eta_\ast \, \rhot (\varphi)}
{\left[ \varphi^2 + Y_\ast'' (\varphi) \right]^2\,}
\,\, \delta Y_k'' (\varphi).
\end{split}
\end{align}
\end{subequations}
Here $\eta_\ast \equiv \etan \bigl( g_\ast, [Y_\ast] \bigr)$ and $\delta \etan \equiv \etan \bigl( g_\ast + \delta g_k, [Y_\ast + \delta Y_k] \bigr) - \eta_\ast$. As is stands, the last formula holds for $S^4$. The one for $R^4$ follows by the replacement
\begin{align}\label{4.7}
& & 
\rho (\varphi) & \longrightarrow \tfrac{1}{12} \, \varphi^4,
& 
\rhot (\varphi) & \longrightarrow \tfrac{1}{18} \, \varphi^6.
& & 
\end{align}
Next we apply this analysis to the GFP.
\subsection{Stability Analysis of the GFP ($\boldsymbol{R^4}$ and $\boldsymbol{S^4}$)}\label{s4.3}
For the anomalous dimension given in \eqref{3.50} we obtain at the GFP \eqref{4.3}:
\begin{align}\label{4.8}
& &
\etan \bigl( g_\ast\gf, [Y_\ast\gf] \bigr) & = 0,
& (\delta \etan) \bigl( g_\ast\gf, [Y_\ast\gf] \bigr) & =0.
& & 
\end{align}
This turns \eqref{4.6a} and \eqref{4.6b} into the following two conditions for $\bigl( y_g, \U (\,\cdot\,) \bigr)$ and $\theta$:
\begin{subequations}\label{4.9}
\begin{gather} \label{4.9a}
\left[ 2 + \theta \right] \, y_g =0
\\
\label{4.9b}
\left[ 2 + \theta \right] \, \U (\varphi) - \varphi \, \U' (\varphi)
= \frac{y_g}{2 \pi} \, \frac{\varphi^2 \, \rho (\varphi)}{\varphi^2 + 2\,c}.
\end{gather}
\end{subequations}
The solutions to this linear system are easily found. As \eqref{4.9a} can be satisfied by either $y_g=0$ or $2 + \theta =0$ we have two types of scaling fields at the GFP. They can be summarized as follows:
\begin{enumerate}
\item For every $\theta \in \mathds{R}$ there exists a scaling field $(y_g=0, 
\U = \U_\theta)$ with scaling dimension $\theta$ and\footnote{Here we assume that the exponent $n$ of $\U_\theta \propto \varphi^n$ is real. At this point it is not clear though whether one should impose $n \in \mathds{R}$ or $\mathds{Z}$ or $\mathds{N}$, or one should allow all $n \in \mathds{C}$ even. We are going to discuss this issue of the proper choice for the $\{ Y (\,\cdot\,) \}$-function space in detail in Subsection \ref{s4.5}.}
\begin{align} \label{4.10}
\U_\theta (\varphi) & = \w c_\theta \, \varphi^{2+\theta}
\end{align}
with constants $\w c_\theta$. For $\theta >0$, $\theta =0$, and $\theta <0$ these scaling fields are relevant, marginal, and irrelevant, respectively.
\item There exists a single additional scaling field $(y_g \neq 0, 
\U = \h \U_{-2})$ with scaling dimension $\theta =-2$ and
\begin{align}\label{4.11}
\h \U_{-2} (\varphi) & =
- \frac{y_g}{2 \pi} \, \int \limits^\varphi \!\! \dd \ov{\varphi} ~
\frac{\ov{\varphi} \, \rho (\,\ov{\varphi}\,)}{\,\ov{\varphi}^{\,2} + 2 \, c}.
\end{align}
This scaling field is irrelevant.
\end{enumerate}

The scaling fields of type (i) are remarkable in that a monomial $\varphi^n$, $n \in \mathds{R}$, has the dimension
\begin{align}\label{4.12}
\theta & = n-2.
\end{align}
Relative to a standard scalar field theory, the scaling dimension at the GFP is shifted by $2$ units. This shift can be traced back to the additional factors of $\phi$ in the FRGE which, in turn, originate from the $\chib$-dependence of $\mathcal{R}_k$ which was needed in order to give the desired physical interpretation to the cutoff and to implement ``background independence''. Again we see that the RG behavior of conformally reduced gravity, despite its appearance, is very different from that of a scalar matter field theory.

In Table \ref{tab1} we list the scaling fields with integer exponents $n$ as an example, along with their dimensions. The two entries printed bold-faced are the scaling fields accessible to the conformally reduced Einstein--Hilbert truncation. The scaling field $(y_g=0, \varphi^4)$ with $\theta = +2$ corresponds to the $2$-component eigenvector of the stability matrix for the $(g, \lambda)$-plane which is parallel to the $\lambda$-axis; the other one, $(y_g \neq 0, \h \U_{-2})$ with $\theta =-2$, represents the eigenvector with both a non-vanishing $g$- and $\lambda$-component. This pattern is exactly the same as in the full \cite{frank1,oliver1} and the conformally reduced [\,I\,] Einstein--Hilbert truncation. In fact, $\h \U_{-2}$ is essentially $\propto \varphi^4$, plus corrections which are subleading at large $\varphi$. If we perform the integral \eqref{4.11} with the spectral function $\rho$ of \eqref{4.7},
appropriate for the $R^4$ case, we obtain
\begin{align} \label{4.13}
\h \U_{-2} (\varphi)
& =
- \frac{y_g}{96 \pi} \, \Bigl[ \varphi^4 - 4 \, c \, \varphi^2
+ 8 \, c^2 \, \ln (\varphi^2 + 2 \,c) \Bigr].
\end{align}
If we invoke local $\sigma$-invariance and set $c = c_0 [R^4] =0$ we find that $\h \U_{-2} \propto \varphi^4$ on $R^4$.
%

On $S^4$ the smooth $\rho (\varphi)$ given by the polynomial \eqref{3.36} yields the analogous scaling field (if $c \geq 0$):
\begin{align}\label{4.14}
\begin{split}
\h \U_{-2} (\varphi)
& =
- \frac{y_g}{2 \pi} \, \left[
\frac{1}{48} \, \varphi^4 + \frac{a_3}{3} \, \varphi^3
+ \frac{1}{2} \, \left( a_2 - \frac{c}{6} \right) \, \varphi^2
+ \left( a_1 - 2\, c\, a_3 \right) \, \varphi
\right.
\\
& \phantom{{==}- \frac{y_g}{2 \pi} \, \biggl[}\left.
+ \sqrt{2 \,c \,} \, \left( 2 \, c \, a_3 - a_1 \right) \,
\arctan \bigl( \varphi / \sqrt{2\,c\,}\, \bigr)
\right.
\\
& \phantom{{===}- \frac{y_g}{2 \pi} \, \biggl[}\left.
+ \frac{1}{2} \, \left( 1 - 2\,c\,a_2 + \frac{c^2}{3} \right) \,
\ln (\varphi^2 + 2\,c) + const \right].
\end{split}
\end{align}
We see that $\h \U_{-2}$ becomes a pure $\varphi^4$-term for $\varphi \gg 1$. For $\varphi \to 0$ there would seem to be a singularity in the special case $c=0$. But fortunately the distinguished value is $c = c_0 [S^4] =1$, and for this value $\h \U_{-2}$ is regular for all $\varphi \geq 0$. For $\varphi \to 0$ it approaches a constant.
\begin{table}[t]
\centering
\begin{tabular}{|cc||c|p{1cm}|p{1cm}|p{1cm}|p{1cm}|p{1cm}|p{1cm}|p{1cm}|c|}
\hline
\multicolumn{2}{|c||}{$\theta$} & $\cdots$ & \centering $-3$ & \centering $-2$ & \centering $-1$ & \centering $0$ & \centering $+1$ & \centering $+2$ & \centering $+3$ & $\cdots$ \\
\hline
& & \multicolumn{4}{|c|}{irrelevant} & \centering marg. & \multicolumn{4}{|c|}{relevant} \\
\hline
$y_g=0$,&$\Upsilon=$ & $\cdots$ & \centering $\varphi^{-1}$ & \centering $\varphi^{0}$ & \centering $\varphi^{1}$ & \centering $\varphi^{2}$ & \centering $\varphi^{3}$ & \centering $\boldsymbol{\varphi^{4}}$ & \centering $\varphi^{5}$ & $\cdots$ \\
\hline
$y_g \neq 0$,&$\Upsilon=$ & \multicolumn{2}{|c|}{} & \centering $\boldsymbol{\h \Upsilon_{-2}}$ & \multicolumn{6}{|c|}{} \\
\hline
\end{tabular}
\caption{Scaling fields at the Gaussian fixed point}
\label{tab1}
\end{table}
\subsection{The non-Gaussian Fixed Point}\label{s4.4}
Let us now try to find fixed points $(g_\ast, Y_\ast)$ with $\eta_\ast =-2$. We begin with the flat case.
\subsubsection{The $\boldsymbol{R^4}$ Topology}\label{s.4.4.1}
According to \eqref{3.42} and \eqref{3.50} the coupled system to be solved is, with $\varphi_1$ fixed but unspecified for the time being,
\begin{subequations}\label{4.15}
\begin{gather}\label{4.15a}
\varphi \, Y_\ast' (\varphi)
=
- \frac{g_\ast}{18 \pi} \,
\frac{\varphi^6}
{\varphi^2 + Y_\ast'' (\varphi)}
\\
\label{4.15b}
\frac{g_\ast}{48 \pi} \,
\frac{\Bigl[ \varphi_1^3 \, Y_\ast''' (\varphi_1) \Bigr]^2}
{\Bigl[ \varphi_1^2 + Y_\ast'' (\varphi_1) \Bigr]^4\,}
= 1.
\end{gather}
\end{subequations}

In order to analyze \eqref{4.15a} it is convenient to introduce $h (\varphi) \equiv Y_\ast' (\varphi)$ which satisfies the first order equation
\begin{align}\label{4.16}
- \left( \frac{g_\ast}{18 \pi} \right) \,
\frac{\varphi^5}
{\varphi^2 + h' (\varphi)}
& = h (\varphi).
\end{align}
An asymptotic analysis of this equation reveals that, to leading order, its solution has essentially the same behavior for $\varphi \ll 1$ and $\varphi \gg 1$:
\begin{align} \label{4.17}
h (\varphi) & \approx
\begin{cases}
- \tfrac{2}{3} \, L \, \varphi^3 & \text{if } \varphi \ll 1
\\
- \tfrac{2}{3} \, \w L \, \varphi^3 & \text{if } \varphi \gg 1
\end{cases}
\end{align}
Here $L$ and $\w L$ are constants which must satisfy the same quadratic equation,
\begin{align} \label{4.18}
12 \pi \, L \, \left( 1 - 2 \, L \right) & = g_\ast,
\end{align}
and likewise for $\w L$. This equation can have two different real solutions, so $L$ and $\w L$ can be different in principle. The behavior \eqref{4.17} motivates introducing a function $W (\varphi)$ for $0 \leq \varphi < \infty$ by
\begin{align}\label{4.19}
h (\varphi) & \equiv - \frac{2}{3} \, L \, \varphi^3 \, W (\varphi).
\end{align}
The new function satisfies the boundary conditions
\begin{align}\label{4.20}
& & 
W (0) & = 1,
& W (\infty) & = \w L / L.
& & 
\end{align}
Inserting \eqref{4.19} into \eqref{4.16} we find that the relevant differential equation has the structure
\begin{align}\label{4.21}
\varphi \, \frac{\dd}{\dd \varphi} \, W (\varphi)
& =
\mathcal{B} \bigl( W (\varphi) \bigr)
\end{align}
with the ``beta function''
\begin{align}\label{4.22}
\mathcal{B} (W) & =
\frac{3}{2 \, L} \, \Bigl[ 1 - 2\,L\,W - \left(1 - 2\,L \right) / W \Bigr].
\end{align}
Being first order in $\varphi$, the equation \eqref{4.21} with both boundary conditions \eqref{4.19} imposed is overdetermined and the existence of a solution is questionable a priori. If we start the integration of \eqref{4.21} at $\varphi =0$ with the initial condition $W(0)=1$ then the resulting solution $W (\varphi)$ has no reason to approach $\w L/L$ for $\varphi \to \infty$, at least for a generic $\mathcal{B} (W)$. However, the function \eqref{4.22} has a special property which implies the existence of a solution. It has zeros at $W=1$ and $W = \w L/L$:
\begin{align*}
& & 
\mathcal{B} (1) & = 0,
& \mathcal{B} (\w L/L) & = 0.
& & 
\end{align*}
(To verify the second zero one must exploit that $L$ and $\w L$ satisfy \eqref{4.18}.) Obviously $W=1$ is a ``fixed point'' of the ``flow equation'' \eqref{4.21}: If we impose the initial condition $W(0)=1$ and integrate towards larger $\varphi$'s we obviously get
\begin{align}\label{4.23}
W (\varphi) =1 \qquad \text{for} \quad 0 \leq \varphi < \infty.
\end{align}
This solution also satisfies the second boundary condition $W (\infty) = \w L/L$ if $\w L$ and $L$ are the same solution of the above quadratic equation. (Starting from the other ``fixed point'' $W = \w L/L$ and integrating backward from $\varphi = \infty$ leads to the same conclusion.)

Thus, integrating \eqref{4.19} with \eqref{4.23}, we have shown that the fixed point potential is of the form
\begin{align} \label{4.24}
Y_\ast (\varphi) & = y_\ast - \tfrac{1}{6} \, L \, \varphi^4
\end{align}
where $L$ is a solution of \eqref{4.18}, and $y_\ast$ is a constant.

The second condition $(g_\ast, Y_\ast)$ must satisfy is \eqref{4.15b}. Inserting the result \eqref{4.24} for $Y_\ast(\varphi)$ it boils down to
\begin{align}\label{4.25}
g_\ast \, \frac{L^2}{\left( 1 - 2\,L \right)^4\,} & = 3 \pi.
\end{align}
Note that, thanks to the specific form of $Y_\ast$, the expansion point $\varphi_1$ dropped out of this equation.

What remains to be done is to solve the two coupled algebraic equations \eqref{4.18} and \eqref{4.25} for the constants $g_\ast$ and $L$. Actually, with the identification $L \equiv \lambda_\ast$, these two equations are exactly the same as those we encountered in Subsection 5.2 of [\,I\,] where we determined the NGFP $(g_\ast, \lambda_\ast)$ within the CREH approximation (using the ``kinetic'' version of the anomalous dimension, $\etan^{\text{(kin)}}$). As a result, the much more general LPA yields the same fixed point values for $g$ and $\lambda$ as the conformally reduced Einstein--Hilbert truncation: $g_\ast = g_\ast\cc$, $L \equiv \lambda_\ast = \lambda_\ast\cc$. Recalling the values found in [\,I\,] we can summarize our result for the NGFP as follows:
\begin{subequations}\label{4.26}
\begin{align}\label{4.26a}
Y_\ast (\varphi) & = y_\ast - \frac{1}{6} \, \lambda_\ast \, \varphi^4
\\
\label{4.26b}
\lambda_\ast & = \frac{1}{2} \, \frac{2^{1/3}}{\left( 1+2^{1/3} \right)}
\approx 0.279
\\
\label{4.26c}
g_\ast & = 6 \pi \, \frac{2^{1/3}}{\left( 1+2^{1/3} \right)^2\,}
\approx 4.650
\end{align}
\end{subequations}
The constant $y_\ast$ is not determined by the flow equation. Except for this constant, the equations have not taken advantage of the possibility to generalize the fixed point potential beyond the functional form of the CREH approximation. This result is quite remarkable. It might be related to the impressive robustness and stability properties the full Einstein--Hilbert truncation is known to possess \cite{oliver1,oliver2}.
\subsubsection{The $\boldsymbol{S^4}$ Topology}\label{s.4.4.2}
In the case of $S^4$ the condition $\etan=-2$, again, translates to eq.\ \eqref{4.15b}, while the other fixed point condition $\beta_Y=0$, by \eqref{3.39}, becomes
\begin{align} \label{4.27}
Y_\ast' (\varphi) & = - \frac{g_\ast}{\pi} \,
\frac{\varphi \, \rhoe (\varphi)}{\varphi^2 + Y_\ast'' (\varphi)}
\end{align}
with the effective $\rho$-function
\begin{align}\label{4.28}
\rhoe (\varphi) & \equiv 
\rho (\varphi) - \frac{\rhot (\varphi)}{2 \, \varphi^2\,}.
\end{align}
For $\varphi \to \infty$, $\rhoe$ approaches $\varphi^4/18$ so that \eqref{4.27} coincides with its $R^4$ counterpart, eq.\ \eqref{4.15a}, in this limit. This implies that the asymptotic form of the solution to \eqref{4.27} agrees with that of the $R^4$ solution:
\begin{align}\label{4.29}
Y_\ast (\varphi)
& \xrightarrow[~\varphi \to \infty~]{}
- \tfrac{1}{6} \, \lambda_\ast \, \varphi^4
\equiv Y_{\text{asym}} (\varphi).
\end{align}
Inserting $Y_\ast = Y_{\text{asym}}$ into \eqref{4.27} and letting $\varphi \to \infty$ we obtain a first relation among the constants $g_\ast$ and $\lambda_\ast$; it is given by eq.\ \eqref{4.18} with $L \equiv \lambda_\ast$.

Furthermore, the fixed point condition $\etan=-2$ has the explicit form \eqref{4.15b}. We adopt the choice $\varphi_1 \to \infty$ here. As a result, we may replace $Y_\ast$ by $Y_{\text{asym}}$ in \eqref{4.15b}. For $\varphi_1 \to \infty$ the point $\varphi_1$ actually drops out as before, and what remains is a second relation among $g_\ast$ and $\lambda_\ast$; it is given by eq.\ \eqref{4.25} with $L \equiv \lambda_\ast$. The conditions \eqref{4.18} and \eqref{4.25} determine $g_\ast$ and $\lambda_\ast$ uniquely. Therefore we can conclude that, exactly as for $R^4$, the constants $\lambda_\ast$ and $g_\ast$ assume precisely their CREH values \eqref{4.26b} and \eqref{4.26c}, respectively.

So, what is left to be done is to solve the ordinary differential equation \eqref{4.27} for the by now known value of $g_\ast$ subject to initial conditions which involve $\lambda_\ast$. We require that for large $\varphi$ the function $Y_\ast$ agrees with $Y_{\text{asym}}$, and likewise their derivatives. We start the integration at some $\h \varphi$ where we impose the initial conditions
\begin{align}\label{4.30}
& & 
Y_\ast (\h \varphi\,) & = Y_{\text{asym}} (\h \varphi\,)
& Y_\ast' (\h \varphi\,) & = Y_{\text{asym}}' (\h \varphi\,).
& & 
\end{align}
From $\h \varphi$ we integrate backward towards smaller $\varphi$. Ultimately we are interested in the limit $\h \varphi \to \infty$.

The existence of a fixed point potential is by no means guaranteed. It could happen that the function one obtains by integrating \eqref{4.27}, \eqref{4.30} towards smaller $\varphi$ develops an unacceptable singularity at some point. Then we would have to conclude that there is no fixed point for $Y_k$. The potentially dangerous feature of the differential equation \eqref{4.27} is the denominator $\varphi^2 + Y_\ast'' (\varphi)$ which could possible vanish at some $\varphi$.

Let us assume this does not happen and there exists a solution $Y_\ast$ which is well defined for $0 \leq \varphi \leq \h \varphi \to \infty$. Then this function is necessarily monotonically decreasing, i.\,e.\ $Y_\ast' (\varphi) <0$ everywhere. The reason is as follows. A non-singular solution has $\varphi^2 + Y_\ast'' (\varphi)>0$ everywhere, and also the function $\rhoe (\varphi)$ is found to be positive for any $\varphi$. Since $g_\ast >0$, this entails that the RHS of the fixed point equation \eqref{4.27} is negative.

Unfortunately it is not possible to solve the differential equation for $Y_\ast$ analytically in closed form. We shall have to resort to numerical techniques therefore. It is easy, however, to find the leading behavior for $\varphi \ll 1$. Making a power series ansatz and working out the coefficients one gets
\begin{align}\label{4.31}
Y_\ast (\varphi) = y_0 - y_1 \, \varphi
- \frac{g_\ast}{6 \pi \, y_1} \, \varphi^3
+ \order{\varphi^4}.
\end{align}
The constants of integration $y_0$ and $y_1$ are undetermined at this stage; ultimately they must get fixed by matching \eqref{4.31} with the large-$\varphi$ solution. The monotonicity $Y_\ast' (\varphi) <0$ requires $y_1 >0$. Eq.\ \eqref{4.31} implies that $Y_\ast'' (\varphi) = \order{\varphi}$ which vanishes for $\varphi \to 0$.

\begin{figure}[ht]
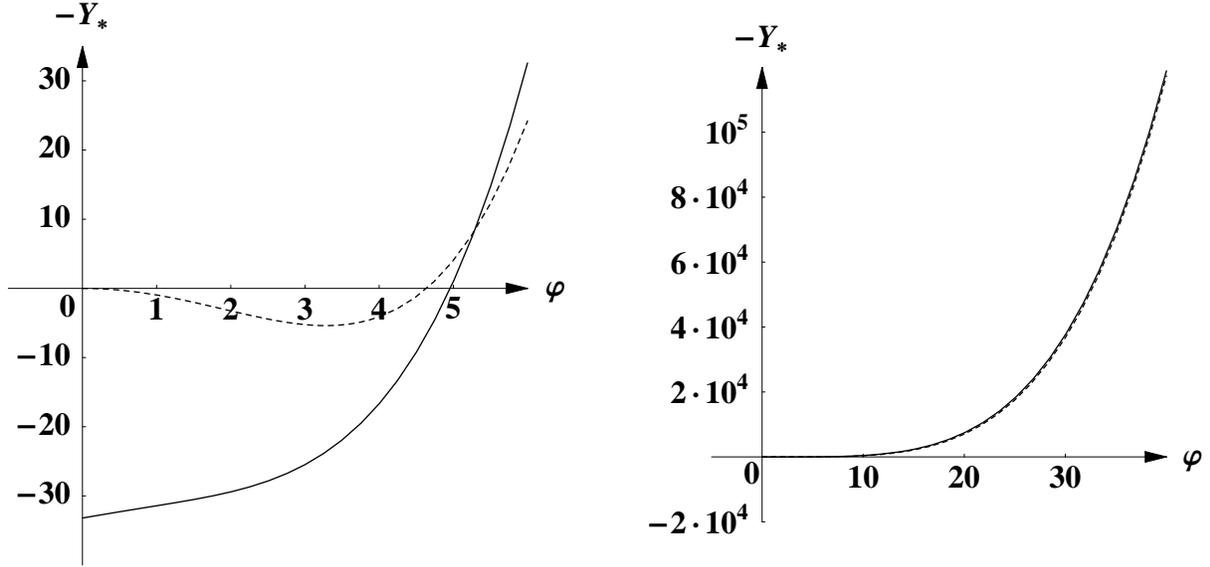

\centering
\pgfuseimage{fixpot-s}
\hfill
\pgfuseimage{fixpot-s-1}
\caption{The negative fixed point potential $-Y_\ast$ for $S^4$. The dashed line is the CREH potential with the same value for $\lambda_\ast$. The figure on the LHS is a detail of the one on the RHS; the latter shows that $Y_\ast$ coincides with the CREH potential asymptotically.}
\label{fixpot-s}
\end{figure}

The numerical analysis of \eqref{4.27} shows that there does indeed exist a solution which is regular for all $\varphi \geq 0$. It is displayed in Fig.\ \ref{fixpot-s} where it is compared to the CREH potential $Y_\ast\bcc (\varphi) = \varphi^2 - (\lambda_\ast /6) \, \varphi^4$ with the same value of $\lambda_\ast$ (which is the same as $Y_{\text{asym}}$, of course). In the Figure we actually plot the \textit{negative} fixed point potential $-Y_\ast$ since, up to a positive factor, it agrees with the true potential $U_\ast$, the non-derivative term in die Euclidean $\Gamma_k$.
As expected, $Y_\ast (\varphi)$ is indeed a monotonic function, contrary to the CREH potential. We also observe that $Y_\ast$ is approximately linear for small $\varphi$, in accord with \eqref{4.31}.

It is important to observe that $Y_\ast (\varphi)$ and $Y_\ast'' (\varphi)$ are finite for $\varphi \searrow 0$. The latter result means that the $S^4$ fixed point potential has a vanishing ``sing'' part in the sense of Subsection \ref{s3s.4}. The same is also true for $R^4$ as the explicit formula \eqref{4.24} shows.
\subsection{Stability Analysis of the NGFP}\label{s4.5}
In this subsection we perform a linear stability analysis of the non-Gaussian fixed point. According to the asymptotic safety idea, the infinite cutoff limit of QEG is to be taken at this fixed point. As a consequence, the structure of the NGFP's ultraviolet critical manifold $\mathscr{S}_{\text{UV}}$ determines the physical properties of the theory in an essential way. In particular its dimensionality $\Delta_{\text{UV}} \equiv \dimension \mathscr{S}_{\text{UV}}$ equals the number of parameters it can depend on and which are not fixed by the requirement of a ``safe'' UV behavior.
\subsubsection{Scaling Fields and the Condition $\boldsymbol{n'<4}$}\label{s4.5.1}
We restrict ourselves to the $R^4$ topology here. Then the relevant fixed point solution is \eqref{4.26} with the CREH values of $g_\ast$ and $\lambda_\ast$. The first condition for the scaling fields and dimensions is \eqref{4.6a}, and the second one is \eqref{4.6b} with the spectral functions \eqref{4.7} inserted.

Thanks to $\eta_\ast =-2$ the \textbf{first condition} assumes the form
\begin{align}\label{4.50}
(\delta \etan) \bigl( g_\ast, [Y_\ast] \bigr) & =
- \theta \, \delta g_k / g_\ast.
\end{align}
The anomalous dimension is given in eq.\ \eqref{3.50}. Varying it with respect to $g$ and $Y(\,\cdot\,)$ while keeping $\varphi_1$ fixed we obtain after inserting the fixed point values:
\begin{align}\label{4.51}
(\delta \etan) \bigl( g_\ast, [Y_\ast] \bigr)
& =
- 2 \, \frac{\delta g_k}{g_\ast}
+ \frac{g_\ast}{3 \pi} \, \frac{\lambda_\ast}{\left( 1-2\,\lambda_\ast \right)^4\,} \,
\left[ \frac{\delta Y_k'''(\varphi_1)}{\varphi_1}
+ \frac{8 \, \lambda_\ast}{1-2\,\lambda_\ast} \,
\frac{\delta Y_k''(\varphi_1)}{\varphi_1^2\,} \right].
\end{align}
Using \eqref{4.51} together with \eqref{4.5} in \eqref{4.50} we arrive at a condition relating the ``components'' of the scaling field, $\bigl( y_g, \U (\,\cdot\,) \bigr)$, and its dimension, $\theta$. It is convenient to distinguish the cases $y_g=0$ and $y_g \neq 0$. For $y_g=0$ the condition boils down to the $\theta$-independent relation
\begin{align}\label{4.52}
\frac{\U''' (\varphi_1)}{\varphi_1}
+ \frac{8 \, \lambda_\ast}{1-2\,\lambda_\ast} \,
\frac{\U'' (\varphi_1)}{\varphi_1^2\,}
& = 0.
\end{align}
For $y_g \neq 0$ the analogous condition reads
\begin{align}\label{4.53}
\theta & =
2 - \frac{g_\ast^2}{3 \pi \, y_g} \,
\frac{\lambda_\ast}{\left( 1-2\,\lambda_\ast \right)^4\,}
\, \left[ \frac{\U''' (\varphi_1)}{\varphi_1}
+ \frac{8 \, \lambda_\ast}{1-2\,\lambda_\ast} \,
\frac{\U'' (\varphi_1)}{\varphi_1^2\,} \right]
\end{align}

In the case at hand the \textbf{second condition} is found to be
\begin{gather}\label{4.54}
(-\theta) \, \bigl[ \U (\varphi) - y_\ast \, y_g / g_\ast \bigr]
+ \varphi \, \U'(\varphi)
- \alpha \, \varphi^2 \, \U'' (\varphi)
= y_g \, \left( \gamma_1 \, \theta - \gamma_2 \right) \, \varphi^4
\end{gather}
with the useful abbreviations
\begin{align}\label{4.55}
\alpha & \equiv
\frac{g_\ast}{18 \pi \, \left( 1-2\,\lambda_\ast \right)^2\,}
= \frac{2^{1/3}}{3} \approx 0.41997\cdots
\\
\label{4.56}
\gamma_1 & \equiv
\frac{1}{6} \, \left[ \frac{\lambda_\ast}{g_\ast}
- \frac{1}{24 \pi \, \left( 1-2\,\lambda_\ast \right)} \right]
\\
\label{4.57}
\gamma_2 & \equiv
\frac{1}{18 \pi \, \left( 1-2\, \lambda_\ast \right)}.
\end{align}
In the special case $y_g=0$ we get the homogeneous Eulerian differential equation
\begin{gather}\label{4.58}
\alpha \, \varphi^2 \, \U'' (\varphi)
- \varphi \, \U'(\varphi)
+ \theta \, \U (\varphi) =0.
\end{gather}

We shall now analyze the cases $y_g=0$ and $y_g \neq 0$ in turn.
\paragraph*{(a) Scaling fields with $\boldsymbol{y_g=0}$.}
In order to solve the differential equation \eqref{4.58} we make a power law ansatz $\U (\varphi) \propto \varphi^n$, admitting arbitrary complex exponents $n \in \mathds{C}$ a priori. The ansatz is a solution if $\theta = \theta (n)$ where
\begin{align}\label{4.59}
\theta (n) & =
\left( 1 + \alpha \right) \, n - \alpha \, n^2.
\end{align}
Solving for $n$ we get the two possible values
\begin{align}\label{4.60}
n_\pm (\theta) & =
\omega \pm \sqrt{\omega^2 - \theta / \alpha\,}
\end{align}
where
\begin{align}\label{4.61}
\omega \equiv \frac{1+\alpha}{2 \alpha}
\approx 1.690
\end{align}
So, if $\theta \neq \alpha \, \omega^2$, i.\,e.\ $n_+ \neq n_-$, the general solution of \eqref{4.58} for a given $\theta$ reads
\begin{align}\label{4.62}
\U (\varphi) & =
h_+ \, \varphi^{n_+ (\theta)} + h_- \, \varphi^{n_- (\theta)}
\end{align}
where $h_+$ and $h_-$ are arbitrary complex constants. In the exceptional case $\theta = \alpha \, \omega^2$ the general solution is instead ($h_{1,2} \in \mathds{C}$)
\begin{align}\label{4.63}
\U (\varphi) & =
\varphi^\omega \, \left[ h_1 + h_2 \, \ln \varphi \right].
\end{align}

Generically both $\theta$ and $n$ will be complex. We set $\theta \equiv \theta'+ i \theta''$ and $n \equiv n' + i n''$ with real and imaginary parts $\theta'$, $n'$ and $\theta''$, $n''$, respectively. Then \eqref{4.59} decomposes as
\begin{subequations}\label{4.64}
\begin{align}\label{4.64a}
\theta' & =
\left( 1+\alpha \right) \, n'- \alpha \, {n'}^2
+ \alpha \, {n''}^2
\\
\label{4.64b}
\theta'' & =
n'' \, \left[ 1 +\alpha -  2 \, \alpha \, n' \right].
\end{align}
\end{subequations}
We shall also need the inverse of these relations. After some algebra one obtains the following results for $n_\pm'$ and $n_\pm''$ as functions of $\theta'$ and $\theta''$. Three cases are to be distinguished:

\noindent\textbf{(i) The case $\boldsymbol{\theta'' \neq 0}$:}
\begin{subequations}\label{4.65}
\begin{align}\label{4.65a}
\begin{split}
n_\pm' (\theta', \theta'') & =
\omega \left( 1 \pm \frac{1}{\sqrt{2\,}\,} \,
\left[ \sqrt{\left( 1 - \frac{\theta'}{\alpha\,\omega^2\,} \right)^2 
+ \left( \frac{\theta''}{\alpha\,\omega^2\,} \right)^2\,}
+ \left( 1 - \frac{\theta'}{\alpha\,\omega^2\,} \right)
\right]^{1/2} \right)
\\
n_\pm'' (\theta', \theta'') & =
\mp \frac{\omega}{\sqrt{2\,}\,} \, \sign (\theta'') \,
\left[ \sqrt{\left( 1 - \frac{\theta'}{\alpha\,\omega^2\,} \right)^2 
+ \left( \frac{\theta''}{\alpha\,\omega^2\,} \right)^2\,}
- \left( 1 - \frac{\theta'}{\alpha\,\omega^2\,} \right)
\right]^{1/2}
\end{split}
\end{align}

\noindent\textbf{(iia) The case $\boldsymbol{\theta'' =0}$ and $\boldsymbol{\theta'\leq \alpha \, \omega^2}$:}
\begin{align} \label{4.65b}
\begin{split}
n_\pm' (\theta', \theta'') & =
\omega \, \left[ 1 \pm \sqrt{1 - \theta'/ \alpha \, \omega^2\,} \right]
\\
n_\pm'' (\theta', \theta'') & =
0
\end{split}
\end{align}

\noindent\textbf{(iib) The case $\boldsymbol{\theta'' =0}$ and $\boldsymbol{\theta' > \alpha \, \omega^2}$:}
\begin{align} \label{4.65c}
\begin{split}
n_\pm' (\theta', \theta'') & = \omega 
\\
n_\pm'' (\theta', \theta'') & =
\pm \omega \, \sqrt{\theta'/\alpha \, \omega^2 -1 \,}
\end{split}
\end{align}
\end{subequations}

By now we imposed only one of the two conditions scaling fields must meet. The other one, for $y_g=0$, is eq.\ \eqref{4.52}. If $\U (\varphi) \propto \varphi^n$ it reads
\begin{align}\label{4.66}
n \, (n-1) \, \left[ n-2+12\,\alpha \right] \, \varphi_1^{n-4} =0
\end{align}
where we used that $8 \, \lambda_\ast / (1-2\,\lambda_\ast) = 12 \, \alpha \approx 5.0395$. Eq.\ \eqref{4.66} is satisfied if at least one of the following conditions holds: $n=0$, $n=1$, $n=2-12\,\alpha$, $\varphi_1^{n-4}=0$. The first three of them provide us with finitely many scaling fields only, while we expect infinitely many, of course. Therefore we should demand that $\varphi_1^{n-4}=0$ in an appropriate sense. This condition implies that either $\varphi_1=0$ with $\re n >4$, or that $\varphi_1 \to \infty$ with $\re n <4$. The first case is clearly excluded; as the metric with $\varphi=0$ is singular one should not use $\varphi_1=0$ as the expansion point in the computation of $\etan$. It remains the option $\varphi_1 \to \infty$ which we had actually advocated for a different reason already: If $\varphi$ is large enough, $\etan$ becomes actually independent of it.

Thus we conclude on the basis of the stability analysis that our present approximation can lead to a consistent picture only with the choice $\varphi_1 \to \infty$. For the scaling fields with $y_g=0$, $\U (\varphi) \propto \varphi^n$ this entails that the exponent is constrained by
\begin{align} \label{4.67}
\re n (\theta) & \equiv n'(\theta', \theta'') <4.
\end{align}

\begin{figure}[ht]
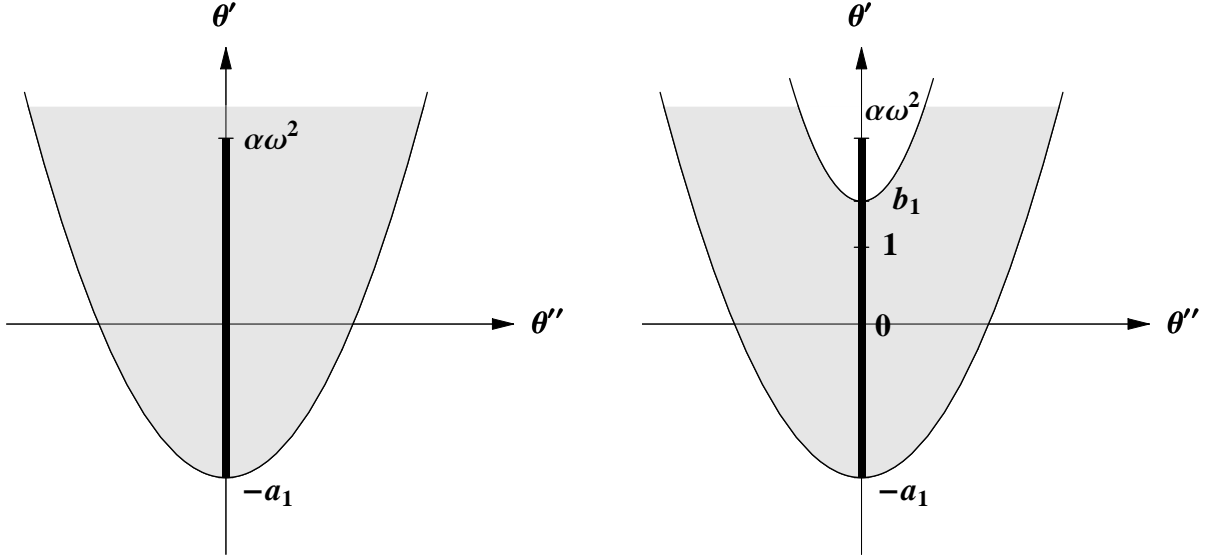

\centering
\pgfuseimage{para}
\hfill
\pgfuseimage{para-rechts}
\caption{Left figure: The points in the shaded region correspond to the scaling dimensions $(\theta', \theta'')$ realized by the ``$n_+$-family''. The bold-faced part of the vertical axis represents the scaling fields with $n''_+=0$. The points above the parabola satisfy $n_+' <4$. Right figure: The condition $n_+' \geq 2$ is imposed in addition. It is satisfied by all points below the upper parabola.}
\label{para}
\end{figure}

In order to find the spectrum of scaling dimensions we now combine the two conditions scaling fields must meet. The allowed values of $(\theta', \theta'')$ are such that the real parts $n_\pm'$ given by the expressions \eqref{4.65} are strictly smaller than $4$. After some tedious algebra one finds that there exist the following two families of scaling fields $(y_g, \U) = (0, \varphi^n)$:

\noindent\textbf{The ``$\boldsymbol{n_+}$-family'': } There exists a scaling field $(y_g, \U) = (0, \varphi^{n_+'+in_+''})$ of complex scaling dimension $\theta \equiv \theta'+i \theta''$ for every point $(\theta',\theta'')$ in the complex $\theta$-plane such that
\begin{align}\label{4.68}
\theta' & > - a_1 + a_2 \, {\theta''}^2
\end{align}
where $a_1 \equiv 4 \, (2^{1/3} -1) \approx 1.0397$ and $a_2 \equiv 3 \cdot 2^{1/3} \, (7 \cdot 2^{1/3}-3)^{-2} \approx 0.1116$. The associated exponents $n'+in''$ are given by the eqs.\ \eqref{4.65} with the upper sign. The domain of allowed scaling dimensions in the $\theta$-plane is bounded by the parabola depicted in Fig. \ref{para}.

\noindent\textbf{The ``$\boldsymbol{n_-}$-family'': } There exists a scaling field $(y_g, \U) = (0, \varphi^{n_-'+in_-''})$ of complex scaling dimensions $\theta \equiv \theta'+i \theta''$ for \textit{every} point $(\theta',\theta'')$ in the $\theta$-plane. The associated exponents $n'+in''$ are given by the eqs.\ \eqref{4.65} with the lower sign.

In either family scaling fields with $\theta'>0$, $\theta'=0$, $\theta'<0$ are relevant, marginal, and irrelevant, respectively.

We also found the exceptional solution \eqref{4.63} with $\theta'=\alpha\,\omega^2 \approx 1.2$, $\theta''=0$. Since $\omega \approx 1.69$ it grows more slowly than $\varphi^4$ asymptotically and defines a relevant scaling field therefore.

All scaling dimensions come in complex conjugate pairs. If $(y_g, \U)$ is a complex solution of the linearized flow equation for $(\theta',\theta'')$, so is $(y_g^\ast, \U^\ast)$ with $(\theta',-\theta'')$. Forming linear combinations we obtain the following two real solutions:
\begin{align}\label{4.69}
\begin{split}
\re 
\begin{pmatrix}
\delta g_k \\ \delta Y_k (\varphi)
\end{pmatrix}
& =
\vare \, \lvert h \rvert \, 
\begin{pmatrix}
0 \\ 1
\end{pmatrix}\,
\varphi^{n'} \, \ee^{-\theta' t} \,
\cos \bigl( n'' \ln \varphi - \theta'' t + \delta \bigr)
\\
\im 
\begin{pmatrix}
\delta g_k \\ \delta Y_k (\varphi)
\end{pmatrix}
& =
\vare \, \lvert h \rvert \, 
\begin{pmatrix}
0 \\ 1
\end{pmatrix}\,
\varphi^{n'} \, \ee^{-\theta' t} \,
\sin \bigl( n'' \ln \varphi - \theta'' t + \delta \bigr).
\end{split}
\end{align}
Here we wrote $\U (\varphi) = h \, \varphi^n$ and allowed for a complex amplitude $h = \lvert h \rvert \, \ee^{i \delta}$. In \eqref{4.69} the complex exponent $(n', n'')$ stands for either $(n_+', n_+'')$ or $(n_-', n_-'')$ and is given in terms of $(\theta', \theta'')$ by the expressions \eqref{4.65}. Obviously all linear solutions with $n'' \neq 0$ are non-polynomial in $\varphi$, and their dependence on both $\varphi$ and the scale is oscillatory. When $\varphi$ approaches the singular point of vanishing metric, $\varphi=0$, the solutions oscillate infinitely rapidly. Solutions with $n''=0$ do not oscillate, but are still non-polynomial unless $n'\in \mathds{N}$.
\paragraph*{(b) Scaling fields with $\boldsymbol{y_g \neq 0}$.}
Now we switch on the inhomogeneity in eq.\ \eqref{4.54}, the terms proportional to $y_g$. The general solution of the inhomogeneous equation is of the form $\U = \U_{\text{in}} + \U_{\text{hom}}$ where $\U_{\text{in}}$ is a special solution of the inhomogeneous, and $\U_{\text{hom}}$ the general solution of the homogeneous equation (with the same parameter $\theta$). Making a $\varphi^4$-ansatz one easily finds the following special inhomogeneous solution:
\begin{align}\label{4.70}
\U_{\text{in}} (\varphi) &=
\frac{y_\ast \, y_g}{g_\ast} \,
+ y_g \, \frac{\gamma_1 \, \theta -\gamma_2}{4 - \theta - 12 \, \alpha} \,
\varphi^4.
\end{align}
The second condition to be satisfied is \eqref{4.53}. Inserting $\U = \U_{\text{in}} + \U_{\text{hom}}$ with \eqref{4.70} we are led to
\begin{align} \label{4.71}
\theta & = 2 - \kappa \, 
\frac{\left( \gamma_1 \, \theta -\gamma_2 \right)}
{\left( 4 - \theta - 12 \, \alpha \right)}
+ \vartheta_{\text{hom}}
\end{align}
with
\begin{align}\label{4.72}
\vartheta_{\text{hom}} & \equiv
- \frac{g_\ast^2}{3 \pi \, y_g} \,
\frac{\lambda_\ast}{\left( 1-2\,\lambda_\ast \right)^4\,} \,
\left[ \frac{\U_{\text{hom}}'''(\varphi_1)}{\varphi_1}
+ \frac{8 \, \lambda_\ast}{1-2\,\lambda_\ast} \, 
\frac{\U_{\text{hom}}''(\varphi_1)}{\varphi_1^2\,} \right]
\end{align}
and the $\theta$-independent constant
\begin{align}\label{4.73}
\kappa & \equiv
\frac{8 \, g_\ast^2}{\pi} \, 
\frac{\lambda_\ast \, \left( 1+2\,\lambda_\ast \right)}{\left( 1-2\,\lambda_\ast \right)^5\,}
= 288 \pi \, \left( 1+2\,\lambda_\ast \right).
\end{align}
In \eqref{4.72} it is implied that $\varphi_1 \to \infty$, as before. The equation \eqref{4.71} determines the values of $\theta$ for which scaling fields with $y_g \neq 0$ exist. It can be solved in the following way:
\begin{enumerate}
\item We assume that its solutions $\theta_{1,2}$ are such that $\vartheta_{\text{hom}} =0$ for the function $\U_{\text{hom}}$ solving the homogeneous equation \eqref{4.58} with $\theta = \theta_{1,2}$.
\item We determine $\theta_{1,2}$ by solving the simplified equation in which $\vartheta_{\text{hom}} =0$:
\begin{align}\label{4.74}
\theta & = 2 - \kappa \, 
\frac{\left( \gamma_1 \, \theta -\gamma_2 \right)}
{\left( 4 - \theta - 12 \, \alpha \right)}.
\end{align}
\item We prove selfconsistency by showing that the values $\theta_{1,2}$ obtained do indeed belong to a homogeneous solution with $\vartheta_{\text{hom}} =0$.
\end{enumerate}

Eq.\ \eqref{4.74} is a quadratic equation for $\theta$. Recalling the definitions of the various constants one finds that the solutions are a complex conjugate pair $\theta_{1,2} = \theta' \pm i \theta''$ with
\begin{align}\label{4.75}
\begin{split}
\theta' & = 4
\\
\theta'' & = 2 \sqrt{2\,}\, \sqrt{1 + 3 \cdot 2^{1/3}\,}
\approx 6.1837
\end{split}
\end{align}

As for step (iii), the general homogeneous solution is given by ($h_\pm \in \mathds{C}$)
\begin{align}\label{4.76}
\U_{\text{hom}} &=
h_+ \, \varphi^{n_+ (\theta', \theta'')} +
h_- \, \varphi^{n_- (\theta', \theta'')}.
\end{align}
Using \eqref{4.65} we find that for the critical exponents \eqref{4.75} the real parts of the exponents $n_+$ and $n_-$ are given by
\begin{align} \label{4.77}
& & 
n_+' (\theta', \theta'') & \approx 3.87,
& n_-' (\theta', \theta'') & \approx -0.488
& & 
\end{align}
Since both of them are smaller than $4$, we indeed obtain $\vartheta_{\text{hom}} =0$ when we insert $\U_{\text{hom}}$ into \eqref{4.72} and let $\varphi_1 \to \infty$. This proves the consistency of the procedure.

We observe that the scaling field $(y_g \neq 0, \U_{\text{in}})$ is the only one which grows $\propto \varphi^4$ asymptotically; all the others have a weaker growth and are subdominant for $\varphi \to \infty$. This complex scaling field, or the two real ones equivalent to it, is exactly the one which is accessible to the Einstein--Hilbert truncation. In fact, the numbers \eqref{4.75} are exactly the critical exponents we found in the CREH approximation [\,I\,]. They describe the spiraling of the trajectories near the NGFP on the $(g, \lambda)$-theory space. In the present context this $2$-dimensional space is to be regarded as a subspace of the infinite dimensional $\bigl( g, Y (\,\cdot\,) \bigr)$-theory space.

To summarize, we can describe the linear flow near the NGFP on $\bigl( g, Y (\,\cdot\,) \bigr)$-space as follows. There are two scaling fields with a non-vanishing $g$-component. Their only other non-vanishing component is in the $\varphi^4$-direction of the function space $\{ Y(\,\cdot\,)\}$ which we identify with the $\lambda$-direction of the CREH truncation. Those two scaling fields are relevant, have the critical exponents $\theta'=4$, $\theta'' \approx 6.18$, and coincide \textit{exactly} with those of the CREH approximation (cf.\ the discussion in Subsection \ref{s3s.3}). All other scaling fields have vanishing $g$-component and correspond to perturbations of the potential alone. There are both relevant and irrelevant fields of this type; generically they correspond to non-polynomial oscillatory functions of $\varphi$.
\subsubsection{UV Critical Manifold and Subsidiary Conditions}\label{s.4.5.2}
In the simplest case, when the eigenvectors of the stability matrix at the NGFP form a complete set, the subset of eigenvectors belonging to eigenvalues $- \theta$ with $\re \theta >0$ spans the tangent space to the UV critical manifold $\mathscr{S}_{\text{UV}}$. Its dimensionality $\Delta_{\text{UV}}$ equals the number of scaling fields with $\re \theta >0$ then. They are relevant in the sense that they grow when $k$ is lowered. Since every complete trajectory inside $\mathscr{S}_{\text{UV}}$ defines a possible asymptotically safe quantum theory, there exists a $\Delta_{\text{UV}}$-parameter family of such theories if the solutions of the linearized flow equations generalize to solutions of the full nonlinear equations which extend down to $k=0$. If this latter condition is not satisfied for all scaling solutions the number of free parameters is smaller than $\Delta_{\text{UV}}$.

What the above stability analysis shows is that, to be precise, \textit{for a specific definition of the function space $\mathit{\{ Y(\,\cdot\,)\}}$} there are infinitely many directions in the truncated theory space along which the NGFP is UV attractrive \textit{at the linearized level}. Whether this result is directly relevant for the dimensionality of $\mathscr{S}_{\text{UV}}$ in full quantum gravity and its degree of predictivity is not clear yet.
The actual number $\Delta$ of free parameters the quantum theory has is decided at the nonlinear level only. This number can well be different from $\Delta_{\text{UV}}$ which refers to (the tangent space to) $\mathscr{S}_{\text{UV}}$ at the NGFP. Even in the full (untruncated) theory it can happen, for instance, that not all trajectories starting there can be continued to $k=0$, whence $\Delta < \Delta_{\text{UV}}$. Moreover, even if they reach $k=0$, most trajectories will in general be unacceptable according to additional physical criteria (existence of a classical regime, etc.) which also can decrease $\Delta$. This issue could be analyzed only by a comprehensive numerical analysis of the partial differential equation for $Y_k$. We shall not embark on this analysis here, in particular as it is not clear how the number $\Delta$ of the conformally reduced theory relates the corresponding number in the space of functionals depending on the full metric.

Let us now discuss the other issue mentioned above, the precise definition of theory space. Already at the linear level there exists an ambiguity which has an enormous impact on the number of relevant directions one finds, namely the choice of the function space $\{ Y(\,\cdot\,)\}$ in which all admissible potential terms and hence scaling fields are supposed to ``live''. In the above analysis the only condition which we imposed (besides differentiability) was an asymptotic growth not faster than $\varphi^4$. To show that the precise specification of the space $\{ Y(\,\cdot\,)\}$ is crucial we shall now consider various plausible choices. For the time being we do not know which one will be ultimately correct (in the sense of closest to full QEG).\vspace{\baselinestretch pt}

\subsubsection*{(a) Subsidiary condition $\boldsymbol{Y\sr \equiv 0}$: $\boldsymbol{n' \geq 2}$}
\noindent
In Subsection \ref{s3s.4} we saw that the RG flow does not generate contributions to $Y_k\sr$ if $Y_k\sr =0$ initially: $Y_k\sr =0 \Longrightarrow \p_k \, Y_k\sr =0$. Furthermore, in Subsection \ref{s4.4} we pointed out that, for both topologies considered, the fixed point potential has no singular part: $Y_\ast\sr =0$. This means that the space of functions $\{ Y \, \vert \, Y\sr =0 \} \equiv \mathcal{Y}$ contains the fixed point and is invariant under the RG flow: No trajectory starting on it will ever leave it.

As a consequence, we may consistently impose the subsidiary condition $Y\sr =0$ and define the theory on the smaller function space $\mathcal{Y}$. Imposing the condition $Y\sr =0$ implies a corresponding condition on allowed basis vectors in the tangent spaces of $\mathcal{Y}$, in particular at the NGFP: Allowed scaling fields have vanishing components in the ``sing'' directions. For the scaling fields of the type $(y_g =0, \U = \varphi^n)$ this condition is met when $\U'' = n \, (n-1) \, \varphi^{n-2}$ does not blow up in the limit $\varphi \to 0$, i.\,e.\ when $n' \equiv \re n \geq 2$ and in the special cases $n=0$ and $n=1$.

Imposing $n_\pm' \geq 2$ on the $(0, \varphi^n)$-scaling fields has the following consequences: (i) The $n_-$-family is eliminated completely; no $\theta \in \mathds{C}$ satisfies $n_-' (\theta', \theta'') \geq 2$. (ii) As for the $n_+$-family, $n_+' (\theta', \theta'') \geq 2$ is satisfied if 
\begin{align}\label{4.100}
\theta' & \leq b_1 + b_2 \, {\theta''}^2
\end{align}
with $b_1 = 2 \, (1 - 2^{1/3}/3) \approx 1.160$ and $b_2 = \tfrac{1}{3} \, 2^{1/3} / (2^{1/3}-1)^2 \approx 6.217$.
The scaling dimensions $(\theta', \theta'')$ of allowed scaling fields correspond to points in the $\theta$-plane which lie below the parabola $\theta' = b_1 + b_2 \, {\theta''}^2$ and, because of the $n_+' <4$ constraint, above the parabola \eqref{4.68}. The corresponding region is shown in the right diagram of Fig.\ \ref{para}.\vspace{\baselinestretch pt}

\subsubsection*{(b) Integer exponents: $\boldsymbol{n \in \mathds{Z}}$}
\begin{figure}[ht]
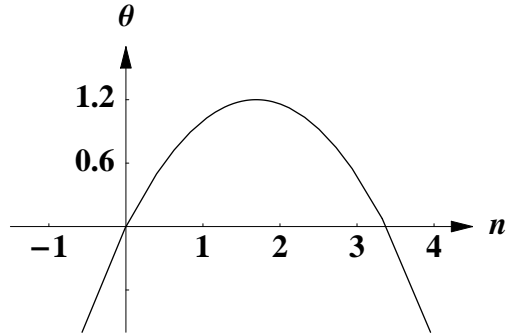

\centering
\pgfuseimage{theta}
\caption{The function $\theta (n)$ of \eqref{4.59} for $n$ real. It is positive only between $n=0$ and $n=2 \, \omega \approx 3.38$.}
\label{theta}
\end{figure}

%
\noindent
As another example, we assume that $\{ \varphi^n,~n \in \mathds{Z} \}$ is a ``basis'' on $\{ Y(\,\cdot\,)\}$ so that every function can be expanded in a Laurent series: $Y (\varphi) = \sum_{n \in \mathds{Z}} a_n \, \varphi^n$. The $y_g =0$-scaling fields are still the monomials $\varphi^n$, but $n$ is restricted to be real and integer now. As a consequence, the associated scaling dimension $\theta (n)$ is also real. It is given by eq.\ \eqref{4.59}. The function $\theta = \theta (n)$ for $n$ real is plotted in Fig.\ \ref{theta}.

\subsubsection*{(c) Real exponents: $\boldsymbol{n \in \mathds{R}}$}
\noindent
As a last example, let us assume that for some reason the oscillatory scaling fields are to be excluded so that we must impose $n'' =0$. Using \eqref{4.65} we can show that \textit{the scaling fields $\mathit{\varUpsilon \propto \varphi^n}$ with $\mathit{n_\pm'' =0}$ have dimensions $\mathit{\{ (\theta', \theta''=0) \,\vert\, -\infty < \theta' \leq \alpha \, \omega^2 \}}$ and real exponents $\mathit{n \equiv n_\pm' = \omega \, [ 1 \pm \sqrt{1-\theta'/\alpha\,\omega^2\,}\, ]}$.} In the complex $\theta$-plane the allowed dimensions lie on a part of the $\theta'$-axis. If, on top of $n''=0$, the constraint $n' <4$ is imposed we find:
\begin{itemize}
\item In the $n_-$-family, all scaling fields with $n_-''=0$ also satisfy $n_-'<4$.
\item In the $n_+$-family, the scaling fields which satisfy both $n_+''=0$ and $n_+'<4$ have scaling dimensions
\begin{align*}
\{ (\theta', \theta''=0) \,\vert\, -a_1 < \theta' \leq \alpha \, \omega^2 \}
\end{align*}
and real exponents $n \equiv n_+' = \omega \, [ 1 \pm \sqrt{1-\theta'/\alpha\,\omega^2\,}\, ]$. In Fig.\ \ref{para} these scaling fields correspond to the bold-faced interval on the vertical axis.
\end{itemize}

For the time being it is not clear how to choose the function space $\{ Y(\,\cdot\,) \}$ optimally, but it is obvious that the count of relevant directions crucially depends on it. The optimal choice would be the one for which the LPA mimics full quantum gravity as well as possible. In order to find it one would have to gain a better understanding of the invariants of the full metric $g_{\mu \nu}$, their importance for the flow, and their relation to the terms that could possibly appear in the conformally reduced action.

\vspace{2\baselinestretch pt}
We close this subsetion with several remarks.

\noindent\textbf{(i) } Also the NGFP found by the method of symmetry reduction \cite{max} where only metrics with two Killing vectors are quantized has infinitely many relevant directions. In fact, \textit{all} scaling fields accessible to this approximation are relevant. Their relation to the invariants of the full $4$-dimensional metric is likewise unclear.

\noindent\textbf{(ii) } For the unconstrained $\{ Y(\,\cdot\,) \}$-space the real form of the $y_g =0$-scaling field is given in \eqref{4.69}. These solutions are non-polynomial in $\varphi$ (if $n \notin \mathds{N}$). In a sense, they are analogous to the so-called \textit{Halpern--Huang scaling fields} \cite{hh1,hh2} in standard scalar matter field theories. Halpern and Huang analyzed their Gaussian fixed point by means of the LPA and found, quite unexpectedly, infinitely many relevant directions. They had been overlooked by perturbation theory because the eigenpotentials depend on the field in a non-polynomial way. As to yet, their status is still somewhat unclear. In the literature \cite{morris-hh} arguments have been put forward suggesting that one should eliminate them by an appropriate choice of the function space. Likewise discarding them in gravity we are back to only three relevant directions, $\varphi$, $\varphi^2$, and $\varphi^3$.

\noindent\textbf{(iii) } The above analysis applies to the $R^4$ topology. For $S^4$ it is not possible to obtain comparably transparent results in analytic form since the fixed point potential is known only numerically. However, we do not expect qualitatively new features to arise for $S^4$.
%
%
%
\section{Transition to a Phase with Unbroken \\Diffeomorphism Invariance}\label{s5}
An exploration of the RG flow and in particular the structure of $\mathscr{S}_{\text{UV}}$ at the nonlinear level has to rely upon numerical methods mostly. There are essentially two strategies for finding trajectories in $\mathscr{S}_{\text{UV}}$. The first one consists in a downward evolution (i.\,e.\ towards smaller values of $k$) starting at $Y_\ast$ plus a relevant scaling field. We tried this method for $R^4$ where all scaling fields at the NGFP are known explicitly; it turned out however that it is very difficult to implement it in a numerically reliable way because of the densely winding spirals near the fixed point.

The second strategy consists in guessing an initial point in theory space and evolving upward (towards larger $k$) from this point. If the point happens to lie on $\mathscr{S}_{\text{UV}}$ the trajectory will hit the fixed point for $k \to \infty$. In practice one will vary the initial point until the trajectory stays close to the NGFP for a very long RG time. Small errors in the guessed initial point will always cause the trajectory to run away from the NGFP ultimately. However, by finetuning the initial point one can construct a numerical approximation to a trajectory in $\mathscr{S}_{\text{UV}}$ at any desired level of accuracy. By systematically changing the initial point it is possible to trace out the nonlinear structure of $\mathscr{S}_{\text{UV}}$ in this way, at least in principle. A complete numerical analysis of this kind is beyond the scope of the present paper. We shall rather present some typical trajectories found by this trial and error method and describe their general properties.

Before turning to the numerical solutions we discuss a rather special, but instructive class of trajectories which can be found analytically.
\subsection{An Analytic Solution ($\boldsymbol{R^4}$ case)}\label{s5.1}
In the case of the $R^4$ topology there exists for every given trajectory of the Einstein--Hilbert truncation (here always taken of type IIIa) a $1$-parameter family of trajectories in $\mathscr{S}_{\text{UV}}$ which can be found in closed form:
\begin{align}\label{5.20}
Y_k (\varphi)
& =
C_1 \, \frac{g_k}{k} \, \varphi
- \frac{1}{6} \, \lambda_k \, \varphi^4.
\end{align}
It is easy to see that \eqref{5.20} solves eq.\ \eqref{3.42}. Here $C_1$ is a free parameter with the dimension of a mass. It is obvious that these trajectories lie inside $\mathscr{S}_{\text{UV}}$: for $k \to \infty$ the potentials \eqref{5.20} approach the fixed point $- \lambda_\ast \, \varphi^4 /6$. Consistent with the linear analysis the $\varphi$-term is relevant, it grows when $k$ is lowered. (Eq.\ \eqref{5.20} is valid beyond the linearization though.) In dimensionful units the corresponding potential has a $k$-independent $\phi$-term:
\begin{align}\label{5.21}
U_k (\phi)
&=
- \frac{3}{4 \pi} \, \left( C_1 \, \phi
- \frac{1}{6} \, \frac{\Lambda_k}{G_k} \, \phi^4 \right).
\end{align}
If $C_1 >0$ the functions $- Y_k (\varphi)$ and $U_k (\phi)$ have a critical point, a global minimum, at nonzero values of the field:
\begin{align}\label{5.22}
\varphi_0 (k)
& =
\left[ \frac{3}{2} \, C_1 \, \frac{g_k}{\lambda_k} \, \frac{1}{k} \right]^{1/3}
\\
\label{5.23}
\phi_0 (k)
& =
\left[ \frac{3}{2} \, C_1 \, \frac{G_k}{\Lambda_k} \right]^{1/3}
= \varphi_0 (k) / k.
\end{align}
If $C_1<0$ both $- Y_k (\varphi)$ and $U_k (\phi)$, for any $k$, assume their global minimum at vanishing $\varphi$ and $\phi$, respectively.

In the case $C_1>0$, $\varphi_0 (k)$ vanishes only in the limit $k \to \infty$ where $\varphi_0 (k) \propto k^{-1/3}$. While we lower $k$ from infinity down to the turning point scale $\kT$, the minimum $\varphi_0 (k)$ moves towards \textit{larger} $\varphi$-values. Slightly below $\kT$ the type IIIa trajectory enters the classical regime in which $G_k$, $\Lambda_k$, and $\phi_0 (k)$ are approximately $k$-independent. In this regime $\varphi_0 (k)$ is heading for \textit{smaller} values of $\varphi$ when $k$ is lowered further: $\varphi_0 (k) = k \, \phi_0 (k) \propto k$.

Fig.\ \ref{ssb1} displays the exact trajectory \eqref{5.20} for $C_1=+1$; the underlying trajectory $(g_k, \lambda_k)$ of the Einstein--Hilbert truncation was taken to be the type IIIa trajectory with $g_{\text{T}} = 10^{-14}$ at the turning point.
\begin{figure}
\centering
\pgfuseimage{i-a}
\hfill
\pgfuseimage{i-b}
\\
(a) \hfill (b)
\\[1\baselinestretch pt]
\pgfuseimage{i-c}
\hfill
\pgfuseimage{i-d}
\\
(c) \hfill (d)
\\[1\baselinestretch pt]
\pgfuseimage{i-e}
\hfill
\pgfuseimage{i-f}
\\
(e) \hfill (f)
\caption{The trajectory discussed in Subsection \ref{s5.1}. The diagrams a and b (c and d) show the potential for different values of $k$ above (below) $\kT$, as indicated in the schematic diagram e. The dotted (dashed) line is the fixed point (turning point) potential. The plot b (d) zooms into the diagram a (c) at small $\varphi$. The plot f shows the dimensionful minimum $\phi_0$ as a function of $k < \kT$. Dimensionful quantities are in units of $\kT$. The gray level of the various curves changes from black to gray for decreasing values of $k$.}\label{ssb1}
\end{figure}
We picked this tiny value because it leads to a long classical regime, see \cite{h3} and \cite{cosmofrank}. The first four plots in Fig.\ \ref{ssb1} show the potential $- Y_k (\varphi)$ for different values of $k$. The plots (a) and (b) correspond to scales above the turning point, $\kT < k < \infty$, while the diagrams (c) and (d) refer to scales between the turning point and the termination scale, $\kT > k > \kt$, cf. Fig.\ \ref{ssb1}(e). Here and in the following dimensionful quantities are always plotted in units of $\kT$, and the dotted (dashed) line represents the fixed point (turning point) potential. The gray level of the potential curves changes from black to gray for decreasing values of $k$. We always plot \textit{minus} $Y_k (\varphi)$ because it is this function that corresponds to the actual potential $U_k (\phi)$ by a rescaling with positive factors. The plots \ref{ssb1}(b) and (d) are ``zooms'' into the small-$\varphi$ region of the diagrams (a) and (c), respectively. We observe the following pattern which is generic for $C_1 >0$: When $k$ is lowered from ``$k=\infty$'' the potential $- Y_k (\varphi)$ immediately assumes its minimum at a \textit{nonzero} $\varphi$-value. Between $k = \infty$ and $k = \kT$ this minimum moves towards larger $\varphi$-values, while it returns to smaller $\varphi$-values once $k$ has passed $\kT$. It is instructive to look at the latter regime in dimensionful units. Fig.\ \ref{ssb1}(f) displays the position of the minimum $\phi_0 (k) = \varphi_0 (k) / k$ below the turning point. We see that $\phi_0$ is perfectly constant all the way down to the termination scale. (It is at about $\kt \approx 10^{-7}$ for the trajectory chosen.)

It is tempting to interpret this RG trajectory as describing a kind of continuous (``second order'') phase transition with respect to the scale, taking place for $k \to \infty$: At $k$``$=$''$\infty$ the potential $- Y_k (\varphi)$ has its global minimum at $\varphi =0$; for all $k < \infty$ it is located at a nonzero $\varphi = \varphi_0 (k) \neq 0$. Likewise the minimum $\phi_0 (k)$ of $U_k (\phi)$ continuously approaches zero for $k \to \infty$ and is nonzero for all $k < \infty$. The significance of those minima becomes clear when we search for (running) solutions of the (running) effective field equations, i.\,e.\ look for stationary points of $\Gamma_k$.

In fact, the vacuum expectation value of the metric can be found by solving the effective field equation $\delta \Gamma_k / \delta \phi =0$. Therefore, within the LPA, the $k$-dependent metric expectation value is $\langle g_{\mu \nu} \rangle_k = \phi_0 (k)^2 \,\, \h g_{\mu \nu}$. If $\varphi_0=0$ this expectation value vanishes according to the LPA. Hence the phase the system is in at $k$``$=$''$\infty$ is a phase of unbroken diffeomorphism invariance \cite{witten,floper}. In this phase the metric has a vanishing expectation value, with non-trivial fluctuations about $\langle g_{\mu \nu} \rangle_k =0$, but no ``metric condensate'' spontaneously breaks diffeomorphism invariance. For $k<\infty$ a nonzero (and within the present approximation therefore necessarily non-degenerate) metric $\langle g_{\mu \nu} \rangle_k$ spontaneously breaks the diffeomorphism group down to the stability group of the ground state metric. For $\langle g_{\mu \nu} \rangle_k = \phi_0 (k)^2 \,\, \h g_{\mu \nu}$ with $\h g_{\mu \nu}$ the flat metric on $R^4$ this is the (Euclidean) Poincar\'e group.

If $\phi_0$ is approximately $k$-independent over a wide range of $k$-values the ``frozen out'' metric $\langle g_{\mu \nu} \rangle_k = \phi_0^2 \,\, \h g_{\mu \nu}$ defines an approximately classical spacetime. In the above example we saw that this is indeed the case for $k$ between $\kT$ and $\kt$. We also know \cite{h3,cosmofrank} that the classical regime predicted by the Einstein--Hilbert truncation is the longer the closer the turning point is to the GFP. If $\phi_0$ does not depend on $k$ we may introduce units such that $\phi_0=1$ so that, with the $R^4$ reference metric $\h g_{\mu \nu} = \delta_{\mu \nu}$, the expectation value equals $\langle g_{\mu \nu} \rangle_k = \delta_{\mu \nu}$ within the classical range of $k$-values. In this way the emergence of an approximately classical, flat spacetime can be understood as a condensation and symmetry breaking phenomenon similar to those in standard scalar field theories.

While we are familiar with gravity in the classical regime of the broken phase, it seems to be much more the exception than the rule. In the present setting it requires a very small\footnote{If one tentatively applies the discussion to the real Universe one has $g_{\text{T}} \approx 10^{-60}$ \cite{h3,entropy}.} value of $g_{\text{T}}$ and a positive constant $C_1$ in \eqref{5.20}. For $C_1<0$ the global minimum of $- Y_k (\varphi)$ stays always at $\varphi_0 =0$. So gravity is always in its ``symmetric phase'' and the metric develops no expectation value, let alone a scale independent one.
\subsection{First order-type Transitions ($\boldsymbol{R^4}$ case)}\label{s5.2}
\begin{figure}
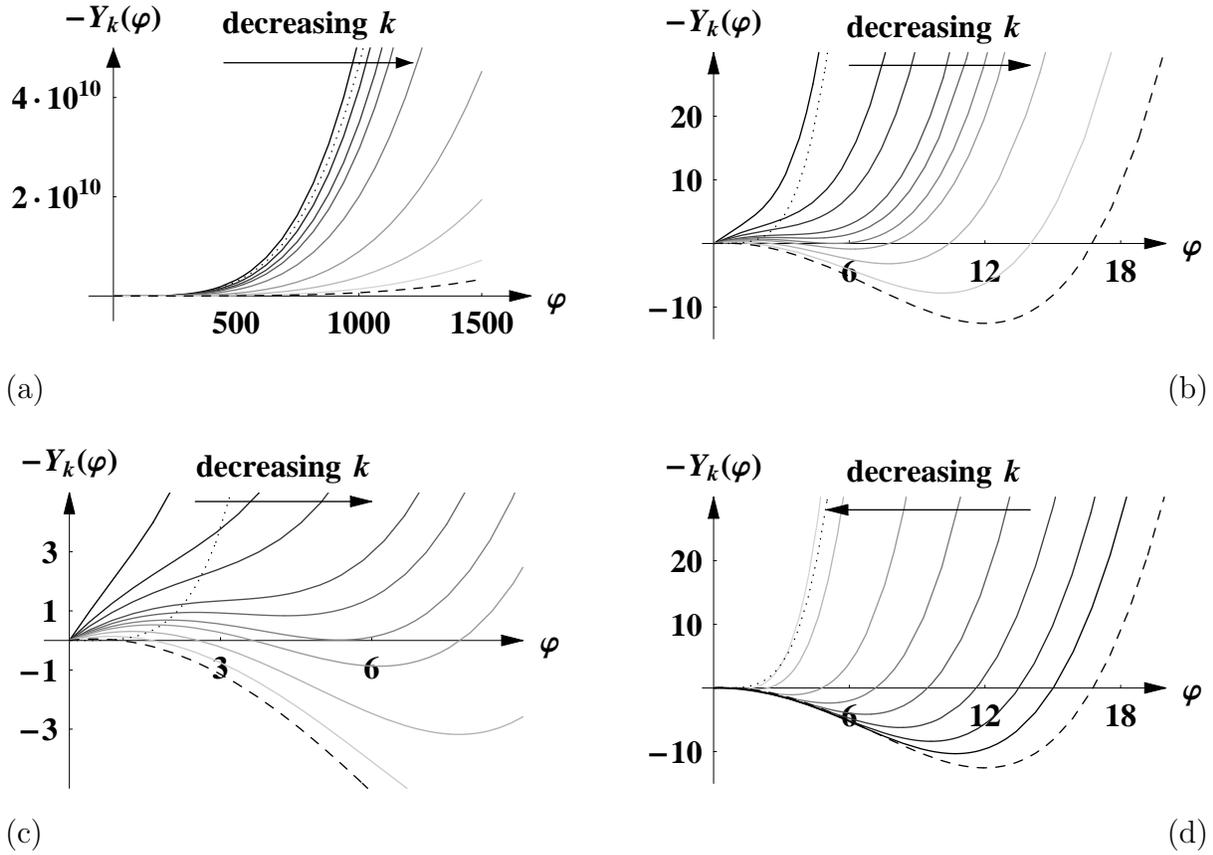

\centering
\pgfuseimage{ii-a}
\hfill
\pgfuseimage{ii-b}
\\
(a) \hfill (b)
\\[1\baselinestretch pt]
\pgfuseimage{ii-c}
\hfill
\pgfuseimage{ii-d}
\\
(c) \hfill (d)
\caption{The trajectory discussed in Subsection \ref{s5.2}. The diagrams a, b, and c show the potential for different values of $k$ above $\kT$, the diagram d for $k$-values below $\kT$. The dotted (dashed) line is the fixed point (turning point) potential. The plots b and c are zooms into the diagram a at smaller $\varphi$ values.}\label{ssb2}
\end{figure}

In Fig.\ \ref{ssb2} we show a trajectory in $\mathscr{S}_{\text{UV}}$, for $R^4$, which was obtained numerically by the trial and error method.
The pertinent Einstein--Hilbert trajectory has $g_{\text{T}} = 10^{-1}$. The initial scale for the upward integration of the partial differential equation was chosen as $\ki = \kT/2$ where $\kT \equiv 1$ in computer units. Assuming that the initial potential $Y_{\ki}$ is a polynomial of the form $\varphi + \varphi^2 + \varphi^4$ we finetuned the coefficients of the $\varphi$- and the $\varphi^2$-term in such a way that $Y_k (\varphi)$ gets close to $Y_\ast (\varphi)$ and stays there for a long RG time. The successful initial potential reads
\begin{align} \label{5.30}
Y_{\ki} (\varphi)
& =
-0.10 \, \varphi + 0.20 \, \varphi^2
- \tfrac{1}{6} \, \lambda_{\ki} \, \varphi^4.
\end{align}
To obtain the complete trajectory we also evolved \eqref{5.30} downward, from $\ki$ to $\kt$. The Figs.\ \ref{ssb2}(a), (b), and (c) show the running potential for various values of $k$ above $\kT$, while Fig.\ \ref{ssb2}(d) shows representative potentials with $k$ below $\kT$. In Fig.\ \ref{ssb2}(a) we see that $Y_k (\varphi)$ approaches the (dotted) fixed point potential for large values of $k$. We verified that the approach to $Y_\ast (\varphi)$ is oscillatory, as a consequence of the spirals in the $g$-$\lambda$--plane. The Figs.\ \ref{ssb2}(b) and (c) show that at a certain value of $k$ the potential first develops an inflection point and then a new local minimum and maximum. At an even lower, ``critical'' scale $k_{\text{c}}$ the new minimum becomes the global one. Above $\kT$ the new global minimum moves towards larger $\varphi$-values, then towards smaller ones, see Fig.\ \ref{ssb2}(d). The latter regime is the classical one again, with an approximately constant $\phi_0 = \varphi_0 / k$. Making $g_{\text{T}}$ smaller one can obtain an arbitrarily long (in $k$) regime with an essentially classical, flat spacetime.

As compared to the analytic trajectory of the previous subsection two features of the phase transition are new here: it happens at a \textit{finite transition scale} $k_{\text{c}} < \infty$, and it is \textit{discontinuous}. For $k > k_{\text{c}}$ the global minimum is at $\phi_0 =0$, at $k = k_{\text{c}}$ it jumps to a value $\phi_0 >0$. In analogy with the phase transitions in standard scalar theories we call this a first order transition.

According to this trajectory the expectation value $\langle g_{\mu\nu} \rangle_k$ vanishes for all $k > k_{\text{c}}$. One might wonder therefore whether the field modes with momenta above $k_{\text{c}}$ are correctly integrated out. The answer is in the affirmative. To see why we recall that a RG trajectory is a running functional $\Gamma_k [\,\cdot\,]$ defined over a certain space of functions $\{ \phi (\,\cdot\,) \}$. The global minimum of $\Gamma_k$ is merely a single point in this space, $\phi_0 (\,\cdot\,)$. The flow equation has the structure ``$k\p_k \, \Gamma_k [\phi (\,\cdot\,)] = \text{RHS} \bigl( \phi (\,\cdot\,) \bigr)$ \textit{for all} $\phi (\,\cdot\,)$'' so the point $\phi_0 (\,\cdot\,)$ plays no special role. Only after having constructed and solved the flow equation which involves arbitrary ``off-shell'' arguments $\phi (\,\cdot\,)$ we can go ``on-shell'', i.\,e.\ look for solutions to the effective field equation $(\delta \Gamma_k / \delta \phi) [\phi_0 (\,\cdot\,)] =0$. Only here it is decided whether the metric has a nonzero expectation value or not.

In the general case of the full functional $\Gamma_k [g_{\mu \nu}, \ov{g}_{\mu\nu}, \cdots]$ there exists a subtlety, however, which is absent in the present setting for conformally reduced QEG. In the original construction of the full gravitational average action \cite{mr} it is assumed that $g_{\mu \nu}$ and $\ov{g}_{\mu \nu}$ are \textit{non-degenerate} because the inverses $g^{\mu \nu}$ and $\ov{g}\,^{\mu \nu}$ are needed there. Hence, at best, $g_{\mu \nu}=0$ and/or $\ov{g}_{\mu \nu}=0$ are singular boundary points of the space $\{ g_{\mu \nu} (\,\cdot\,), \ov{g}_{\mu \nu} (\,\cdot\,), \cdots\}$ over which $\Gamma_k$ is defined. As a result, the original average action approach can describe the symmetric phase of gravity only as a singular limit. (In the full Einstein--Hilbert truncation, for instance, $\langle g_{\mu\nu} \rangle_k \to 0$ for $k \to \infty$.) The problem of generalizing the original construction to a theory space $\big\{ \Gamma [g_{\mu \nu}, \ov{g}_{\mu\nu}, \cdots] \big\}$ where $\Gamma$ is defined for non-invertible $g_{\mu \nu}$ and $\ov{g}_{\mu \nu}$, too, will be addressed elsewhere. Here we only mention that in the present reduced setting this problem does not arise since we may use the reference metric $\h g_{\mu \nu}$ which is always non-degenerate in order to formulate the flow equation.

A similar analysis of spontaneous symmetry breaking but in a perturbatively renormalizable model of quantum gravity has been performed in ref.\ \cite{floper}. The scaling properties of the potential and the average metric are quite different there as the theory is not asymptotically safe. In particular $\langle g_{\mu\nu} \rangle_k$ increases rather than decreases for $k \to \infty$.
\subsection{Phase Transitions in the $\boldsymbol{S^4}$ case}\label{s5.3}
In the previous subsection we observed that in the $R^4$ topology a phase transition \textit{can} occur at finite $k$. It is easy to see that, in the case of $S^4$, a phase transition, if it occurs at all, \textit{must} take place at a finite value of $k$. For $S^4$ the fixed point potential $- Y_\ast (\varphi)$ has the shape plotted in Fig.\ \ref{fixpot-s}. Contrary to the simple $-Y_\ast \propto \varphi^4$ on $R^4$ this function is \textit{structurally stable} in the sense that a smooth infinitesimal deformation of $-Y_\ast$ cannot give rise to a new local or global minimum. This was different for $- Y_\ast \propto \varphi^4$ where an infinitesimal term $\propto \varphi$ can shift the global minimum from $\varphi_0 =0$ to $\varphi_0 \neq 0$.

As for phase transitions on $S^4$, both a first order and a second order scenario are possible. Assume, for instance, we start the downward evolution from the $-Y_\ast (\varphi)$ pictured in Fig.\ \ref{fixpot-s} with an infinitesimal relevant perturbation added. Then it might happen that for a certain period of RG time, for $k_0 < k < \infty$, say, no critical points form at $\varphi>0$, but that the slope of $-Y_k (\varphi)$ at the origin decreases from the originally positive value $-Y_\ast' (0) >0$ to zero at $k=k_0$: $-Y_{k_0}' (0) =0$. If the slope continues to decrease for $k$ below $k_0$ then a new global minimum forms at $\varphi_0 (k) >0$, and $\varphi_0 (k)$ moves away from the origin in a continuous way. This, then, is a ``second order'' phase transition taking place at $k=k_0$. The scale $k_0$ is necessarily finite ($k_0 < \infty$) since the initial slope $-Y_\ast'(0)$ is strictly positive; hence a nonzero running time is needed in order to reduce it to zero. One can find trajectories which indeed realize this scenario.

A possible first order scenario proceeds as follows. During the downward evolution from $-Y_\ast (\varphi)$ at $k$``$=$''$\infty$ first an inflection point forms at some $\varphi_{\text{infl}}>0$ and then a new local minimum and a maximum arise from it. Upon further downward evolution the new minimum becomes the global minimum at a critical scale $k_{\text{c}}$. This behavior corresponds to a discontinuous phase transition; the global minimum suddenly jumps from $\varphi_0=0$ to $\varphi_0 \neq 0$. Also in this scenario the phase transition happens at a finite scale $k_{\text{c}} < \infty$. The reason is that the slope of $-Y_\ast$ is strictly positive everywhere. Hence, again, we need a nonzero running time until $Y_k' (\varphi)$ vanishes at $\varphi_{\text{infl}} \neq 0$. 

\begin{figure}[ht]
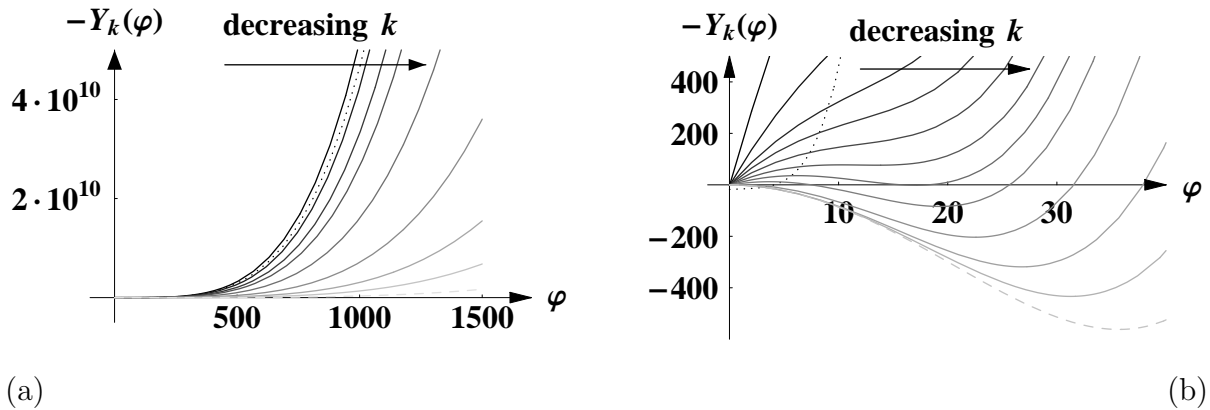

\centering
\pgfuseimage{iii-a}
\hfill
\pgfuseimage{iii-b}
\\
(a) \hfill (b)
\caption{The $S^4$ trajectory discussed in Subsection \ref{s5.3}. The diagram a shows the potential for different values of $k$ above $\kT$. The plot b zooms into diagram a for small $\varphi$ and shows the first order phase transition. The dotted (dashed gray) line is the fixed point (initial) potential. The gray level of the various curves changes from black to gray for decreasing values of $k$.}\label{ssb3}
\end{figure}

In Fig.\ \ref{ssb3} we display a trajectory, found numerically by the trial and error method, which realizes this kind of first order scenario. The respective Einstein--Hilbert trajectory has $g_{\text{T}} = 10^{-3}$. At the initial scale for the upward integration of the partial differential equation, $\ki = 10 \, \kT$, we chose an initial potential $Y_{k_{\text{intl}}}$ of the type $\varphi^0 + \varphi^2 + \varphi^3 + \varphi^4$. After finetuning the $\varphi^0$, $\varphi^2$-, and $\varphi^3$-coefficients so that $Y_k (\varphi)$ gets close to $Y_\ast (\varphi)$ and stays there for a very long RG time we obtained approximately
\begin{align}
Y_{k_{\text{intl}}} (\varphi)
& =
0.3845 + 0.9 \, \varphi^2 - 0.001 \, \varphi^3 
- \tfrac{1}{6} \, \lambda_{k_{\text{intl}}} \, \varphi^4.
\end{align}
As for the smoothing scheme, the spectral functions used in the numerical integration were $\rho (\varphi) = 1 + \tfrac{1}{12} \, \varphi^4$ and $\w \rho (\varphi) = \tfrac{1}{18} \, \varphi^6$.
%
%
%
\section{Summary and Conclusions}\label{s6}
In this paper we explored the ``background independent'' renormalization group flow of the effective average action for QEG on an infinite dimensional theory space. Considering conformally reduced gravity we quantized only the fluctuations of the conformal factor, employing the Local Potential Approximation for its effective average action. The relative simplicity of the resulting system of flow equations, a partial differential equation coupled to an ordinary one, allowed us detailed investigations which at present are prohibitively complicated in the full theory. While at first sight reminiscent of the RG equations for standard scalar theories, the requirement of a ``background independent'' quantization results in crucial differences. In the infinite dimensional space of potential functions we found both a Gaussian and a non-Gaussian fixed point; they generalize the fixed points known from the (conformally reduced) Einstein--Hilbert truncation. The scaling dimensions at the GFP were found to be very different from those in a scalar theory, as a consequence of the built-in ``background independence''. The results on the structure of the RG flow provide further non-trivial evidence for the viability of the asymptotic safety scenario.

We studied in detail the linearized RG flow near the NGFP and we found that for some choices of the space of scaling fields infinitely many relevant directions can arise. The corresponding scaling fields are non-polynomial functions similar to the Halpern--Huang directions at the GFP of conventional scalar field theory. As to yet, their status and implications for the actual predictivity of the theory is not understood, not even in the case of standard scalars. At least in gravity the LPA is not sufficient in order to meaningfully address this question; in particular it is unclear whether the Halpern--Huang-like scaling fields can descend from invariants built from the full metric which occur in the fixed point action.

In QEG the beta functions, by construction, are independent of any specific metric. They do depend on the topology of spacetime, however. We illustrated this point by considering the cases $S^4$ and $R^4$ in parallel.

Analyzing the general properties of the RG flow and in particular of the UV critical manifold associated with the NGFP we saw that on $\mathscr{S}_{\text{UV}}$ the potential $Y_k (\varphi)$ behaves as $\varphi^4$ for $\varphi \to \infty$. Recalling that for $S^4$ every term in $\Gamma_k [g_{\mu \nu}]$ which has the structure $\int \dd^4 x \sqrt{g\,} \, (\text{curvature})^n$ contributes to the $\varphi^{4-2n}$-monomial we can conclude that, in the conformally reduced theory, we see no ``shadows'' of invariants with negative powers of the curvature. Leaving aside the possibility of cancellations among invariants of the same dimension, this result suggests that terms of the form $\int \dd^4 x \sqrt{g\,} \, R^{-n}$, $n=1,2,3,\cdots$, should not occur in RG trajectories on $\mathscr{S}_{\text{UV}}$. In fact, recently Machado and Saueressig \cite{frankmach} have analyzed $3$-parameter truncations where a single term of this type was added to the Einstein--Hilbert truncation. And indeed for moderate $n$ it turned out that on this $3$-dimensional theory space no satisfactory UV fixed point could be found. In the light of the present results the natural explanation which suggests itself is that the trajectories of \cite{frankmach} are not on $\mathscr{S}_{\text{UV}}$ when the non-local invariant is included. The same remark applies to $\int \dd^4 x \sqrt{g\,} \, \ln R$ which produces a term $\varphi^4 \, \ln \varphi$ which likewise grows faster than $\varphi^4$ and, exactly as expected, spoils the NGFP. (The actual motivation in \cite{frankmach} for including these invariants was the hope that they would improve the IR behavior, and this indeed turned out to be the case.)

In the opposite limit $\varphi \to 0$, the NGFP potential $Y_\ast (\varphi)$ was found to be regular. We also saw that after appropriately constraining the scaling fields even the running potential $Y_k (\varphi)$ is regular along all trajectories in $\mathscr{S}_{\text{UV}}$. We can interpret this regularity by saying that $Y_\ast$ contains no ``shadows'' of the invariants $I_n \equiv \int \dd^4 x \sqrt{g\,} \, (\text{curvature})^n \propto \varphi^{4-2n}$ for $n=3,4,\cdots$. The monomials $I_0 \propto \varphi^4$, $I_1 \propto \varphi^2$ and $I_2 = const$ are regular; they represent the cosmological constant, the Einstein--Hilbert term, and the $(\text{curvature})^2$-terms, and they have a nonzero fixed point coupling\footnote{The $I_2$-coupling , the constant piece in $Y_\ast$, is nonzero at least for $S^4$, see Fig.\ \ref{fixpot-s}.}. For the higher ones $I_3 \propto \varphi^{-2}$, $I_4 \propto \varphi^{-4}$, $\cdots$ the fixed point coupling is zero.

Here we see that in the conformally reduced framework the Goroff--Sagnotti term $\int \dd^4 x \sqrt{g\,}\, R_{\mu \nu \alpha \beta} \, R^{\alpha \beta}_{\phantom{\alpha\beta} \rho \sigma} \, R^{\rho \sigma \mu \nu}$ which contributes to $\varphi^{-2}$ does not play any special role for the UV behavior of the theory. Imposing the subsidiary condition $Y\sr =0$ it is even absent all along the trajectories in $\mathscr{S}_{\text{UV}}$. This term is known to spoil the perturbative renormalizability of full quantum gravity at the two-loop level \cite{sagnotti}. According to the general theory \cite{livrev} and well known examples \cite{gakup} perturbative nonrenormalizability can well coexist with asymptotic safety. The present results are fully consistent with this picture.

Solving the flow equation for the potential numerically we obtained examples of RG trajectories in the NGFP's ultraviolet critical hypersurface $\mathscr{S}_{\text{UV}}$. The quantum theories based upon some of them show phase transitions from a phase in which the global minimum of $U_k (\phi)$ is at some $\phi \neq 0$ to a phase where it is at $\phi =0$.
In the former situation the conformal factor has a nonzero expectation value, while it vanishes in the latter. We interpret this as an indication that at some large (but in some cases finite) value of $k$ there occurs a transition from the familiar low energy phase of gravity with a nonzero expectation value of the metric to a new phase in which $g_{\mu \nu} \equiv \langle \gamma_{\mu \nu} \rangle =0 $. In the ``low energy'' phase diffeomorphism invariance is spontaneously broken down to the stability group of the metric condensate $\langle \gamma_{\mu \nu} \rangle \neq 0$, while it is unbroken in the ``high energy'' phase. There also exist trajectories along which the symmetry is always unbroken, i.\,e.\ the metric never develops an expectation value. The corresponding quantum theories have no classical regime.
%
%
%
\pagebreak
\appendix
\noindent {\Large \textbf{APPENDIX}}
\section{Generalizations for $\boldsymbol{d}$ Spacetime Dimensions}\label{AppA}
In this appendix we tabulate the key formulas of the main text generalized for an arbitrary spacetime dimensionality $d$. 

The LPA truncation ansatz generalizing the average action \eqref{3.3} is given by
\begin{align}\label{a.1}
\begin{split}
& \Gamma_k \bigl[ \,\ov{f}; \chib, \h g_{\mu \nu} 
\bigr( S^d (\h r\,) \bigr) \bigr]
\\
& \phantom{{==}} =
- \frac{d-1}{2 \pi \, (d-2)} \, \frac{1}{G_k} \, \int \!\!\dd^d x \, \sqrt{\h g\,}~
\bigg\{ - \tfrac{1}{2} ( \chib + \ov{f}\,) \,
\h \Box \, ( \chib + \ov{f} \,)
\\
& \phantom{{====}- \frac{d-1}{2 \pi \, (d-2)} \, \frac{1}{G_k} \, \int \!\!\dd^d x \, \sqrt{\h g\,}~
\bigg\{ }
+ \h r^{\,-d} \, F_k \bigl( \h r^{1/\nu} \, ( \chib + \ov{f} \, )
\bigr) \bigg\}
\end{split}
\end{align}
where $\nu \equiv 2 / (d-2)$. (Cf.\ ref.\ \cite{creh1}.) The running potential of the CREH truncation reads
\begin{align}\label{a.2}
F_k\bcc (x) &=
c_0 \, \frac{d (d-2)}{8} \, x^{(d-2) \nu}
- \frac{d-2}{4 (d-1)} \, \Lambda_k \, x^{d \nu}.
\end{align}
The cutoff operator $\mathcal{R}_k$ must be chosen as
\begin{align}\label{a.3}
\mathcal{R}_k [\chib; \h g_{\mu \nu}]
& =
- \frac{d-1}{2 \pi \, (d-2)} \, \frac{1}{G_k} \, \chib^{2 \nu} \, k^2 \,
R^{(0)} \bigl( - \tfrac{\h \Box}{\chib^{2\nu} \, k^2\,} \bigr)
\end{align}
so that the cutoff action \eqref{3.11} becomes
\begin{align}\label{a.4}
\Delta_k S [f; \chib, \h g_{\mu \nu}]
& =
- \frac{d-1}{4 \pi \, (d-2)} \, \frac{1}{G_k} \, \chib^{2 \nu} \, k^2 \,
\int \!\! d^d x \, \sqrt{\h g\,}~
f (x) \, R^{(0)} \bigl( - \tfrac{\h \Box}{\chib^{2 \nu} \, k^2\,} \bigr)
\, f(x).
\end{align}
In terms of the new field variables
\begin{align}\label{a.5}
\w \chi_{\text{B}} \equiv \h r\,^{1/\nu} \, \chib, \qquad
\w{\ov{f}} \equiv \h r\,^{1/\nu} \, \ov{f}, \qquad
\w \phi \equiv \h r\,^{1/\nu} \, \phi
\end{align}
the generating equation for the RG flow on $S^d$ reads
\begin{align}\label{a.6}
\begin{split}
&
- \frac{d-1}{4 \pi \, (d-2)} \, 
\int \!\! \text{d}^d x \, \sqrt{\h g\,} ~
\left \{
k \p_k \left( \frac{1}{G_k} \right) \,
\frac{1}{2} \, \ov{f} (x)\, (- \h \Box) \, \ov{f} (x)
+ k \p_k \frac{F_k \bigl( \chib + \ov{f} (x) \bigr)}{G_k}
\right \}
\\
& \phantom{{==}} =
\chib^{2\nu} \, k^2 \, \tr \, \left[ \frac{
\left \{ 1 - \frac{\etan}{2} \right \} \, 
R^{(0)} \left( - \frac{\h \Box}{\chib^{2\nu} \, k^2\,} \right)
- \left( - \frac{\h \Box}{\chib^{2\nu} \, k^2\,} \right) \, 
R^{(0) \prime} \left( - \frac{\h \Box}{\chib^{2\nu} \, k^2\,} \right)}
{- \h \Box
+ \chib^{2\nu} \, k^2 \, R^{(0)}
\left( - \frac{\h \Box}{\chib^{2\nu} \, k^2\,} \right)
+ F_k'' \bigl( \chib + \ov{f} (x) \bigr)}
\right].
\end{split}
\end{align}
Eq.\ \eqref{a.6} reduces to \eqref{3.21} in $4$ dimensions.

The negative eigenvalues and degeneracies of the Laplace--Beltrami operator on a unit $d$-sphere \cite{rubin} are, respectively,
\begin{gather} \label{a.7}
\mathcal{E}_{n;d} = n \, (n+d-1)
\quad \text{and} \quad
D_{n;d} = \frac{(2n+d-1) \, (n+d-2)!}{n! \, (d-1)!}.
\end{gather}
The spectral functions of \eqref{3.30} generalize to $\rho_d (\varphi) \equiv \sum_{n=0}^\infty D_{n;d} \, \theta (\varphi^{2\nu} - \mathcal{E}_{n;d}) = J_d \bigl( n_{\text{max}} (\varphi;d) \bigr)$ and $\w \rho_d (\varphi) \equiv \sum_{n=0}^\infty \mathcal{E}_{n;d} \, D_{n;d} \, \theta (\varphi^{2\nu} - \mathcal{E}_{n;d}) = \w J_d \bigl( n_{\text{max}} (\varphi;d) \bigr)$ with the sums
\begin{align} \label{a.8}
\begin{split}
J_d (N)
& =
\frac{1}{(d-1)!} \, \sum_{n=0}^N
\frac{(2n+d-1) \, (n+d-2)!}{n!}
\\
\w J_d (N)
& =
\frac{1}{(d-1)!} \, \sum_{n=0}^N
\frac{(n+d-1) \, (2n+d-1) \, (n+d-2)!}{(n-1)!}
\end{split}
\end{align}
where $N \equiv n_{\text{max}} (\varphi;d)$ is the largest positive integer satisfying $\mathcal{E}_{N;d} = N (N+d-1) < \varphi^{2 \nu}$. For $\varphi \gg 1$ we may approximate $\rho_d (\varphi) \approx \frac{2}{d!} \, \varphi^{d \nu}$ and $\w \rho_d (\varphi) \approx \tfrac{2d(d + 1)}{(d+2)!} \, \varphi^{(d+2) \nu}$ which boils down to \eqref{3.34} for $4$ dimensions. 

The partial differential equations for the potential $F_k (\phi)$ and its dimensionless analog $Y_k (\varphi) \equiv k^{d-2} \, F_k (\varphi / k^{1/\nu})$ read in arbitrary dimensions, respectively,
\begin{subequations} \label{a.9}
\begin{align} \label{a.9a}
\begin{split}
k\p_k \,  F_k (\phi)- \etan \, F_k (\phi)
& =
- \frac{2 \pi (d-2)}{(d-1) \, \sigma_d} \, \, G_k
\frac{
\left \{ 1 - \frac{\etan}{2} \right \} \, k^2 \phi^{2 \nu} \, 
\rho_d (k^{1/\nu} \phi) 
+ \frac{\etan}{2} \, \w \rho_d (k^{1/\nu} \phi)}
{k^2 \phi^{2 \nu} + F_k'' (\phi)}
\end{split}
\end{align}
and, with $\varphi \equiv k^{1/\nu} \, \phi$,
\begin{align} \label{a.9b}
\begin{split}
& k\p_k \, Y_k (\varphi)
+ \left( 2-d-\etan \right) \, Y_k (\varphi)
+ \tfrac{1}{\nu} \, \varphi \, Y_k' (\varphi)
\\
& \phantom{{==}} =
- \frac{2 \pi (d-2)}{(d-1) \, \sigma_d} \, \, g_k \,
\frac{
\left \{ 1 - \frac{\etan}{2} \right \} \, \varphi^{2 \nu} \, 
\rho_d (\varphi) 
+ \frac{\etan}{2} \, \w \rho_d (\varphi)}
{\varphi^{2 \nu} + Y_k'' (\varphi)}.
\end{split}
\end{align}
\end{subequations}
Their counterparts in $4$ dimensions are given by eqs.\ \eqref{3.28} and \eqref{3.39}, respectively. Here, $\sigma_d \equiv 2 \pi^{(d+1)/2} / \Gamma \bigl( (d+1)/2 \bigr)$.

With the optimized shape function the anomalous dimension $\etan$ for the $R^d$ topology assumes the form
\begin{align}\label{a.10}
\etan
& =
- \frac{8 \pi \, v_d \, (d-2)}{d (d-1)} \, \, G_k \,
F_k''' (\phi_1)^2 \, \frac{k^{d+2} \, \phi_1^{(d+2) \nu}}{\left[ k^2 \phi_1^{2 \nu}
+ F_k'' (\phi_1) \right]^4\,}
\end{align}
which reduces to \eqref{3.49} in $4$ dimensions. In terms of the dimensionless quantities $g_k$ and $Y_k$ the equation \eqref{a.10} reads, with $\varphi_1 = k^{1/\nu} \, \phi_1$,
\begin{align} \label{a.11}
\etan \bigl( g_k, [Y_k] \bigr)
& =
- \frac{8 \pi \, v_d \, (d-2)}{d (d-1)} \, \, g_k \, Y_k''' (\varphi_1)^2 \, \frac{\varphi_1^{(d+2) \nu}}{\left[ \varphi_1^{2 \nu}
+ Y_k'' (\varphi_1) \right]^4\,}
\end{align}
with the usual abbreviation $v_d \equiv \bigl[ 2^{d+1} \, \pi^{d/2} \, \Gamma (d/2) \bigr]^{-1}$.
%
%
%
%
%
%
%
\pagebreak


\begin{thebibliography}{99}
\bibitem{kiefer}
For a general introduction see C.~Kiefer, \textit{Quantum Gravity}, Second Edition, \\
Oxford Science Publications, Oxford (2007).
%
\bibitem{bron}
L.~Rosenfeld, 
Ann.\ der Physik 5 (1930) 113; Z.\ f\"{u}r Physik 65 (1930) 589; \\
M.~Bronstein, 
Phys.\ Zeitschrift der Sowjetunion 9 (1936) 140.
%
\bibitem{A}
A.~Ashtekar, 
\textit{Lectures on non-perturbative canonical gravity},\\
World Scientific, Singapore (1991); \\
A.~Ashtekar and J.~Lewandowski, 
Class.\ Quant.\ Grav.\ 21 (2004) R53.
%
\bibitem{R}
C.~Rovelli, 
\textit{Quantum Gravity}, 
Cambridge University Press, Cambridge (2004).
%
\bibitem{T}
Th.~Thiemann, \textit{Modern Canonical Quantum General Relativity},\\
Cambridge University Press, Cambridge (2007).
%
\bibitem{wein}
S.~Weinberg 
in \textit{General Relativity, an Einstein Centenary Survey},\\
S.W.~Hawking and W.~Israel (Eds.), 
Cambridge University Press (1979);\\
S.~Weinberg,
hep-th/9702027.
%
\bibitem{mr}
M.~Reuter, Phys.\ Rev.\ D 57 (1998) 971 and
hep-th/9605030.
%
\bibitem{percadou}
D.~Dou and R.~Percacci, 
Class.\ Quant.\ Grav.\ 15 (1998) 3449.
%
\bibitem{oliver1}
O.~Lauscher and M.~Reuter, 
Phys.\ Rev.\ D 65 (2002) 025013 and hep-th/0108040.
%
\bibitem{frank1}
M.~Reuter and F.~Saueressig, 
Phys.\ Rev.\ D 65 (2002) 065016 and hep-th/0110054.
%
\bibitem{oliver2}
O.~Lauscher and M.~Reuter, 
Phys.\ Rev.\ D 66 (2002) 025026 and hep-th/0205062.
%
\bibitem{oliver3}
O.~Lauscher and M.~Reuter, 
Class.\ Quant.\ Grav.\ 19 (2002) 483 and hep-th/0110021.
%
\bibitem{oliver4}
O.~Lauscher and M.~Reuter, 
Int.\ J.\ Mod.\ Phys.\ A 17 (2002) 993 and hep-th/0112089.
%
\bibitem{souma}
W.~Souma,
Prog.\ Theor.\ Phys.\ 102 (1999) 181.
%
\bibitem{frank2}
M.~Reuter and F.~Saueressig, 
Phys.\ Rev.\ D 66 (2002) 125001 and hep-th/0206145;
Fortschr.\ Phys.\ 52 (2004) 650 and hep-th/0311056.
%
\bibitem{prop}
A.~Bonanno and M.~Reuter,
JHEP 02 (2005) 035 and hep-th/0410191.
%
\bibitem{oliverbook}
For reviews on QEG see: 
M.~Reuter and F.~Saueressig, arXiv:0708.1317 [hep-th]; \\
O.~Lauscher and M.~Reuter in \textit{Quantum Gravity}, B.~Fauser, \\
J.~Tolksdorf and E.~Zeidler (Eds.), Birkh\"auser, Basel (2007) and hep-th/0511260;\\
O.~Lauscher and M.~Reuter in \textit{Approaches to Fundamental Physics}, \\
I.-O.~Stamatescu and E.~Seiler (Eds.), Springer, Berlin (2007).
%
\bibitem{perper1}
R.~Percacci and D.~Perini,
Phys.\ Rev.\ D 67 (2003) 081503;\\
Phys.\ Rev.\ D 68 (2003) 044018;
Class.\ Quant.\ Grav.\ 21 (2004) 5035.
%
\bibitem{codello}
A.~Codello and R.~Percacci,
Phys.\ Rev.\ Lett.\ 97 (2006) 221301;\\
A.~Codello, R.~Percacci and C.~Rahmede,
Int.\ J.\ Mod.\ Phys. A23 (2008) 143.
%
\bibitem{litimgrav}
D.~Litim,
Phys.\ Rev.\ Lett.\ 92 (2004) 201301;
AIP Conf.\ Proc.\ 841 (2006) 322;\\
P.~Fischer and D.~Litim, 
Phys.\ Lett.\ B 638 (2006) 497;\\
AIP Conf.\ Proc.\ 861 (2006) 336.
%
\bibitem{frankmach}
P.~Machado and F.~Saueressig,
Phys.\ Rev.\ D 77 (2008) 124045.
%
\bibitem{creh1}
M.~Reuter and H.~Weyer,
Phys.\ Rev.\ D 79 (2009) 105005 and 
arXiv:0801.3287 [hep-th].
%
\bibitem{oliverfrac}
O.~Lauscher and M.~Reuter,
JHEP 10 (2005) 050 and hep-th/0508202.
%
\bibitem{jan1}
M.~Reuter and J.-M.~Schwindt,
JHEP 01 (2006) 070 and hep-th/0511021.
%
\bibitem{jan2}
M.~Reuter and J.-M.~Schwindt,
JHEP 01 (2007) 049 and hep-th/0611294.
%
\bibitem{je1}
J.-E.~Daum and M.~Reuter,
preprint arXiv:0806.3907 [hep-th].
%
\bibitem{neuge}
F.~Neugebohrn,
arXiv:0704.3205 [hep-th]
%
\bibitem{max}
P.~Forg\'acs and M.~Niedermaier, 
hep-th/0207028; \\
M.~Niedermaier,
JHEP 12 (2002) 066;
Nucl.\ Phys.\ B 673 (2003) 131;\\
Class.\ Quant.\ Grav.\ 24 (2007) R171.
%
\bibitem{livrev}
For detailed reviews of asymptotic safety in gravity see:\\
M.~Niedermaier and M.~Reuter,
Living Reviews in Relativity 9 (2006) 5;\\
R.~Percacci, arXiv:0709.3851 [hep-th].
%
\bibitem{back}
L.F.~Abbott,
Nucl.\ Phys.\ B 185 (1981) 189;\\
B.S.~DeWitt, 
Phys.\ Rev.\ 162 (1967) 1195;\\
M.T.~Grisaru, P.\ van Nieuwenhuizen and C.C.~Wu,
Phys.\ Rev.\ D 12 (1975) 3203;\\
D.M.~Capper, J.J.~Dulwich and M.\ Ramon Medrano,
Nucl.\ Phys.\ B 254 (1985) 737;\\
S.L.~Adler, 
Rev.\ Mod.\ Phys.\ 54 (1982) 729.
%
\bibitem{avact}
C.~Wetterich,
Phys.\ Lett.\ B 301 (1993) 90.
%
\bibitem{ym}
M.~Reuter and C.~Wetterich, \\
Nucl.\ Phys.\ B 417 (1994) 181,
Nucl.\ Phys.\ B 427 (1994) 291, \\
Nucl.\ Phys.\ B 391 (1993) 147, 
Nucl.\ Phys.\ B 408 (1993) 91; \\
M.~Reuter, 
Phys.\ Rev. D 53 (1996) 4430, 
Mod.\ Phys.\ Lett.\ A 12 (1997) 2777.
%
\bibitem{avactrev}
J.~Berges, N.~Tetradis and C.~Wetterich,
Phys.\ Rep.\ 363 (2002) 223;\\
C.~Wetterich,
Int.\ J.\ Mod.\ Phys.\ A 16 (2001) 1951.
%
\bibitem{ymrev}
For reviews of the effective average action in Yang--Mills theory see:\\
M.~Reuter, hep-th/9602012;
J.~Pawlowski, Ann. Phys. 322 (2007) 2831;\\
H.~Gies, hep-ph/0611146.
%
\bibitem{bh}
A.~Bonanno and M.~Reuter, 
Phys.\ Rev.\ D 62 (2000) 043008 and hep-th/0002196;
Phys.\ Rev.\ D 73 (2006) 083005 and hep-th/0602159;\\
Phys.\ Rev.\ D 60 (1999) 084011 and gr-qc/9811026.
%
\bibitem{erick1}
M.~Reuter and E.~Tuiran,
hep-th/0612037.
%
\bibitem{cosmo1}
A.~Bonanno and M.~Reuter,
Phys.\ Rev.\ D 65 (2002) 043508 and hep-th/0106133.
%
\bibitem{cosmofrank}
M.~Reuter and F.~Saueressig, 
JCAP 09 (2005) 012 and hep-th/0507167.
%
\bibitem{cosmo2}
A.~Bonanno and M.~Reuter,
Phys.\ Lett.\ B 527 (2002) 9 and astro-ph/0106468; \\
Int.\ J.\ Mod.\ Phys.\ D 13 (2004) 107 and astro-ph/0210472;\\
E.~Bentivegna, A.~Bonanno and M.~Reuter,\\
JCAP 01 (2004) 001 and astro-ph/0303150.
%
\bibitem{entropy}
A.~Bonanno and M.~Reuter,
JCAP 08 (2007) 024 and arXiv:0706.0174 [hep-th].
%
\bibitem{esposito}
A.~Bonanno, G.~Esposito and C.~Rubano,
Gen.\ Rel.\ Grav.\ 35 (2003) 1899;\\
Class.\ Quant.\ Grav.\ 21 (2004) 5005;\\
A.~Bonanno, G.~Esposito, C.~Rubano and P.~Scudellaro,\\
Class.\ Quant.\ Grav.\ 23 (2006) 3103 and 24 (2007) 1443.
%
\bibitem{h1}
M.~Reuter and H.~Weyer, 
Phys.\ Rev.\ D 69 (2004) 104022
and hep-th/0311196.
%
\bibitem{h2}
M.~Reuter and H.~Weyer,
Phys.\ Rev.\ D 70 (2004) 124028
and hep-th/0410117.
%
\bibitem{h3}
M.~Reuter and H.~Weyer,
JCAP 12 (2004) 001 and hep-th/0410119.
%
\bibitem{girelli}
F.~Girelli, S.~Liberati, R.~Percacci and C.~Rahmede,\\
Class.\ Quant.\ Grav.\ 24 (2007) 3995.
%
\bibitem{litim}
D.~Litim and T.~Plehn,
Phys.\ Rev.\ Lett.\ 100 (2008) 131301.
%
\bibitem{mof}
J.~Moffat,
JCAP 05 (2005) 2003; \\
J.R.~Brownstein and J.~Moffat, 
Astrophys.\ J.\ 636 (2006) 721;\\
Mon.\ Not.\ Roy.\ Astron.\ Soc.\ 367 (2006) 527.
%
\bibitem{ajl1}
J.~Ambj\o{}rn, J.~Jurkiewicz and R.~Loll,
Phys.\ Rev.\ Lett.\ 93 (2004) 131301.
%
\bibitem{ajl2}
J.~Ambj\o{}rn, J.~Jurkiewicz and R.~Loll,
Phys.\ Lett.\ B 607 (2005) 205.
%
\bibitem{ajl34}
J.~Ambj\o{}rn, J.~Jurkiewicz and R.~Loll,
Phys.\ Rev.\ Lett.\ 95 (2005) 171301;\\
Phys.\ Rev.\ D 72 (2005) 064014;
Contemp.\ Phys.\ 47 (2006) 103.
%
\bibitem{polyakov}
A.M.~Polyakov,
Yad.\ Fiz.\ 64 (2001) 594 \\
$[$English Translation: Phys.\ Atom.\ Nucl.\ 64 (2001) 540$]$.
%
\bibitem{jackiw}
R.~Jackiw, C.~N\'u\~nez and S.-Y.~Pi,
Phys.\ Lett.\ A 347 (2005) 47.
%
\bibitem{floper}
R.~Floreanini and R.~Percacci,
Nucl.\ Phys.\ B 436 (1995) 141;\\
Phys.\ Rev.\ D 46 (1992) 1566.
%
\bibitem{syman}
K.~Symanzik,
Nuovo Cim.\ Lett.\ 6 (1973) 77.
%
\bibitem{hist}
For a historic account see: 
G.\ 't Hooft,
Nucl.\ Phys.\ B 254 (1985) 11.
%
\bibitem{kincond}
O.~Lauscher, M.~Reuter and C.~Wetterich,\\
Phys.\ Rev.\ D 62 (2000) 125021 and hep-th/0006099.
%
\bibitem{narpad}
J.V.~Narlikar and T.~Padmanabhan,\\
\textit{Gravity, Gauge Theories and Quantum Cosmology},
D.~Reidel, Dordrecht (1986), Chapter 12 and references therein.
%
\bibitem{sagnotti}
M.H.~Goroff and A.~Sagnotti,
Phys.\ Lett.\ B 160 (1985) 81; \\
A.E.M.~van de Ven, 
Nucl.\ Phys.\ B 378 (1992) 309.
%
\bibitem{witten}
E.~Witten,
Nucl.\ Phys.\ B 311 (1988) 46.
%
\bibitem{rubin}
M.A.~Rubin and C.R.~Ord\'o\~nez,
J.\ Math.\ Phys.\ 25 (1984) 2888;
26 (1985) 65.
%
\bibitem{opt}
D.~Litim,
Phys.\ Lett.\ B 486 (2000) 92;
Phys.\ Rev.\ D 64 (2001) 105007;\\
Int.\ J.\ Mod.\ Phys.\ A 16 (2001) 2081.
%
\bibitem{hh1}
K.~Halpern and K.~Huang,
Phys.\ Rev.\ Lett.\ 74 (1995) 3526;\\
Phys.\ Rev\ D 53 (1996) 3252.
%
\bibitem{hh2}
A.~Bonanno,
Phys.\ Rev.\ D 62 (2000) 027701; \\
H.~Gies,
Phys.\ Rev.\ D 63 (2001) 065011.
%
\bibitem{morris-hh}
T.R.~Morris,
Phys.\ Rev.\ Lett.\ 77 (1996) 1658;
Phys.\ Lett.\ B 334 (1994) 355.
%
\bibitem{gakup}
G.~Parisi,
Nucl.\ Phys,\ B 100 (1975) 368;
Nucl.\ Phys,\ B 254 (1985) 58;
\\
K.~Gawedzki and A.~Kupiainen,
Nucl.\ Phys.\ B 262 (1985) 33;\\
Phys.\ Rev.\ Lett.\ 54 (1985) 2191;
Phys.\ Rev.\ Lett.\ 55 (1985) 363; \\
B.~Rosenstein, B.J.~Warr and S.H.~Park,
Phys.\ Rept.\ 205 (1991) 59; \\
C.~de Calan, P.A.~Faria da Veiga, L.~Magnen and R.~S\'en\'eor,\\
Phys.\ Rev.\ Lett.\ 66 (1991) 3233.
%
%
\end{thebibliography}
\end{document}